\documentclass[a4paper]{article}
\usepackage[a4paper,margin=25mm]{geometry}    
\usepackage{setspace} 
\usepackage{amsfonts,amssymb,amsmath}         
\usepackage{graphicx}                         
\graphicspath{{./figs/}}
\pdfpageattr{/Group << /S /Transparency /I true /CS /DeviceRGB>>}
\usepackage[colorlinks=true, allcolors=black]{hyperref}
\usepackage{hhline}
\usepackage{subcaption}
\let\oldbibliography\thebibliography
\renewcommand{\thebibliography}[1]{%
  \oldbibliography{#1}%
  \setlength{\itemsep}{0pt}%
}
\DeclareMathOperator*{\argmin}{arg\,min}
\usepackage[toc]{multitoc}
\newcommand\numberthis{\addtocounter{equation}{1}\tag{\theequation}}
\title{\vspace{-50pt}\huge\bfseries Developing and Implementing a CubeSat's Equations of Motion}

\author{Liam Wheen \\ Supervised by Dr O.\ Benjamin}

\date{\today}
\begin{document}
\begin{titlepage}
\maketitle
\hrule
\vspace{-10pt}
\begin{abstract}
\noindent
As part of the Bristol PROVE mission, a nano satellite in low Earth orbit will
be required to track a ground based target during a 400 second flyover. This
requires agile attitude control that will be achieved using a system of
flywheels. To calculate the necessary torque from these flywheels, 
a controller was designed. Using newly derived equations of motion for
the system, an expression to optimise the gains was produced. With this
controller, simulations were run to evaluate the largest causes of
error in target pointing. Disturbance torques were safely handled by the
controller, but led to a 12\% increase in wheel speeds, reaching 8325 rpm. 
This higher speed led to an increased gyroscopic torque, reaching $10^{-7}$\,Nm
in the worst case. However since the flywheels can deliver
$10^{-5}$\,Nm of torque, the controller could also correct for this.
Hardware performance was then varied to assess the effect of each component
on pointing accuracy. Attitude sensor noise was found to increase pointing error by
1.9$^\circ$ in the worst case. Minimum performance requirements were then 
determined for each component in order to maintain an acceptable pointing
accuracy.
\end{abstract}
\hrule
\end{titlepage}
\newpage
\pagenumbering{gobble}
\tableofcontents
\newpage
\pagenumbering{arabic}
\section{Introduction}

\subsection{Project Outline}
This project aims to extend the previous work done on attitude control within
the \textit{Pointable Radiometer for Observation of Volcanic Emissions} (PROVE)
mission run within the University of Bristol. This mission involves sending a small
satellite, known as a CubeSat, into low Earth orbit. The CubesSat will have an
infrared camera fixed to its base that must image a ground target whilst
passing above. This requires the satellite to rotate in a sharp arc about the
ground target, meaning that attitude control must be particularly accurate for
this mission. The attitude will predominantly be controlled by a system of
flywheels which are accelerated to generate torque within the CubeSat.

The key elements that are focussed on in this project are
\begin{itemize}
  \item Accurately representing the full dynamics of the rotational system
  \item Understanding the gyroscopic effects of the rotating flywheels
  \item The potential advantage of adding a fourth flywheel
  \item A control law that improves on previous work
  \item An analytical approach to optimise controller gains
  \item A robust method of testing the controller with simulated noise
\end{itemize}

To accurately represent the full dynamics of the satellite during attitude
control, a more comprehensive derivation of its equations of motion will be
necessary. So far within the PROVE mission, the motion of the satellite has
been described using Euler's equation of motion for a rigid body;
\[
  \mathbf{I} \dot{\boldsymbol{\omega}} = 
  \boldsymbol{\tau}_{\mathrm{Tot}} -
  \boldsymbol{\omega}\times\mathbf{I}
  \boldsymbol{\omega},\numberthis \label{eq:euler}
\]
where $\mathbf{I}$ is the satellite's moment of inertia tensor,
$\boldsymbol{\omega}$ is its angular velocity, and
$\boldsymbol{\tau}_{\mathrm{Tot}}$ is the net torque acting on it. This would
include the torque generated by flywheels within the satellite.

This equation treats the satellite as a single rigid body.
This is not accurate as there will be independently spinning flywheels
within the body contributing to the angular momentum of the system. The
omission of these terms thus far has meant that gyroscopic effects within the
satellite could not be quantified. Hence (\ref{eq:euler}) will be
re-derived, now with consideration for the independently rotating flywheels,
allowing for the gyroscopic interactions to be analysed. This will give an
approximate magnitude of the unintentional torques that will arise from non
parallel axes of rotation in the flywheels and the satellite as a whole.

The number of flywheels to use for the PROVE mission CubeSat has not yet
been settled upon. Hence the potential benefit of a four wheel configuration
will be assessed. The orientation and positions of the flywheels within the
satellite are also subject to change. For this reason the derivation outlined
in Section \ref{sec:eqs_of_motion} remains as general as possible to remain
applicable for any outcome.

Controllers in previous projects used error vectors for
orientation and angular velocity to calculate a corrective torque
vector that would be generated by the flywheels. This is known as a
proportional-derivative controller as it responds to both the attitude error and
the rate that it is changing. The addition of an integrated term is considered in this
project, as the increase in data given to the controller, allowing it to
remember previous errors, could improve its performance.

With this new controller, a method by which to automatically tune the gains
based on system parameters is explored. Because the inertia values and
computational capabilities of the controller are not yet known, these must be
included in the expression for controller gains. This could allow the
work done in this project to remain relevant despite changes to the satellite's
properties.

The controller will then be assessed with simulated noise added to the input
sensors. This will provide safe margins of operation for the CubeSat whilst
gaining an understanding of the most significant factors that will affect its
pointing accuracy.

\subsection{Bristol PROVE Mission}
The Eyjafjallajökull volcano erupted in 2010, resulting in an ash cloud that
covered most of Northern Europe, causing a complete air-traffic 
shutdown for over a week. This grounded over 10 million people
and cost an estimated \pounds 130 million per day to the airline industry
\cite{volcano_cost}. 

Understanding the properties of these ash clouds and where they are safe to
pass through will benefit both airlines and consumers in the event of another
eruption. Bristol has been working on the PROVE mission since 2015. The mission
aims to attain infrared images of the ash cloud from different angles as the
CubeSat passes over it.

\begin{figure}
\begin{subfigure}{.5\textwidth}
  \centering
  \includegraphics[width=0.62\textwidth]{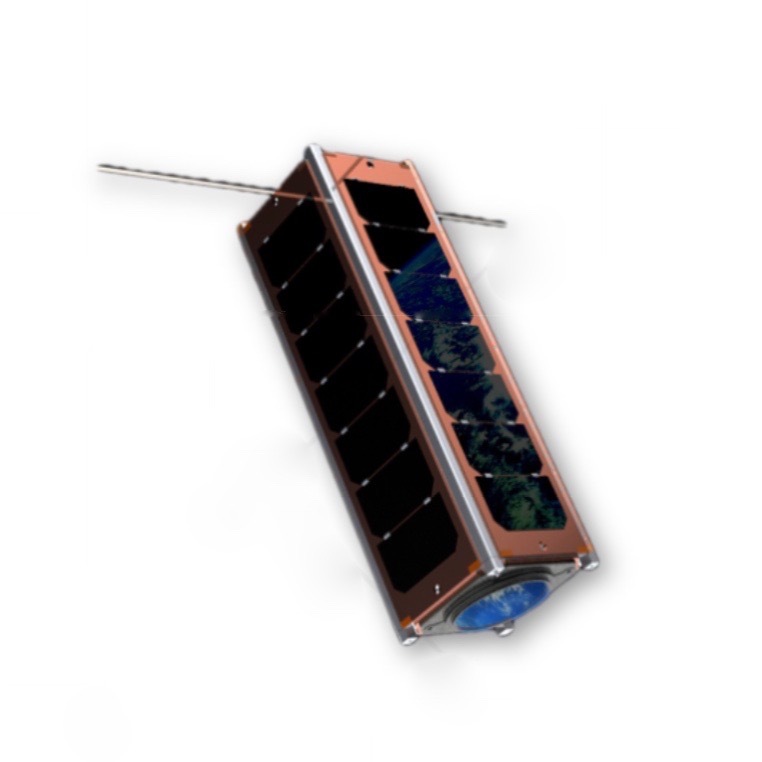}
  \caption{Artistic rendering of a 3 unit CubeSat \cite{cube_pic}.}
  \label{fig:cube_pic}
\end{subfigure}%
\begin{subfigure}{.5\textwidth}
  \centering
  \includegraphics[width=0.91\textwidth]{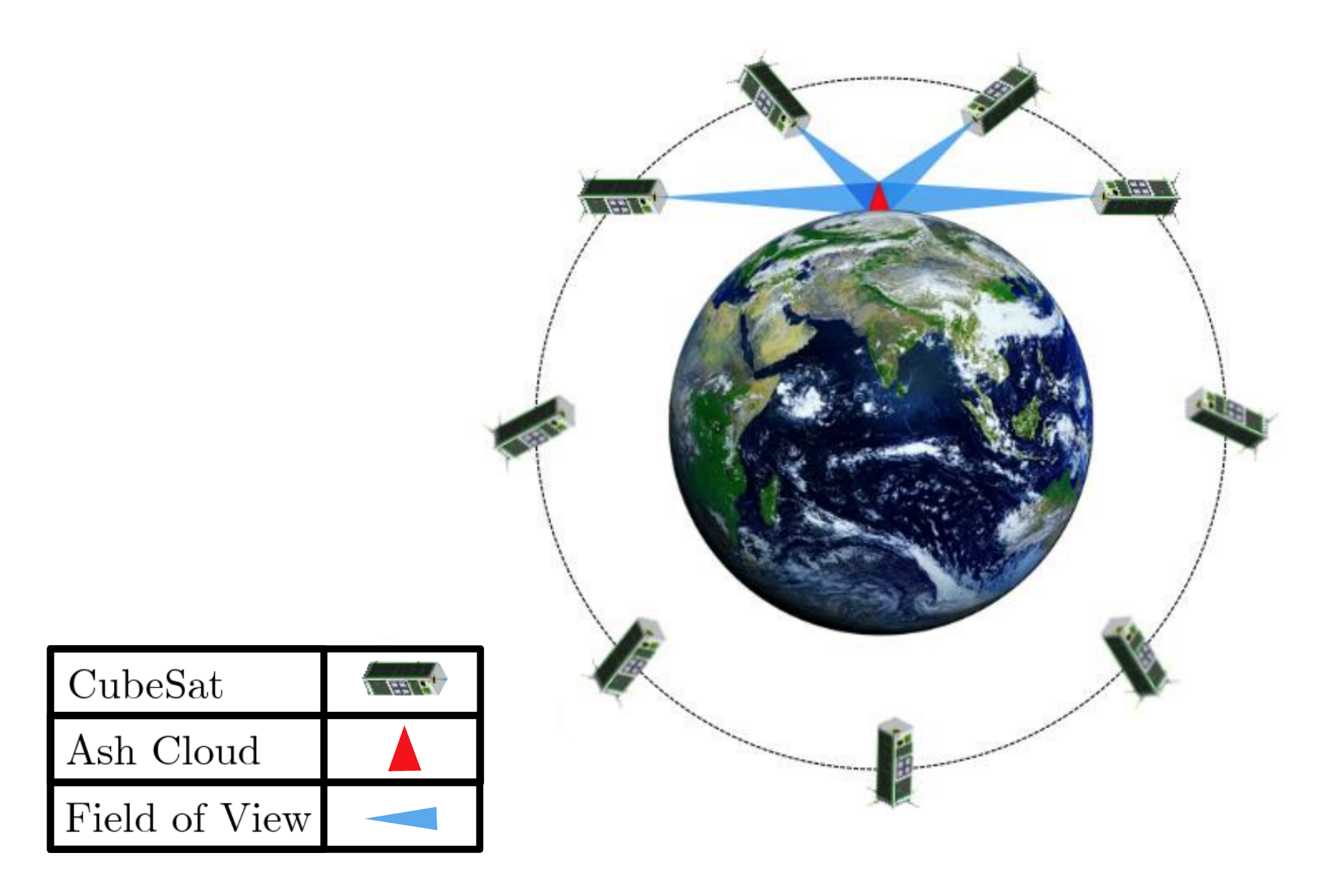}
  \caption{Intended operation of PROVE mission CubeSat.}
  \label{fig:orbit_diagram}
\end{subfigure}%
\caption{Two perspectives of the CubeSat in operation.}
\label{fig:cubesats}
\end{figure}

A CubeSat is formed of cubic units that are $10$cm$\times 
10$cm$\times10$cm in size as shown in Figure \ref{fig:cube_pic}.
This satellite will use a thermal infrared camera, sending the images back to
the PROVE mission lab for processing. With this data, a tomographic 
view of a volcano's ash cloud can be generated. This is a series of cross sectional images
through the vertical axis of the cloud, similar to those attained from an MRI
scan. During its flyover of the volcano, the satellite will pass approximately
300\,km above the ash cloud. By rotating itself during this flyover, it can maintain a
constant alignment with the ash cloud, providing a range of angles from which
to capture it. A diagram of this manoeuvre
is shown in Figure \ref{fig:orbit_diagram}. Due to the low altitude of the
satellite, it will be able to see the ash cloud from a side view as well as
above. Adding this third dimension to the images would provide a significant
improvement to images taken from directly above by larger satellites. 
This could allow planes to be rerouted through areas of low ash concentration,
avoiding the complete shutdown of a large airspace. 

\begin{figure}
  \centering
  \includegraphics[width=0.7\textwidth]{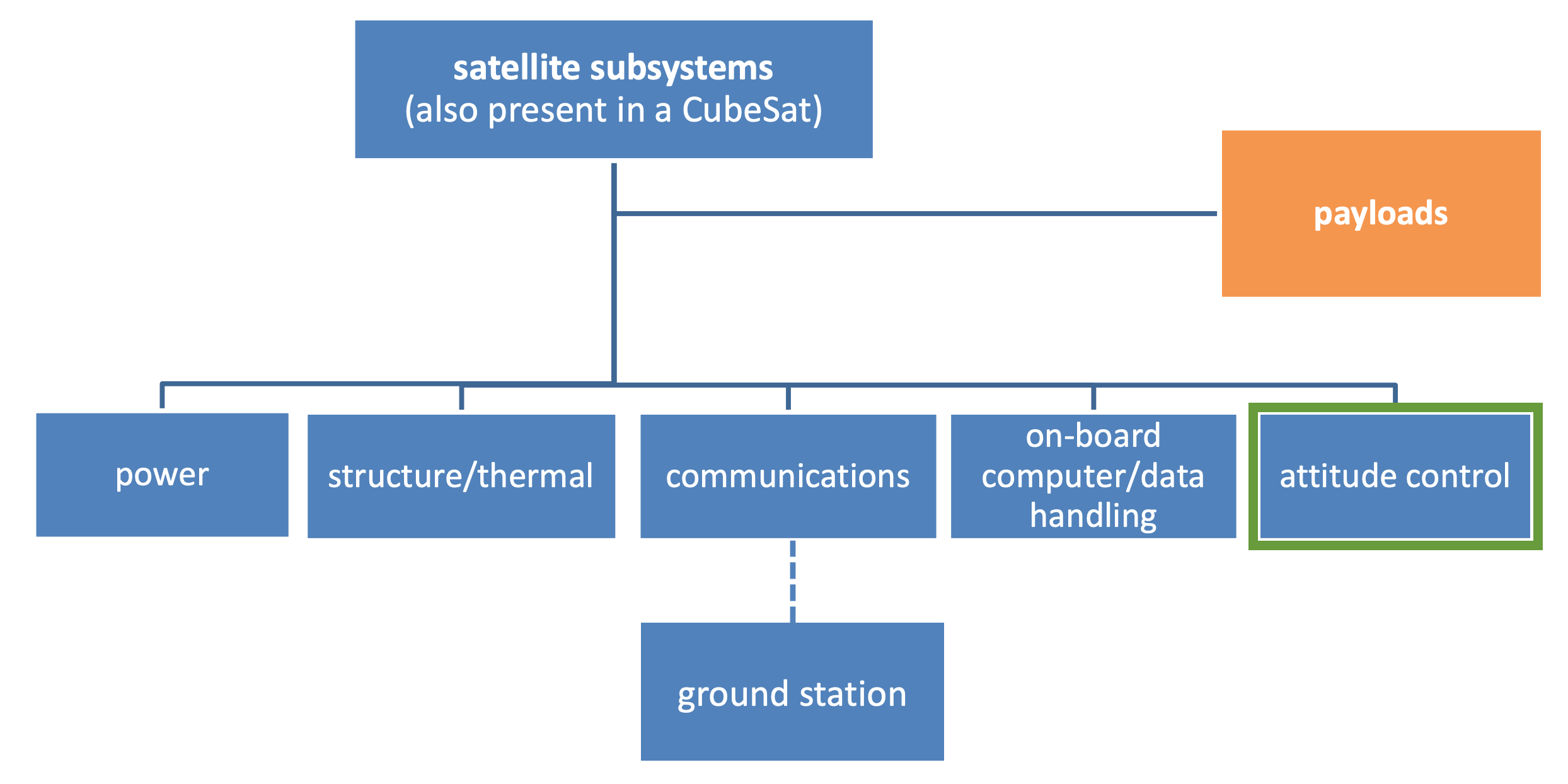}
  \caption{Subsystem structure for a generic CubeSat \cite{cubesatshop}.}
  \label{fig:cube_components}
\end{figure}
The subsystems that are required for the PROVE mission are outlined in Figure 
\ref{fig:cube_components}. This project is focussed on the attitude
control component as highlighted.

Thus far, work has been done on each one of these subsystems, as well as preliminary
work on implementing multiple subsystems together through the use of a
`FlatSat'. The FlatSat is a method for trialling the core components of
the satellite within a lab environment. Components such as communication, computer and data
handling, and power supply are assembled on a workbench to assess how they will
function cooperatively in different configurations. Understanding
and improving the system at this fundamental level has been shown to yield a
higher rate of success in similar projects \cite{cubesat_survey}. This work is
primarily focussed on the electronic hardware within the satellite. An equally
important aspect of the PROVE mission is the set of external systems that have been
developed. This includes the CubeSat lab, fitted with a clean room for component
assembly, that has housed various projects working on CubeSat hardware. The lab
is also linked to the CubeSat communications system that uses a pair of
antennas, capable of tracking the satellite, to send orbital data
as well as receive the images taken by the on-board infrared camera. 

It is through this communication system that the satellite will be sent
its position and orbital trajectory in the form of a `two-line
element' (TLE). This is a data format that encodes a satellite's orbit 
and can be used to estimate the position of the satellite as a function of
time. The data is attained through \url{www.space-track.org}, and periodically
broadcast to the CubeSat when its orbit aligns with the PROVE mission antennas
that are fixed to the roof of the Queen's building in Bristol. Although this is
one of the most accurate methods of orbit tracking, the cumulative effect 
of the forces discussed in Section \ref{sec:external torq} will, over time,
result in the satellite diverging from its predicted path.
For a 3U CubeSat in low Earth orbit (below 350\,km) one day after a TLE has been calculated,
the position estimates have a standard deviation between 10\,km and
30\,km, however this can be significantly reduced by factoring in previous TLEs \cite{tle}. The satellite's orbit will coincide with the antennas every
12 hours. Since there is no on-board GPS system, due to cost restrictions, this
is the only method through which to communicate to the satellite where it is in
its orbit. Accurate knowledge of the satellite's position is crucial for the Attitude
Determination and Control Systems (ADCS). 
\subsection{ADCS}
The currently proposed measurement instruments for attitude determination in
the PROVE mission are a sun sensor and a magnetometer. To control the attitude,
a set of flywheels and a magnetorquer will be used.

The magnetorquer is a 3-axis system of electromagnets that can
interact with the Earth's magnetic field to produce an external torque on the
satellite. The magnetorquer used in this mission can create a maximum torque of $4.6\times
10^{-6}$\,Nm \cite{ext_torq_vals}. This is also dependant on the satellite's
orientation within the Earth's magnetic field. Therefore, the magnetorquer
alone will not be sufficient to generate the
angular rate required for the satellite to remain pointing at the
ground target during its 6 minute flyover \cite{magnetorquer_control}.
Therefore in this project, the satellite's magnetorquer will not be included in
calculations that discuss torque produced from the satellite as it functions predominantly as a tool for
desaturating the flywheels and detumbling the satellite immediately after
deployment. By measuring the induced electromotive force through the three components of the
magnetorquer, it can also function as a magnetometer
\cite{magnetorquer_as_magnetometer}. With the measurements over three
dimensions, the direction of the field can be calculated, thus supporting the
sun sensor measurements to determine attitude.

The flywheel system uses at least three rotating wheels at an angle to each
other to give attitude control in each axis. The wheel configuration of this
project has not yet been confirmed. In response to this, the equations of motion
have been left in a general form to be applicable for both 3 and 4 wheel
configurations with the wheels in any orientation. 
For three axis orientation control, only three wheels are needed. However, a 
fourth wheel may be added
as a safety measure, allowing for continued control in the case that one wheel
fails. To demonstrate the operation of flywheels in a 2 dimensional case; a single flywheel
accelerating within a body will produce a torque at the point of rotation,
causing the body to
accelerate in the opposite direction resulting in zero net torque. This is due to
the rotational analogue of Newton's third law such that every torque exerted
results in an equal reaction torque in the opposite direction.
By combining the torque of multiple wheels, more complex rotations can be
performed. However, as flywheel speeds increase, the impact of gyroscopic effects between
the wheels and the body will increase. The significance of this is examined in Section
\ref{sec:gyroscopic}.

The PROVE mission satellite requires an attitude control system with higher
accuracy than most previous CubeSat projects. The satellite will be required
to rotate rapidly in order to maintain a line of sight with the ash cloud. In
researching the PROVE CubeSat's design, L. Hawkins developed a series of
constraints that the ADCS components must adhere to \cite{flywheel_motor}. 
These include operational
requirements, such as having a minimum lifetime of 12 months, and physical
requirements, such as weighing no more than 0.67\,kg and taking up a maximum of
half a unit of space in the CubeSat. These constraints, along with a financial
limit on components, mean that more work is required to achieve high
performance in ADCS. 

To increase the chances of safe operation over a long
period, most components must maintain a safe margin from their peak performance. This
was also addressed by Hawkins when assessing the performance of the flywheels.
The \textit{Faulhaber 2610} was the proposed flywheel motor at the time, which 
had a maximum angular rate of 7000\,rpm. However, Hawkins found that the
proposed flywheels would become at risk of
tangential stress damage from rates beyond 6000\,rpm \cite{flywheel_damage}.
Since this paper, the choice of motor has changed to the \textit{Faulhaber
1509T006B} as it can achieve angular rates of up to 10,000\,rpm. This presents
the need for flywheels that can withstand a higher degree of tangential stress.
Flywheel performance was also investigated by M. Tisaev for the PROVE mission,
who showed there to be a 3\% drop in efficiency for
flywheel speeds of 6395\,rpm compared to 1223\,rpm \cite{flywheel_friction}.

Another constraint on the flywheels occurs due to saturation. This occurs once
the wheel has accelerated in one direction long enough for the maximum speed to
be reached, meaning no more torque can be produced in that axis and
desaturation is required. This is likely to occur due to external torques that
could continuously act in one direction on the CubeSat. This would require a
prolonged correcting torque from the flywheels meaning a constant acceleration
must occur. The magnetorquers can provide an external torque through
interacting with Earth's magnetic field, allowing the flywheels
to reduce their speeds without affecting satellite rotation \cite{desaturation_needed}.

Initially, the attitude determination system consisted of a sun sensor, and
three motion references; an accelerometer, a gyroscope, and the magnetorquer.
The sun sensor approach, investigated by T. Kariniemi-Eldridge, uses sensors on
each face of the CubeSat, orientation is determined by which sensors are
illuminated \cite{sun_sensor}. This component, however, can only function when in direct
sunlight; passing through Earth's umbra renders it useless. Therefore, during
this period, the satellite would need to rely on dead reckoning using the
on-board motion sensors. This could be mitigated by using a star tracker to
determine orientation, although this is a more expensive option due to the
added computation required.

In response to this, R. Biggs assessed the use of an Extended Kalman Filter to
increase the accuracy of attitude estimations. Biggs also evaluated Earth
horizon sensors \cite{kalman_star_tracker}. Kalman Filtering uses
probabilistic modelling to make predictions, based on previous measurements, and update
its model as new data is made available. This method has been implemented in other
satellites and is shown to largely improve attitude determination with minimal
extra cost \cite{kalman_sun_sensor}. Simulations of this approach reduced error
in the sun sensor to $\pm 0.25^\circ$ when used with an Earth horizon
sensor. The more expensive star tracker had an error of
just $\pm 0.1^\circ$. This is compared with unfiltered attitude 
determination, which produced an error of $\pm 0.5^\circ$ \cite{normal_sensors}.
An important result of Biggs' research was showing
that using an Extended Kalman Filter provides enough added accuracy that
a gyroscope may not be needed. This would reduce
computational resources and power within the satellite.

\section{External Torques}
\label{sec:external torq}

There are four potentially significant external sources of torque that can act on a satellite in low
Earth orbit \cite{ext_torqs}. These are
\begin{itemize}
  \item Gravitational gradient
  \item Aerodynamic drag
  \item Environmental radiation 
  \item Magnetic field interaction
\end{itemize}
The physical characteristics of the satellite, and the type of its orbit, will decide how strong each of these
torques are. 

Gravitational torque arises due to the difference in force felt across the
satellite. For a point mass, gravitational force acts at its centre, but for a
body with differently distributed mass, the net force may act at a point
away from the centre of mass. A second factor comes from the non-uniformity of
Earth's gravitational field. Earth is an oblate spheroid, i.e., it has a larger
radius at its equator than it does at its poles, this causes the gravitational
field of Earth to vary in strength during an orbit. Passing
through this field of varying strength can also produce a torque on the
satellite.

The air density at an altitude of 300\,km
is approximately 11 orders of magnitude less than that of the density at sea level
\cite{air_density}. However, this will still produce a small drag force on the
satellite, resulting in a disturbance torque.

Pressure is generated on the satellite's surface when sun light is reflected
and absorbed. This is due to the non-zero momentum of photons which is partially 
transferred on impact. The resultant force of this solar pressure will not
necessarily coincide with the centre of mass of the satellite, thus producing a
torque.

Finally, magnetic torque arises due to Earth’s magnetic field, behaving
like that of a dipole as the satellite passes through it. The magnetic moment of the satellite interacts with this
field and results in a torque as the two fields are drawn into alignment.
However, as discussed, this phenomenon can be exploited by using a magnetorquer
to control the satellite’s magnetic moment, producing an intentional external
torque that can be used for detumbling, or to desaturate the flywheels.

This external torque is given to be
\[
  \boldsymbol{\tau}_{\mathrm{Mag}} = \boldsymbol{\mu}_{\mathrm{Sat}}\times
  \mathbf{B}_{\mathrm{Earth}},\numberthis \label{eq:magnetorquer}
\]
where $\boldsymbol{\tau}_{\mathrm{Mag}}$ is the generated torque, given as the cross product of the
satellite's magnetic moment, $\boldsymbol{\mu}_{\mathrm{Sat}}$, with the
magnetic field vector of Earth, $\mathbf{B}_{\mathrm{Earth}}$. The magnetic
moment can be expressed as $\boldsymbol{\mu} = nI\mathbf{A}$, where $n$ is the
number of turns in the magnetorquer solenoid, $I$ is the current, and
$\mathbf{A}$ is the vector area of the solenoid. The cross product in
(\ref{eq:magnetorquer}) indicates that the torque produced from the
magnetorquer is dependent on the misalignment of the two magnetic fields. Using a
three-axis magnetorquer can ensure that the satellite's magnetic moment is not
parallel to the Earth's magnetic field, however the torque it produces will
always be perpendicular to this field. This limits the versatility and power
of the magnetorquer, hence why the more adept flywheels are used during the
flyover.

For a 3U CubeSat in low Earth orbit, the largest source of disturbance torque
comes from aerodynamic drag \cite{aero_dominates}. An upper limit on the
magnitude of this torque is $6\times 10^{-7}$\,Nm \cite{max_ext_torq}, whilst
the next largest torque comes from gravitational gradient and has an estimated
magnitude of 5.6$\times 10^{-8}$ at an altitude of 300\,km \cite{grav_torq}.
Environmental radiation and magnetic field interaction are negligible in
comparison to these.
During its orbit, the satellite will register these disturbance torques with
the gyro and attitude sensors, and will correct for them with the controller. If this
disturbance torque is constant, the wheels will require frequent desaturation
from the magnetorquer. An upper limit on the frequency that this desaturation
could be required can be calculated.

The frequency that desaturation is required depends on both the moment of
inertia of the flywheels and the maximum speed at which they can spin.
The moment of inertia of the flywheels about their rotation axis is
approximately 1.1$\times 10^{-6}$\,kg\,m$^2$ \cite{mansur_vals}. Therefore, to produce $6\times 10^{-7}$\,Nm 
of torque with one flywheel, it must accelerate at 0.55\,rad/s$^2$,
meaning it would reach its maximum speed of 840\,rad/s within 1540 seconds.
This means that desaturation could be needed every 25 minutes, in a worst-case
scenario. This only poses
a potential problem during the flyover, when the wheels may already be close to
their maximum speed after producing the necessary torque. 

\begin{figure}
    \centering
  \begin{subfigure}{.45\textwidth}
    \centering
    \includegraphics[width=0.9\textwidth]{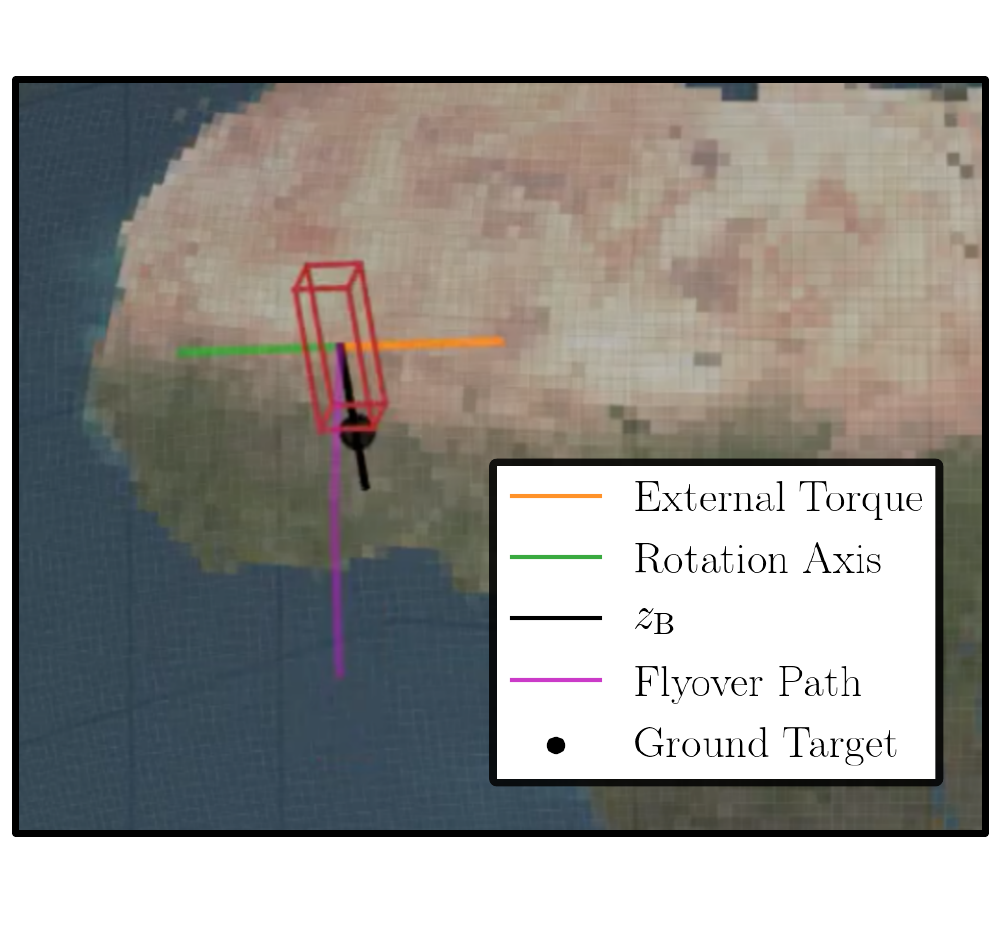}
    \caption{Flyover showing the opposing vectors whilst the satellite points it's $z$-axis at
    the ground target.}
    \label{fig:ext_torq_eg}
  \end{subfigure}%
  \hspace{15pt}
  \begin{subfigure}{.5\textwidth}
    \centering
    \includegraphics[width=\textwidth]{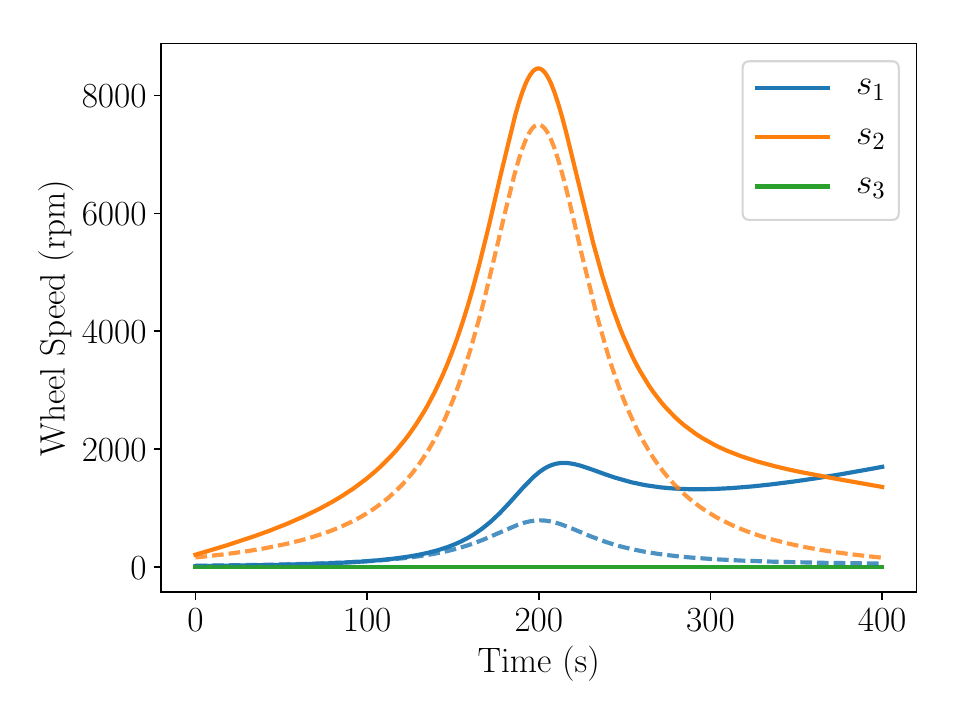}
    \caption{Wheel speeds during a flyover with no
      external torque (dashed) and one with maximum external torque
    (solid).}
    \label{fig:ext_torq_wheel_speed}
  \end{subfigure}%
  \caption{Visualising the effect of external torque during a flyover
    from directly above. All other parameters are set to their default values, shown
  in Appendix \ref{sec:params_table}.}
\label{fig:ext_torq}
\end{figure}

The most strenuous flyover occurs when the satellite passes directly above the
ground target, needing to perform the sharpest turn. When simulating this
flyover in idealised conditions, the
wheel speeds reach a maximum of 7335\,rpm. This assumes that the satellite
starts the flyover with zero flywheel speed. 
Even in this idealised condition, this is close to the
suggested speed limit of 8000\,rpm, indicating that desaturation must occur
before the flyover begins in any case. The worst-case scenario of external
torque occurs when the satellite's rotation is directly opposed. Figure
\ref{fig:ext_torq} shows this case, with a constant torque of magnitude
$6\times 10^{-7}$\,Nm acting along the body-fixed $y$-axis, $y_{\mathrm{B}}$, whilst
rotation occurs in the opposite direction.

The dashed line plots in Figure \ref{fig:ext_torq_wheel_speed} show the wheel
speeds during the flyover if no external torque is present, whilst the solid
lines show their behaviour when the external torque, shown in Figure
\ref{fig:ext_torq_eg}, is present. In the case of no external torque, the wheels display 
symmetrical ramps up and down to match the changing severity of turn required.
However, in the presence of the opposing external torque, these wheels must
continuously accelerate at approximately 0.55\,rad/s. As a result, the maximum
wheel speed increases to 8325\,rpm, impinging on the proposed safety margin
from the actual maximum speed of 10,000\,rpm.
Although this is acceptable for a brief period, higher speeds will
increase the gyroscopic effect produced by the flywheels. This is 
discussed in Section \ref{sec:gyroscopic}.

\section{Quaternions}
\label{sec:quaternions}
For this project, quaternions will be the method used to describe orientations and
rotations.
A quaternion is a hyper complex number that consist of 4 elements; 1 scalar
term and 3 vector terms. The vector terms are named as such since their
basis elements; $\boldsymbol{i}$, $\boldsymbol{j}$, and $\boldsymbol{k}$, can
be interpreted as unit vectors pointing along the three spatial axes.
Quaternions are generally expressed in the form
\[
  q = q_1 + q_2 \boldsymbol{i} + q_3\boldsymbol{j} + q_4\boldsymbol{k},
  \numberthis \label{eq:quat_def}
\]
however, in this application, they may also be written in the vector 
form $q=[q_1,q_2,q_3,q_4]^\top$.

The products of the basis elements of a quaternion are defined by
\[
  \boldsymbol{i}^2 = \boldsymbol{j}^2 = \boldsymbol{k}^2 =
  \boldsymbol{i}\boldsymbol{j}\boldsymbol{k} =  -1,
\]
similar to the imaginary part of a complex number.
The multiplication of two quaternions is noncommutative, meaning
the order of the operands affects the result. The multiplication of
quaternions, $q$ and $p$, is defined as
\[
  q\odot p = \begin{bmatrix}q_1p_1 - q_2p_2 - q_3p_3 - q_4p_4\\
                            q_1p_2 + q_2p_1 + q_3p_4 - q_4p_3\\
                            q_1p_3 - q_2p_4 + q_3p_1 + q_4p_2\\
                          q_1p_4 + q_2p_3 - q_3p_2 + q_4p_1\end{bmatrix}.
\]
The conjugate of a quaternion, much like with complex numbers, is defined as 
\[
  q^* = q_1 - q_2\boldsymbol{i} - q_3\boldsymbol{j} - q_4\boldsymbol{k},
\]
using the definition from (\ref{eq:quat_def}). The
inverse of a quaternion is then written as 
\[
  q^{-1} = \frac{q^*}{|q|^2}.
\]
A unit quaternion, such that $\sqrt{q_1^2+q_2^2+q_3^2+q_4^2} = 1$, can be used
to describe rotations in 3D space.
Orientation can also be described using a rotation from a canonical form, as
shown in Figure \ref{fig:canonical}. Rotations are achieved through
quaternion multiplication. To rotate a vector, $\boldsymbol{v}$, it must first
be represented as a quaternion. This is achieved by setting the quaternion
scalar element to 0 and the vector elements to those of $\boldsymbol{v}$.
Rotating by quaternion $q$ is achieved by first multiplying from the
left by $q$, and then from the right by $q^{-1}$, which is the same as $q^*$
for unit quaternions. This is written as
\[
  q\odot\boldsymbol{v}\odot q^{-1} =
  \begin{bmatrix}q_1\\q_2\\q_3\\q_4\end{bmatrix}\odot\begin{bmatrix}0\\v_1\\v_2\\v_3\end{bmatrix}
  \odot \begin{bmatrix}q_1\\-q_2\\-q_3\\-q_4\end{bmatrix} =
  \begin{bmatrix}0\\v_1'\\v_2'\\v_3' \end{bmatrix}.
\]
The polar decomposition of a unit quaternion is analogous to the polar form of
a complex number. It separates the quaternion into its real, and vector, parts
using trigonometric identities to relate the two components. This is expressed
as
\[
  q = \cos{\frac{\theta}{2}} + \hat{\mathbf{n}}\sin{\frac{\theta}{2}}.
\]
This can be easily interpreted to understand the rotation it will perform on a given
object. $\hat{\mathbf{n}}$ from the vector part,
$\hat{\mathbf{n}}\sin{\frac{\theta}{2}}$, is the unit vector of rotation
whilst the amount it will rotate is $\theta$. This will
prove highly useful in designing a quaternion based controller, discussed in
Section \ref{sec:control}.

The other commonly used method used to describe rotation and
orientation in 3D space is rotation matrices. These are orthogonal matrices
with a determinant of 1. An example of this is
\[
  X(\phi) = \begin{bmatrix}1& 0& 0\\0&\cos{\phi}&-\sin{\phi}\\0& \sin{\phi}&
  \cos{\phi}\end{bmatrix},
\]
which will rotate a vector by $\phi$ about the $x$-axis.

By multiplying three, single axis, rotation matrices together, a composite
rotation matrix can be generated. Although this is an intuitive method 
for describing rotation, it comes with limitations.

Gimbal lock is a phenomenon by which two axes of rotation align, resulting in
the loss of a degree of freedom. This is a significant problem when dealing
with an orbiting body as the orientation is not constrained, meaning an
alignment of rotation axes is inevitable. This suggests the need for a more robust
method of expressing orientation.

And additional benefit of quaternions comes from their invariant constraint,
whereby they must maintain a norm of 1. This is computationally cheap to check
and, if necessary, correct for, by simply rescaling the four elements
proportionally. Comparing this to the
invariants of a rotation matrix, they must maintain a determinant of 1,
as well as remaining orthogonal. If this invariant constraint is not met due to
truncation errors, a single-value decomposition is required to return to
orthogonal, which is far less efficient. Thus, a quaternion based
approach is used for describing orientation and rotation, as well as
in the control law derived in Section \ref{sec:control}.

\section{Equations of Motion}
The equations of motion that have been used in previous projects are written as
\begin{align}
  \mathbf{I} \dot{\boldsymbol{\omega}} &=
  \boldsymbol{\tau}_{\mathrm{Ext}} - \boldsymbol{\tau}_{\mathrm{W}} -
  \boldsymbol{\omega}\times\mathbf{I}
  \boldsymbol{\omega}\label{eq:old_motion}\\
  \dot{q} &= \frac{1}{2}q\odot\boldsymbol{\omega}\label{eq:old_quat}\\
  \boldsymbol{\tau}_{\mathrm{W}}&=\boldsymbol{f}(\mathbf{q}_{\mathrm{err}},\boldsymbol{\omega}_{\mathrm{err}})\label{eq:old_control},
\end{align}
where (\ref{eq:old_motion}) and (\ref{eq:old_quat}) describe the satellite's angular
velocity and orientation respectively whilst (\ref{eq:old_control}) describes
flywheel torque, as generated by the control law. The control law uses the
vector part of the pointing error quaternion, $\mathbf{q}_{\mathrm{err}}$, and
the angular velocity error, $\boldsymbol{\omega}_{\mathrm{err}}$, making this a
proportional-derivative controller.

This will be improved upon by extending (\ref{eq:old_motion}) to treat the
flywheels as separate rotating bodies, giving a more accurate representation of
the system. This can then be used in simulating the performance of the newly
designed controller.

The motion equations that will be used for this project are given as
\begin{align}
  \mathbf{I}_{\mathrm{Tot}} \dot{\boldsymbol{\omega}}_{\mathrm{B}} &=
  \boldsymbol{\tau}_{\mathrm{Ext}} -
  \boldsymbol{\omega}_{\mathrm{B}}\times\mathbf{I}_{\mathrm{Tot}}
  \boldsymbol{\omega}_{\mathrm{B}} -  \boldsymbol{\tau}_{\mathrm{W}} -
  \boldsymbol{\omega}_{\mathrm{B}}\times\mathbf{E}_{\mathrm{W}}\mathbf{s}\label{eq:new_motion}\\
  \dot{q} &=
  \frac{1}{2}q\odot\boldsymbol{\omega}_{\mathrm{B}}\nonumber\\
  \boldsymbol{\tau}_{\mathrm{W}} &=
  \boldsymbol{f}(\mathbf{q}_{\mathrm{err}},\mathbf{g}_{\mathrm{err}},\boldsymbol{\omega}_{\mathrm{err}})\label{eq:new_control},
\end{align}
where the derivation of (\ref{eq:new_motion}) is given in Section
\ref{sec:eqs_of_motion}, whilst the new error term given to the control law,
$\mathbf{g}_{\mathrm{err}}$, will be explained in Section
\ref{sec:control_terms}.
The last term in (\ref{eq:new_motion}) is the cross product of body angular
velocity, $\boldsymbol{\omega}_{\mathrm{B}}$, and the net
angular momentum of the flywheels, $\mathbf{E}_{\mathrm{W}}\mathbf{s}$. This
denotes the previously omitted gyroscopic effect that arises from the flywheels spinning.
The cross product here shows that the gyroscopic torque will act perpendicular to
the net angular momentum of the flywheels which could perturb the satellite
from its rotational path at a high enough magnitude.
\subsection{Definitions}
\label{sec:definitions}
\begin{figure}
  \centering
  \vspace{-90pt}
  \includegraphics[width=\textwidth]{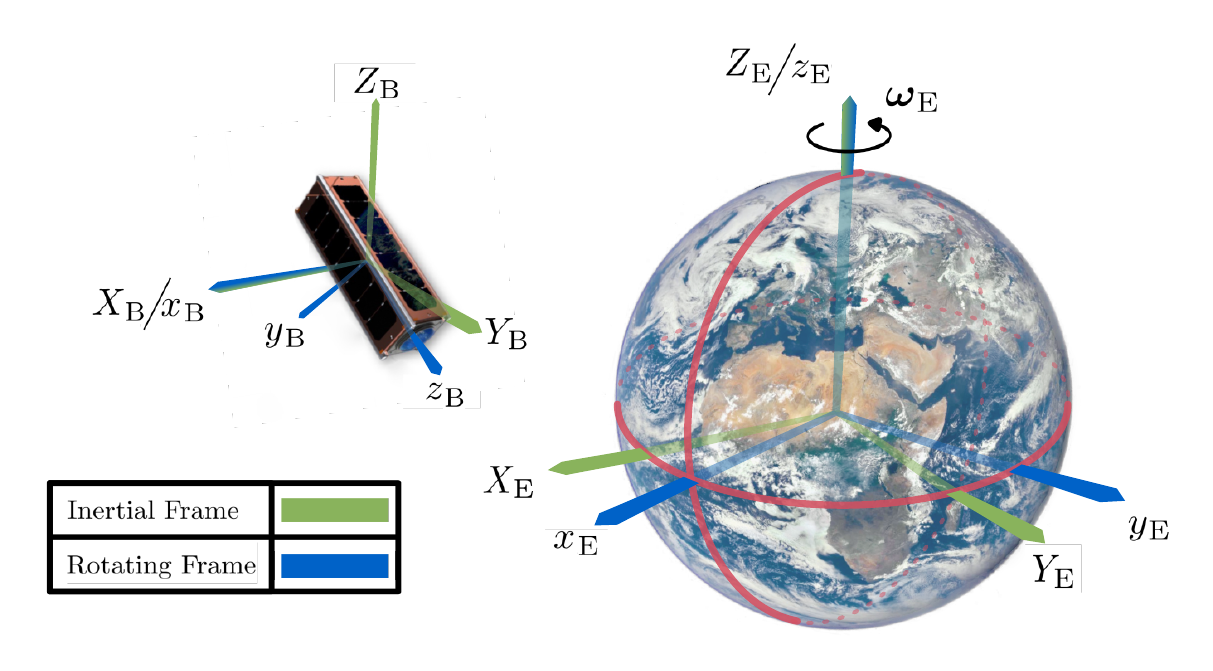}
  \caption{The four reference frames used in this report. The two
    inertial reference frames (green) share the same orientation and do not
    rotate. The Earth fixed frame rotates about
    $\boldsymbol{\omega}_{\mathrm{E}}$, fixing $x_{\mathrm{E}}$ to 
    $(0^\circ,0^\circ)$ which occurs at the intersection of Earth's prime meridian
    and the equator (red). The body-fixed frame keeps the $z$-axis pointing through
  the satellite's camera, here it has rotated $240^\circ$ about its $x$-axis.}
  \label{fig:axis_convention}
  \vspace{+60pt}
  \includegraphics[width=0.6\textwidth]{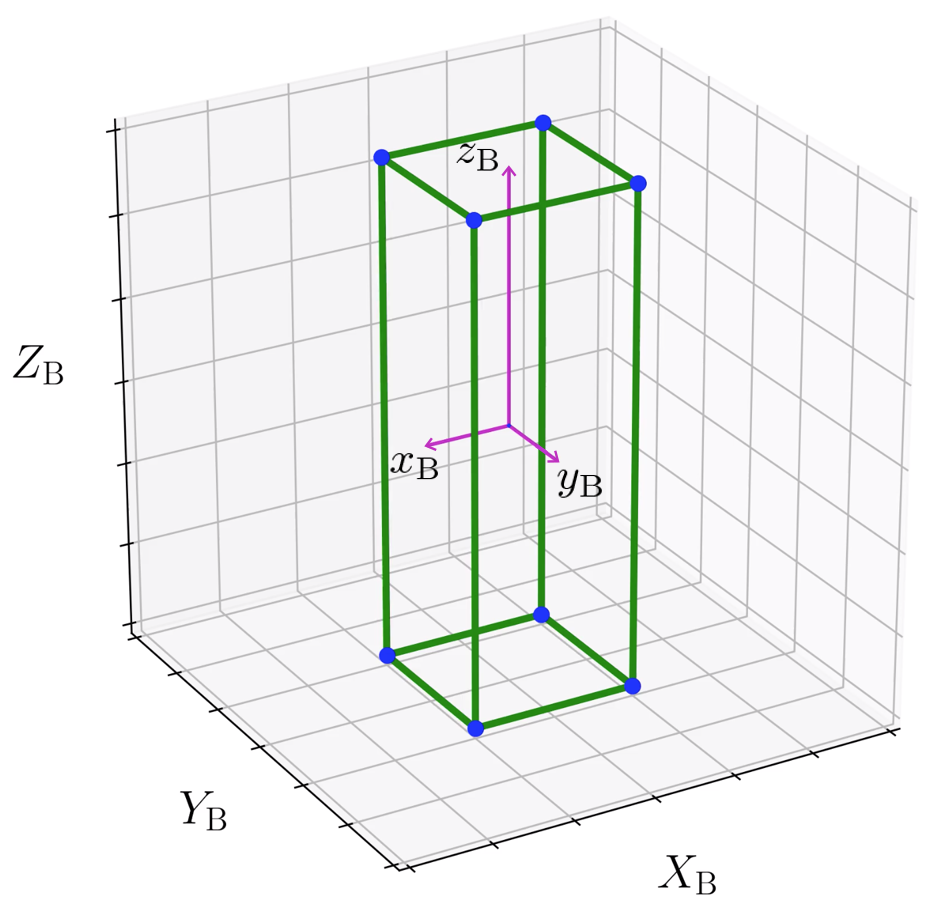}
  \caption{Canonical form of the CubeSat where $q=[1,0,0,0]^\top$. The body-fixed
    axes, $x_{\mathrm{B}}, y_{\mathrm{B}}, z_{\mathrm{B}}$, share an origin and align
  with the inertial frame axes for the body, labelled $X_{\mathrm{B}}, Y_{\mathrm{B}}, Z_{\mathrm{B}}$.}
  \label{fig:canonical}
\end{figure}
There are four key reference frames that will be referred to throughout this
report, all of which are illustrated in Figure \ref{fig:axis_convention}. The
first is the Earth-centred rotating frame. This is a non-inertial reference
frame since it will rotate about its $z$-axis, at an angular rate of 1
revolution per 24 hours.
The convention used is for the $x$-$y$ plane to
intersect the equator, with the $x$-axis intersecting the prime meridian
($0^{\circ}$ longitude). The Earth-centred inertial frame will initially align
with the rotating frame at some arbitrary start time, however this frame will
remain fixed at this orientation for all time. The next reference frame is the
body-centred inertial frame. This is not technically
inertial as its origin is fixed to the centre of mass of the orbiting
satellite. However, this frame will only be used to describe rotations about
its origin, and the orientation of this frame will be fixed to align with the
Earth-centred inertial frame. This means, for all intents and purposes, it can
be considered inertial and will be referred to as such. The final frame that
will be used is the body-centred rotating frame. This frame has the same
angular velocity as the satellite, $\boldsymbol{\omega}_{\mathrm{B}}$, 
keeping it's positive $z$-axis pointing along the sight line of the on-board camera.

Throughout the report, wheel acceleration can be taken to mean the angular
acceleration of the wheels relative to the body-centred rotating frame.

The orientation of the satellite at time $t$ is expressed by the unit quaternion
$q(t)$. This quaternion represents an active rotation of the satellite's
canonical form, as shown in Figure \ref{fig:canonical}. The canonical form of
the satellite is attained by aligning the body-rotating axes with the body-centred
inertial axes. An active rotation then describes the transformation of the
canonical body to a new orientation. This is compared with a passive rotation, in which
the coordinate system itself is rotated
instead, which will not be used here. In this report the
quaternion multiplication operator, $\odot$, will be used between a
quaternion and a regular Euclidean vector. This indicates that the vector is
first transformed into a quaternion following the procedure described in Section
\ref{sec:quaternions}.

In subsequent sections, the satellite will be considered as a composite of its wheels and body.
This is to isolate the separate rigid bodies that will be rotating during attitude
control. When referring to the satellite `wheels', this includes the flywheel
itself and the shaft connecting it to the motor since they rotate as
one body. The satellite's `body' refers to all that remains of the
satellite after excluding the wheels.

\subsection{Derivation}
\label{sec:eqs_of_motion}
The system's angular momentum is first expressed as a sum of the angular momentum of the
satellite body and the angular momenta of the flywheels. 
This gives a top level description of the system's angular momentum to be
\[
\label{eq:ang_mom}
\mathbf{L}_{\mathrm{Tot}} = \mathbf{L}_{\mathrm{B}} + \sum_i
\mathbf{L}_{\mathrm{W}i},
\]
where the subscript, $\mathrm{B}$, denotes the satellite's body and
$\mathrm{W}i$ corresponds with the $i^{\mathrm{th}}$ flywheel.

Angular momentum is an object's angular velocity around its centre of mass,
$\boldsymbol{\omega}$, multiplied by its inertia tensor, $\mathbf{I}$. If the
centre of mass is then also rotating around a different point, this must be incorporated
into the expression for the object's angular momentum. This term is the cross
product of the vector from the centre of rotation to the centre of mass,
$\mathbf{r}$, and the linear momentum of the centre of mass,
$\mathbf{p}=m\mathbf{v}$, where $m$ and $\mathbf{v}$ are the object's mass and
velocity respectively. The vector,
$\mathbf{r}_{\mathrm{B}}$, describes the displacement of the body's centre
of mass from the total centre of mass of the satellite with the mass of the
flywheels included. This gives the body's angular momentum as
\begin{align*}
  \mathbf{L}_{\mathrm{B}} &=
  \mathbf{I}_{\mathrm{B}}\boldsymbol{\omega}_{\mathrm{B}}+
  \mathbf{r}_{\mathrm{B}}\times \mathbf{p}_{\mathrm{B}}\\
  &= \mathbf{I}_{\mathrm{B}}\boldsymbol{\omega}_{\mathrm{B}}+m_{\mathrm{B}}
  \mathbf{r}_{\mathrm{B}}\times\mathbf{v}_{\mathrm{B}}\\
  &= \mathbf{I}_{\mathrm{B}}\boldsymbol{\omega}_{\mathrm{B}}+m_{\mathrm{B}}
  \mathbf{r}_{\mathrm{B}}\times(\boldsymbol{\omega}_{\mathrm{B}}\times 
  \mathbf{r}_{\mathrm{B}}), \numberthis \label{eq:L bod}
\end{align*}
where the relation between linear and angular velocity,
$\mathbf{v}=\boldsymbol{\omega}\times\mathbf{r}$, has been utilised to express
the full angular momentum as a function of angular velocity.
Similarly the angular momentum of the wheels becomes
\begin{align*}
  \sum_i \mathbf{L}_{\mathrm{W}i} &= \sum_i \mathbf{I}_{\mathrm{W}i} \boldsymbol{\omega}_{\mathrm{W}i} +
  \mathbf{r}_{\mathrm{W}i}\times \mathbf{p}_{\mathrm{W}i}\\
  &= \sum_i \mathbf{I}_{\mathrm{W}i} \boldsymbol{\omega}_{\mathrm{W}i} +
  m_{\mathrm{W}}
  \mathbf{r}_{\mathrm{W}i}\times(\boldsymbol{\omega}_{\mathrm{B}}\times
  \mathbf{r}_{\mathrm{W}i}),
\end{align*}
where the angular velocity of the body, $\boldsymbol{\omega}_{\mathrm{B}}$,
also describes the angular velocity of the wheels about the satellite's total centre of
mass.
This derivation aims to produce an equation that will describe the
dynamics of the satellite as a function of varied flywheel speeds.
Hence, it is necessary to isolate the terms that describe the respective wheel
speeds, expressing them from the perspective of the flywheel motor, since this is the angular speed
that the controller must output. For this reason, the angular
velocity of the wheels about their axes of rotation is expressed as the sum
of the wheel's angular velocity from the perspective of the body, and so the
motor, and the angular velocity of the body itself.
This then gives
\begin{align*}
  \sum_i\mathbf{L}_{\mathrm{W}i} &= \sum_i\mathbf{I}_{\mathrm{W}i} \left[
\boldsymbol{\omega}_{\mathrm{B}}+(\boldsymbol{\omega}_{\mathrm{W}i}-\boldsymbol{\omega}_{\mathrm{B}})\right] + 
m_{\mathrm{W}}\mathbf{r}_{\mathrm{W}i}\times
(\boldsymbol{\omega}_{\mathrm{B}}\times\mathbf{r}_{\mathrm{W}i})\\
  &=\sum_i\mathbf{I}_{\mathrm{W}i} \left( \boldsymbol{\omega}_{\mathrm{B}} + s_i
  \mathbf{\hat{e}}_i\right) + m_{\mathrm{W}}
  \mathbf{r}_{\mathrm{W}i}\times(\boldsymbol{\omega}_{\mathrm{B}}\times
\mathbf{r}_{\mathrm{W}i})\\
  &= \sum_i\mathbf{I}_{\mathrm{W}i}\boldsymbol{\omega}_{\mathrm{B}}+s_i
  \mathbf{I}_{\mathrm{W}i}\mathbf{\hat{e}}_i + m_{\mathrm{W}}
  \mathbf{r}_{\mathrm{W}i}\times(\boldsymbol{\omega}_{\mathrm{B}}\times
\mathbf{r}_{\mathrm{W}i}),
\end{align*}
where the wheel's angular velocity, from the perspective of the body, is
expressed as a product of its angular speed and the unit vector $\mathbf{\hat{e}}_i$
that describes the wheel's axis of rotation.

To prepare the equation for factoring, the
triple product $\mathbf{r}_{\mathrm{W}i}\times(\boldsymbol
{\omega}_{\mathrm{B}}\times\mathbf{r}_{\mathrm{W}i})$ is written as a
matrix-vector product
$\mathbf{A}_{\mathrm{W}i}\boldsymbol{\omega}_{\mathrm{B}}$ such that
\[
  \mathbf{A}_{\mathrm{W}i}=\begin{bmatrix}r_y^2+r_z^2 & -r_xr_y & -r_xr_z\\ -r_xr_y &
  r_x^2+r_z^2 &-r_yr_z\\ -r_xr_z & -r_yr_z & r_x^2 + r_y^2\end{bmatrix},
\]
where $\mathbf{r}_{\mathrm{W}i} = (r_x, r_y, r_z)^\top$.

A full derivation of this matrix is provided in Appendix \ref{app:A matrix}.
Using this new representation gives the angular momenta of the wheels to be
\begin{equation}
  \label{eq:L_w with A}
  \sum_i\mathbf{L}_{\mathrm{W}i} = \sum_i\mathbf{I}_{\mathrm{W}i}\boldsymbol{\omega}_{\mathrm{B}}+
  s_i\mathbf{I}_{\mathrm{W}i}\mathbf{\hat{e}}_i + m_{\mathrm{W}}
  \mathbf{A}_{\mathrm{W}i}\boldsymbol{\omega}_{\mathrm{B}}.
\end{equation}
Similarly for the angular momentum of the body, 
\begin{align*}
\mathbf{L}_{\mathrm{B}} &= \mathbf{I}_{\mathrm{B}}\boldsymbol{\omega}_{\mathrm{B}}+m_{\mathrm{B}}\mathbf{r}_{\mathrm{B}}
  \times(\boldsymbol{\omega}_{\mathrm{B}}\times \mathbf{r}_{\mathrm{B}})\\
  &=
  \mathbf{I}_{\mathrm{B}}\boldsymbol{\omega}_{\mathrm{B}}+m_{\mathrm{B}}\mathbf{A}_{\mathrm{B}}\boldsymbol{\omega}_{\mathrm{B}}.
\end{align*}
Since the wheels will be rotating about axes that are independent of the body
frame axes, their inertia tensors, when expressed in the body frame, will not be diagonal.
However, each tensor can be expressed as the diagonal tensor for a hypothetical
wheel that spins about the $x$-axis in the body frame, which is then multiplied
with a rotation matrix. Using one wheel to demonstrate, the second term in
(\ref{eq:L_w with A}) can be written as
\[
  s_i\mathbf{I}_{\mathrm{W}i}\mathbf{\hat{e}}_i = s_i \mathbf{O}_i\mathbf{I}_{\mathrm{W}x}
  \mathbf{O}^{-1}_i\mathbf{\hat{e}}_i, 
\] where \[
  \mathbf{I}_{\mathrm{W}x} =
  \begin{bmatrix} \frac{1}{2} m_{\mathrm{W}} R_{\mathrm{W}}^2 &0&0\\0&
    \frac{1}{12}m_{\mathrm{W}}\left(h_{\mathrm{W}}^2 + 3R_{\mathrm{W}}^2
  \right)&0\\0&0&\frac{1}{12}m_{\mathrm{W}}\left(h_{\mathrm{W}}^2 +
3R_{\mathrm{W}}^2\right)\end{bmatrix},
\]
with $R_{\mathrm{W}}$ and $h_{\mathrm{W}}$ representing the wheel radius and
height respectively. $\mathbf{O}_i$ is the rotation matrix from the $x$-axis to the
$i^{\mathrm{th}}$ wheel's axis of rotation, $\mathbf{\hat{e}}_i$. Therefore the inverse rotation,
$\mathbf{O}_i^{-1}$, will rotate $\mathbf{\hat{e}}_i$ back to the $x$-axis unit
vector to give
\[
   s_i \mathbf{O}_i\mathbf{I}_{\mathrm{W}x} \mathbf{O}^{-1}_i\mathbf{\hat{e}}_i= s_i \mathbf{O}_i\mathbf{I}_{\mathrm{W}x}
  \mathbf{\hat{e}}_x.
\]
Since $\mathbf{\hat{e}}_x = [1,0,0]^\top$, the product, $\mathbf{I}_{\mathrm{W}x}
\mathbf{\hat{e}}_x$ will become $\frac{1}{2}m_{\mathrm{W}}R_{\mathrm{W}}^2
\mathbf{\hat{e}}_x$, giving
\begin{align*}
  s_i \mathbf{O}_i\mathbf{I}_{\mathrm{W}x} \mathbf{\hat{e}}_x &=
  \frac{1}{2}m_{\mathrm{W}}R_{\mathrm{W}}^2 s_i
\mathbf{O}_i  \mathbf{\hat{e}}_x\\
&=\frac{1}{2}m_{\mathrm{W}}R_{\mathrm{W}}^2 s_i \mathbf{\hat{e}}_i,
\end{align*}
where $\mathbf{O}_i  \mathbf{\hat{e}}_x$ has become
$\mathbf{\hat{e}}_i$ from multiplying by the rotation matrix. Reducing this
term to just a scaled unit vector will improve the simplicity of the motion
equations later on. Substituting this term into the equation for the wheels'
angular velocity gives
\begin{align*}
  \sum_i\mathbf{L}_{\mathrm{W}i} &= \sum_i\mathbf{I}_{\mathrm{W}i}\boldsymbol{\omega}_{\mathrm{B}}+
  \frac{1}{2}m_{\mathrm{W}}R_{\mathrm{W}}^2 s_i\mathbf{\hat{e}}_i + m_{\mathrm{W}}
  \mathbf{A}_{\mathrm{W}i}\boldsymbol{\omega}_{\mathrm{B}}\\
  &= \sum_i\left( \mathbf{I}_{\mathrm{W}i}+m_{\mathrm{W}}\mathbf{A}_{\mathrm{W}i}
  \right)\boldsymbol{\omega}_{\mathrm{B}} +
  \frac{1}{2}m_{\mathrm{W}}R_{\mathrm{W}}^2
  s_i\mathbf{\hat{e}}_i.  \numberthis \label{eq:L wheels} 
\end{align*}

Substituting in the expressions for $\mathbf{L}_{\mathrm{B}}$ and
$\sum_i\mathbf{L}_{\mathrm{W}i}$ from (\ref{eq:L bod}) and (\ref{eq:L wheels})
respectively gives the full equation of angular momentum to be 
\begin{align*}
\mathbf{L}_{\mathrm{Tot}} &= \mathbf{L}_{\mathrm{B}} +
\sum_i\mathbf{L}_{\mathrm{W}i}\\
  &=\mathbf{I}_{\mathrm{B}}\boldsymbol{\omega}_{\mathrm{B}}+m_{\mathrm{B}}\mathbf{A}_{\mathrm{B}}\boldsymbol{\omega}_{\mathrm{B}}
  + \sum_i \left( \mathbf{I}_{\mathrm{W}i}+
  m_{\mathrm{W}}\mathbf{A}_{\mathrm{W}i}  \right)\boldsymbol{\omega}_{\mathrm{B}} +  
  \frac{1}{2}m_{\mathrm{W}}R_{\mathrm{W}}^2 s_i\mathbf{\hat{e}}_i\\
  &= \left(\mathbf{I}_{\mathrm{B}}+m_{\mathrm{B}}\mathbf{A}_{\mathrm{B}}+
  \sum_i \mathbf{I}_{\mathrm{W}i}+m_{\mathrm{W}}\mathbf{A}_{\mathrm{W}i} \right) 
  \boldsymbol{\omega}_{\mathrm{B}} + \frac{1}{2}m_{\mathrm{W}}R_{\mathrm{W}}^2 \sum_i
  s_i\mathbf{\hat{e}}_i\\
  &= \mathbf{I}_{\mathrm{Tot}} \boldsymbol{\omega}_{\mathrm{B}} +
  \frac{1}{2}m_{\mathrm{W}}R_{\mathrm{W}}^2 \sum_i s_i\mathbf{\hat{e}}_i, \numberthis \label{eq:full L}
\end{align*}
where many of the factored out terms have been absorbed into a single matrix
$\mathbf{I}_{\mathrm{Tot}}$ multiplying the body's angular velocity.

Using the satellite's total angular momentum, shown in (\ref{eq:full L}),
the equations of motion can be derived.

The relationship between the angular momentum of a system and the torque acting
on it comes from the rotational analogue of Newton’s second law, given as
\[
  \boldsymbol{\tau}=
\left(\dfrac{\mathrm{d}\mathbf{L}}{\mathrm{d}t}\right)_{\mathrm{in}},
\]
where $\boldsymbol{\tau}$ is the net external torque, and
$\left(\dfrac{\mathrm{d}\mathbf{L}}{\mathrm{d}t}\right)_{\mathrm{in}}$
is the time derivative of the total angular momentum of the satellite
in an inertial reference frame. It is important to note that the internal
torques generated from the flywheels are not included in
$\boldsymbol{\tau}$ as they have been in previous projects. The
torques that act from within the satellite must sum to zero as stated by
Newton's third law.

To keep $\mathbf{I}_{\mathrm{Tot}}$ and $\mathbf{\hat{e}}_i$ constant over time, 
the time derivative is taken in the rotating body frame. The general
relationship, when changing the frame of reference for the time derivative of
vector $\mathbf{v}$, is
\[
  \left(\dfrac{\mathrm{d}\mathbf{v}}{\mathrm{d}t}\right)_{\mathrm{in}} =
  \left(\dfrac{\mathrm{d}\mathbf{v}}{\mathrm{d}t}\right)_{\mathrm{rot}} +
  \boldsymbol{\omega}_{\mathrm{F}}\times\mathbf{v}, \label{eq:torq rotating}
  \numberthis
\]
where $\boldsymbol{\omega}_{\mathrm{F}}$ is the angular velocity of the rotating
frame.
Applying the change of reference frame from (\ref{eq:torq rotating}) to this case gives 
\[
\boldsymbol{\tau}_{\mathrm{Ext}} = 
\left(\dfrac{\mathrm{d}\mathbf{L}_{\mathrm{Tot}}}{\mathrm{d}t}\right)_{\mathrm{rot}} 
+ \boldsymbol{\omega}_{\mathrm{B}} \times \mathbf{L}_{\mathrm{Tot}}.
\]
Substituting the expression for $\mathbf{L}_{\mathrm{Tot}}$ in (\ref{eq:full
L}) gives

\[
\boldsymbol{\tau}_{\mathrm{Ext}} =\mathbf{I}_{\mathrm{Tot}}\dot{\boldsymbol{\omega}}_{\mathrm{B}} +
\frac{1}{2}m_{\mathrm{W}}R_{\mathrm{W}}^2 \sum_i \dot{s}_i\mathbf{\hat{e}}_i+
\boldsymbol{\omega}_{\mathrm{B}}\times\mathbf{I}_{\mathrm{Tot}} \boldsymbol{\omega}_{\mathrm{B}} +
\frac{1}{2}m_{\mathrm{W}}R_{\mathrm{W}}^2 \sum_i
  s_i\boldsymbol{\omega}_{\mathrm{B}}\times\mathbf{\hat{e}}_i,
\]
where the time dependent variables, $\boldsymbol{\omega}_{\mathrm{B}}$ and
$s_i$, have become accelerations $\dot{\boldsymbol{\omega}}_{\mathrm{B}}$ and
$\dot{s_i}$ in the first two terms of the equation. Grouping together terms
that relate to either the body, or the wheels, gives 
\[
 \boldsymbol{\tau}_{\mathrm{Ext}} = \mathbf{I}_{\mathrm{Tot}} \dot{\boldsymbol{\omega}}_{\mathrm{B}} +
\boldsymbol{\omega}_{\mathrm{B}}\times\mathbf{I}_{\mathrm{Tot}} \boldsymbol{\omega}_{\mathrm{B}} +
\frac{1}{2}m_{\mathrm{W}}R_{\mathrm{W}}^2 \sum_i\dot{s}_i\mathbf{\hat{e}}_i+
s_i\boldsymbol{\omega}_{\mathrm{B}}\times\mathbf{\hat{e}}_i, \numberthis
\label{eq:tau midway}
\]
which resembles Euler's equations of motion for a rigid body, extended to take 
into account the separate movement of the flywheels.

The sum of flywheel terms can be expressed as a matrix-vector product. This format
will allow for easy calculation of the individual wheel accelerations from the
torque vector calculated in the control law. It will also prove necessary for
the analytical tuning of the control law gains, discussed in Section
\ref{sec:linearisation}. The
matrix is implicitly invertible for the three wheel case due to the required linear independence of 
the flywheels' rotation axes. More work is required for a four wheel system
to make the matrix invertible, this is discussed in Section \ref{sec:four_wheels}.
For the three wheel case, the rotation axis vectors, $\mathbf{\hat{e}}_i$, are
horizontally concatenated and multiplied by the reduced inertia term,
$\frac{1}{2}m_{\mathrm{W}}R_{\mathrm{W}}^2$, giving
\[\mathbf{E}_{\mathrm{W}} = \frac{1}{2}m_{\mathrm{W}}R_{\mathrm{W}}^2
  \begin{bmatrix}\mathbf{\hat{e}}_1 &\mathbf{\hat{e}}_2 &\mathbf{\hat{e}}_3\end{bmatrix} =
  \frac{1}{2}m_{\mathrm{W}}R_{\mathrm{W}}^2 \begin{bmatrix}\mathrm{e}_{11} &
    \mathrm{e}_{21} & \mathrm{e}_{31}\\\mathrm{e}_{12}
  &\mathrm{e}_{22} & \mathrm{e}_{32}\\
\mathrm{e}_{13}&\mathrm{e}_{23}&\mathrm{e}_{33}\end{bmatrix}.
\]
This matrix can be thought of as an inertia-like tensor for the wheels.
The values for each wheel's angular speed are then combined into a
vector such that for a three wheel case, 
\[\mathbf{s}=\begin{bmatrix}s_1\\s_2\\s_3\end{bmatrix},
  \hspace{20pt}\dot{\mathbf{s}} = \begin{bmatrix} \dot{s}_1\\
  \dot{s}_2\\\dot{s}_3 \end{bmatrix}.
\]
This format allows for the net torque and angular momentum of the flywheels to
be treated each as a single term, whilst being able to access the specific wheel
accelerations and speeds of the wheels with multiplication by
$\mathbf{E}_{\mathrm{W}}^{-1}$.
The terms relating to the wheel dynamics in (\ref{eq:tau midway}) can now be expressed as
\[\frac{1}{2}m_{\mathrm{W}}R_{\mathrm{W}}^2 \sum_i\dot{s}_i\mathbf{\hat{e}}_i+
  s_i\boldsymbol{\omega}_{\mathrm{B}}\times\mathbf{\hat{e}}_i =
  \mathbf{E}_{\mathrm{W}}\dot{\mathbf{s}} +
  \boldsymbol{\omega}_{\mathrm{B}}\times\mathbf{E}_{\mathrm{W}}\mathbf{s},
\]
giving the full equation as
\[
  \boldsymbol{\tau}_{\mathrm{Ext}} = \mathbf{I}_{\mathrm{Tot}} \dot{\boldsymbol{\omega}}_{\mathrm{B}} +
  \boldsymbol{\omega}_{\mathrm{B}}\times\mathbf{I}_{\mathrm{Tot}} \boldsymbol{\omega}_{\mathrm{B}} +
  \mathbf{E}_{\mathrm{W}}\dot{\mathbf{s}} +
  \boldsymbol{\omega}_{\mathrm{B}}\times\mathbf{E}_{\mathrm{W}}\mathbf{s}.
\]
Finally, rearranging for $\dot{\boldsymbol{\omega}}_{\mathrm{B}}$ gives 
\[
  \mathbf{I}_{\mathrm{Tot}} \dot{\boldsymbol{\omega}}_{\mathrm{B}} = -  \mathbf{E}_{\mathrm{W}}\dot{\mathbf{s}} 
  +\boldsymbol{\tau}_{\mathrm{Ext}}
  -\boldsymbol{\omega}_{\mathrm{B}}\times\mathbf{I}_{\mathrm{Tot}}
  \boldsymbol{\omega}_{\mathrm{B}}-
  \boldsymbol{\omega}_{\mathrm{B}}\times\mathbf{E}_{\mathrm{W}}\mathbf{s}.
  \numberthis \label{eq:final ode}
\]

This equation describes the rotational dynamics of the satellite, now also
taking account of the gyroscopic effects due to the rotating flywheels. 
The $\mathbf{E}_{\mathrm{W}}\dot{\mathbf{s}}$ term in (\ref{eq:final ode}) is
the torque directly produced by the flywheels, which 
will be calculated from the control law discussed in Section
\ref{sec:control}. The $\boldsymbol{\omega}_{\mathrm{B}} 
\times\mathbf{E}_{\mathrm{W}}\mathbf{s}$ term is the gyroscopic
interaction that occurs between the wheels and the body they rotate within.
From Section \ref{sec:gyroscopic}, the magnitudes of
$\mathbf{I}_{\mathrm{Tot}}\dot{\boldsymbol{\omega}}_{\mathrm{B}}$ and
$\mathbf{E}_{\mathrm{W}}\dot{\mathbf{s}}$
are of order $10^{-7}$\,Nm at the most strenuous point of a flyover, whilst the gyroscopic terms are
at least two orders smaller. This validates that, in the case of no external
torque,
\[
  \mathbf{I}_{\mathrm{Tot}}\boldsymbol{\omega}_{\mathbf{B}} \approx
  -\mathbf{E}_{\mathrm{W}}\mathbf{s}.
\]
This is expected as a result of Newton's third law, stating that
torques generated by the flywheels must be exactly matched by an opposite
torque from the body.
This is an approximate interpretation of the equation but outlines the general
principles that are being utilised to control the satellite's orientation.
More detail will be given, as well as discussion of the gyroscopic terms, in Section
\ref{sec:gyroscopic}.

Another important conclusion to take from (\ref{eq:final ode}) is that the torque
delivered by the flywheels is shown to be independent of their position within the
satellite body. The torque that the flywheels directly produce is represented
by the term, $\mathbf{E}_{\mathrm{W}}\dot{\mathbf{s}}$. This term is a vector of wheel accelerations, multiplied
by the inertia-like matrix, which is comprised of the mass, radius, and
orientations of the wheels within the satellite. The wheel positions
are represented by the $\mathbf{A}_{\mathrm{W}i}$ matrices, from
(\ref{eq:L wheels}). These are then incorporated into the
$\mathbf{I}_{\mathrm{Tot}}$ term as shown in (\ref{eq:full L}). Hence
the positions only factor into the distribution of the satellite's total mass.
This validates the assertion that the wheels can be placed anywhere and it will not impact
the torque that they deliver.

To describe the rate of change of the satellite's orientation, a quaternion
approach is used.
Since quaternion multiplication is noncommutative, the
ordering of the operands results in two different definitions of the quaternion
derivative. For the above case, where the body's angular velocity is expressed in the
rotating frame, the equation is written as
\[
  \dot{q} = \frac{1}{2}q\odot\boldsymbol{\omega}_{\mathrm{B}}. \numberthis
\label{eq:quat_deriv}
\]
A full derivation and explanation of this equation is in Appendix
\ref{sec:quat_deriv}.

\subsection{Four Wheel Configuration}
\label{sec:four_wheels}
The use of either three or four flywheels is being considered for the PROVE mission
satellite. A minimum of three wheels are required, with linearly independent
axes of rotation, to rotate the body in three
dimensions. Using four flywheels, however, provides security in the case that one wheel
fails, allowing the system to maintain three axes of control.

\begin{figure}
  \centering
  \includegraphics[width=0.5\textwidth]{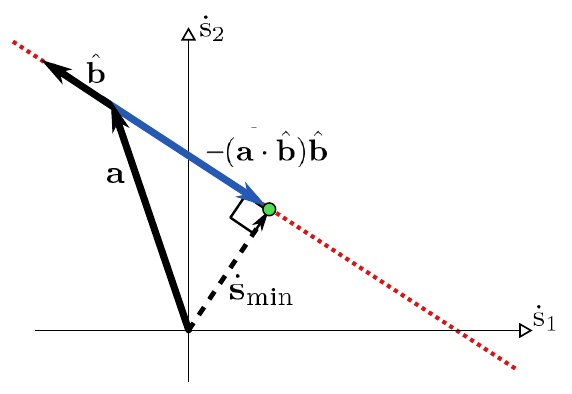}
  \caption{Visualisation of the solution space described by
    $\mathbf{a}+\mu\mathbf{b}$ (red) for $\dot{\mathbf{s}}$ in two
  dimensions, showing how the sum of of its squared components is minimised
when it is closest to the origin (green).}
  \label{fig:four_wheel_sol}
\end{figure}

The equations of motion have been kept in a general form to
accommodate the variable configuration of the flywheels. Their inertia-like 
matrix, $\mathbf{E}_{\mathrm{W}}$, comprises the rotation axes of the
flywheels, multiplied by the moment of inertia about the rotation axes of the
flywheels. The column vectors denoting the axes of rotation are concatenated
together when forming this matrix, hence the four wheel configuration will
result in $\mathbf{E}_{\mathrm{W}}$ being of shape $3\times4$. The
equation that is solved to acquire the wheel accelerations is
\[
  \mathbf{E}_{\mathrm{W}}\dot{\mathbf{s}} = \boldsymbol{\tau}_{\mathrm{W}},
  \numberthis \label{eq:wheel_torq_general}
\]
where $\boldsymbol{\tau}_{\mathrm{W}}$ is the torque vector that will be
calculated from the control law discussed in Section \ref{sec:control}. The
rank of $\mathbf{E}_{\mathrm{W}}$ is 3 in either wheel configuration. When
four wheels are used, the columns representing their axes of rotation will not
be linearly independent. This corresponds with having four variables to solve
for and only three equations with which to do it, hence the solution to the
matrix equation is not unique. This means that the solution for
$\dot{\mathbf{s}}$ will be of the form 
\[
  \dot{\mathbf{s}} = \mathbf{a} + \mu \mathbf{b}. \numberthis \label{eq:s_dot_sol}
\]
This describes a line in 4 dimensions which contains all valid solutions to
(\ref{eq:wheel_torq_general}). Parameters $\mathbf{a}$ and $\mathbf{b}$ are
constant, whilst $\mu$ can be varied to change the distribution of wheel
accelerations. A simplified diagram of this line with the minimised
$\dot{\mathbf{s}}$ vector is shown in Figure \ref{fig:four_wheel_sol}.

This case requires an additional condition to find an explicit
value for wheel accelerations. This condition was for the sum of
squared wheel accelerations to be minimised, written as
\[
  \hspace{120pt}\argmin_{\mu} |\dot{\mathbf{s}}|^2,\hspace{20pt} \text{subject to  } \mathbf{E}_{\mathrm{W}}\dot{\mathbf{s}}=\boldsymbol{\tau}_{\mathrm{W}}
\]
where
\begin{align*}
  |\dot{\mathbf{s}}|^2 &= \dot{\mathbf{s}}\cdot\dot{\mathbf{s}}\\
  &= (\mathbf{a} + \mu \mathbf{b})\cdot(\mathbf{a} + \mu \mathbf{b})\\
        &= \mu^2 |\mathbf{b}|^2 + \mu(2\mathbf{a}\cdot\mathbf{b}) + |\mathbf{a}|^2
\end{align*}

Since $|\dot{\mathbf{s}}|^2$ is a positive quadratic, its minimum is found by
setting
\[
  \dfrac{\mathrm{d}}{\mathrm{d}\mu}|\dot{\mathbf{s}}|^2 = 2\mu|\mathbf{b}|^2 +
  2\mathbf{a}\cdot\mathbf{b}=0,
\]
and then solving for $\mu$, giving
\[
  \mu = -\frac{\mathbf{a}\cdot\mathbf{b}}{|\mathbf{b}|^2}.
\]
Substituting this result into (\ref{eq:s_dot_sol}) gives the optimised
solution
\begin{align*}
  \dot{\mathbf{s}}_{\mathrm{min}} &= \mathbf{a} -
  \left(\frac{\mathbf{a}\cdot\mathbf{b}}{|\mathbf{b}|^2}\right)\,\mathbf{b}\\
  &= \mathbf{a} - (\mathbf{a}\cdot\hat{\mathbf{b}})\,\hat{\mathbf{b}},
\end{align*}
where $\mathbf{b}$ has been rescaled to a unit vector.

The minimum value for $\dot{\mathbf{s}}$ can now be expressed purely in terms
of vectors $\mathbf{a}$ and $\mathbf{b}$. The line of solutions described by these parameters is dependent on both
$\mathbf{E}_{\mathrm{W}}$ and $\boldsymbol{\tau}_{\mathrm{W}}$. However, the
inertia-like matrix, $\mathbf{E}_{\mathrm{W}}$, is fixed. This corresponds to
a fixed gradient for the line, whilst the value of
$\boldsymbol{\tau}_{\mathrm{W}}$ determines its offset from the origin.
Consequently, the gradient, $\mathbf{b}$, need only be calculated once. Solving
the special case of $\mathbf{E}_{\mathrm{W}}\dot{\mathbf{s}}=\mathbf{0}$,
where $\mathbf{0}$ denotes a vector of all zeros,
results in $\mathbf{a}=\mathbf{0}$. This is because, if 
$\boldsymbol{\tau}_{\mathrm{W}}=\mathbf{0}$, then the minimum wheel accelerations to
achieve this torque are $\dot{\mathbf{s}}=\mathbf{0}$, meaning the solution
line has no constant offset from the origin.
Performing Gaussian elimination on the augmented matrix,
$[\mathbf{E}_{\mathrm{W}}|\boldsymbol{\tau}_{\mathrm{W}}]$, yields 
\[
  \left[\begin{array}{@{}cccc|c@{}}1&0&0&\alpha&0\\0&1&0&\beta&0\\0&0&1&\gamma&0\end{array}\right],
\]
where $\alpha$, $\beta$, and $\gamma$ are determined by
$\mathbf{E}_{\mathrm{W}}$.
This result is guaranteed by the system requirement that three of the wheels must have
linearly independent axes of rotation, meaning $\mathbf{E}_{\mathrm{W}}$ will
be of rank 3. Making $\dot{\mathbf{s}}_4$ the varied parameter, $\mu$, gives
the solution
\[
  \dot{\mathbf{s}} =
  \mu\begin{bmatrix}-\alpha\\-\beta\\-\gamma\\1\end{bmatrix}= \mu\mathbf{b}, 
\]
where $\alpha$, $\beta$, and $\gamma$ are constants determined by the elements
of $\mathbf{E}_{\mathrm{W}}$.

With $\mathbf{b}$ calculated, the computation for the optimum four wheel
accelerations is simplified. The fourth element of $\mathbf{a}$ will always be
0 since $\dot{s}_4$ acts as the varied parameter in $\mu\mathbf{b}$. Hence,
the product, $\mathbf{E}_{\mathrm{W}}\mathbf{a}$, gives the same result if the
fourth column of $\mathbf{E}_{\mathrm{W}}$ is omitted. This now gives an
invertible matrix in the equation,
\[
\begin{bmatrix}E_{\mathrm{W}11}&E_{\mathrm{W}12}&E_{\mathrm{W}13}
\\E_{\mathrm{W}21}&E_{\mathrm{W}22}&E_{\mathrm{W}23}
\\E_{\mathrm{W}31}&E_{\mathrm{W}32}&E_{\mathrm{W}33}\end{bmatrix}
\begin{bmatrix}a_1\\a_2\\a_3\end{bmatrix} =
\begin{bmatrix}\tau_{\mathrm{W1}}\\\tau_{\mathrm{W2}}\\\tau_{\mathrm{W3}}\end{bmatrix},
\]
from which $\mathbf{a}$ can be calculated.
This equation is equivalent to solving for $\dot{\mathbf{s}}$ in the three
wheel case, where the solution is uniquely defined.

To summarise this procedure; $\mathbf{b}$ is calculated pre-launch by solving
$\mathbf{E}_{\mathrm{W}}\mathbf{b}=\mathbf{0}$ and setting $b_4=1$. $\mathbf{b}$ is
then rescaled to a unit vector. During operation, the control law will be used
to calculate a correcting torque, $\boldsymbol{\tau}_{\mathrm{W}}$, based on
sensor inputs. $\mathbf{a}$ is calculated by solving
$\mathbf{E}_{\mathrm{W}}^{3\times3}\mathbf{a}^{3\times1} =
\boldsymbol{\tau}_{\mathrm{W}}$, where the fourth dimensions of
$\mathbf{E}_{\mathrm{W}}$ and $\mathbf{a}$ have been omitted. The fourth
dimension of $\mathbf{a}$ is then set to $a_4=0$. Finally, the optimised solution
for $\dot{\mathbf{s}}$ is given as $\dot{\mathbf{s}}_{\mathrm{min}}=\mathbf{a}
- (\mathbf{a}\cdot\hat{\mathbf{b}})\,\hat{\mathbf{b}}$.

An example of this approach being applied is given in Appendix
\ref{sec:four_wheel_plot}. In this demonstration the four wheels
are evenly spaced about the body's $z$-axis, each with a $40^\circ$ incline to
the $x$-$y$ plane. This is the currently proposed configuration by M. Tisaev in his
evaluation of the ADCS hardware \cite{mansur_vals}. A similar configuration was
used for a three wheel case, with the wheels again equally spaced about
the $z$-axis. The sum of the wheel acceleration magnitudes for both cases are
shown to be identical. This supports that the four wheel case has been made as
efficient as the three wheel case, whilst adding a redundancy in the case of a
failure.

\section{Gyroscopic Effects}
\label{sec:gyroscopic}
\begin{figure}
  \begin{subfigure}{.5\textwidth}
    \centering
    \includegraphics[width=\textwidth]{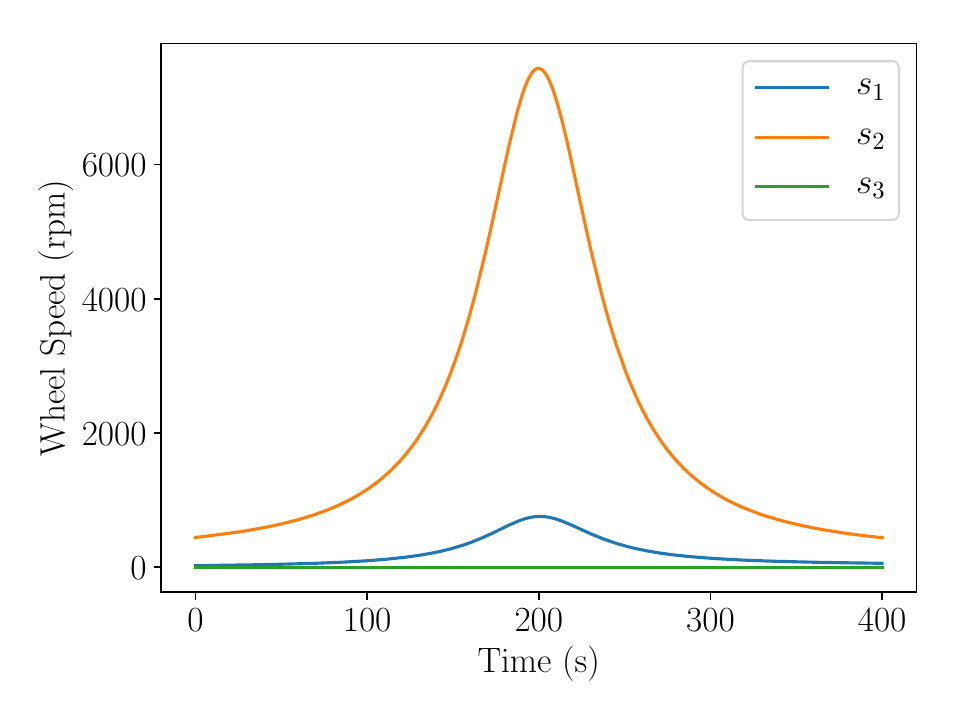}
    \caption{Speed profiles for the three wheels.}
  \end{subfigure}%
  \begin{subfigure}{.5\textwidth}
    \centering
    \includegraphics[width=\textwidth]{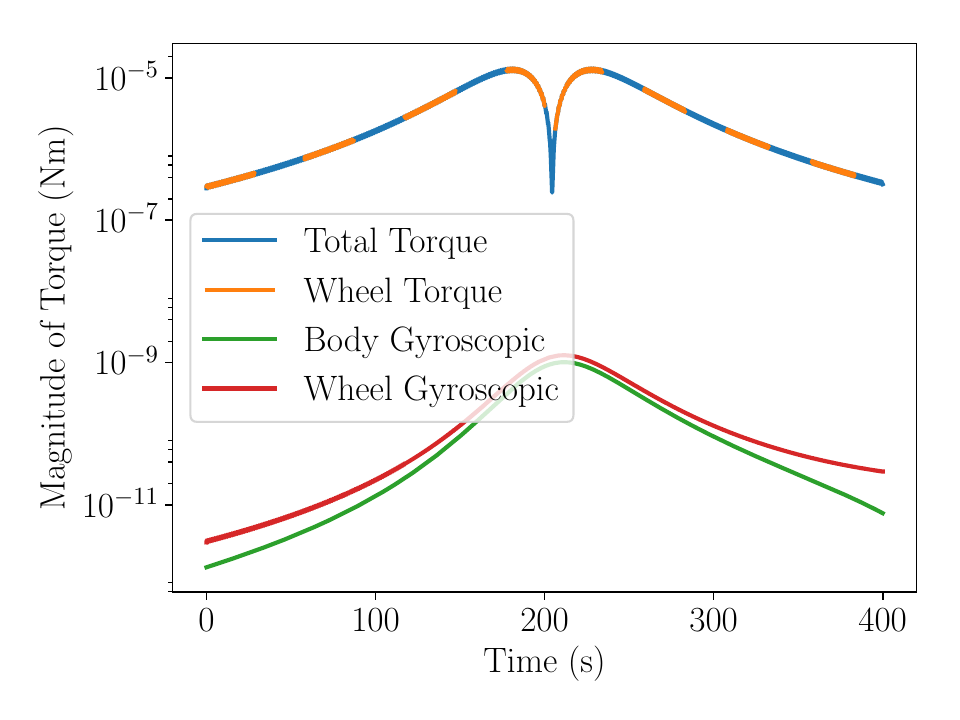}
    \caption{Torque magnitudes from the three internal sources.} 
  \end{subfigure}%
\caption{Time series plots of wheel speeds and the magnitudes of the three
  torques that occur within the
satellite, outlined in (\ref{eq:labelled_torque}). The body and wheel torque
plots have been interspersed for viewing purposes.}%
\label{fig:gyro_mags}
\end{figure}

The gyroscopic effects caused by the flywheels have previously been omitted
within the PROVE mission. However, with the equations of
motion that have now been derived, the dynamics of the flywheels can be
isolated and measured to understand their significance. This is done by
analysing the magnitude of the gyroscopic torque in comparison to the other torques that
will be occurring within the satellite. The internal sources of torque from
(\ref{eq:final ode}) are classified as follows 
\[
  \mathbf{I}_{\mathrm{Tot}}
  \dot{\boldsymbol{\omega}}_{\mathrm{B}} =
  \boldsymbol{\tau}_{\mathrm{Ext}} -
  \underbrace{\boldsymbol{\omega}_{\mathrm{B}}\times\mathbf{I}_{\mathrm{Tot}}\boldsymbol{\omega}_{\mathrm{B}} }_{\text{Gyroscopic
  response of body}}-  
  \underbrace{\mathbf{E}_{\mathrm{W}}\dot{\mathbf{s}}}_{\text{Torque from
  flywheels}} -
  \underbrace{\boldsymbol{\omega}_{\mathrm{B}}\times\mathbf{E}_{\mathrm{W}}\mathbf{s}.}_{\text{Gyroscopic
  response of flywheels}}\numberthis \label{eq:labelled_torque}
\]
By isolating these terms during a simulated flyover, the
relative scale of gyroscopic effects can be understood. 
The magnitude of these terms, and the net torque that they produce, are shown in Figure
\ref{fig:gyro_mags} alongside the corresponding flywheel speeds.
In this case, no external torque
is present and the satellite passes directly above the ground target. 
This simulation shows a large difference in magnitudes
between intentional torque and the gyroscopic reactions that occur. As a
result, the total torque produced appears to exactly match that of the
flywheels. This is expected since external torque was omitted during
this simulation.

\begin{figure}
\centering
  \includegraphics[width=0.9\textwidth]{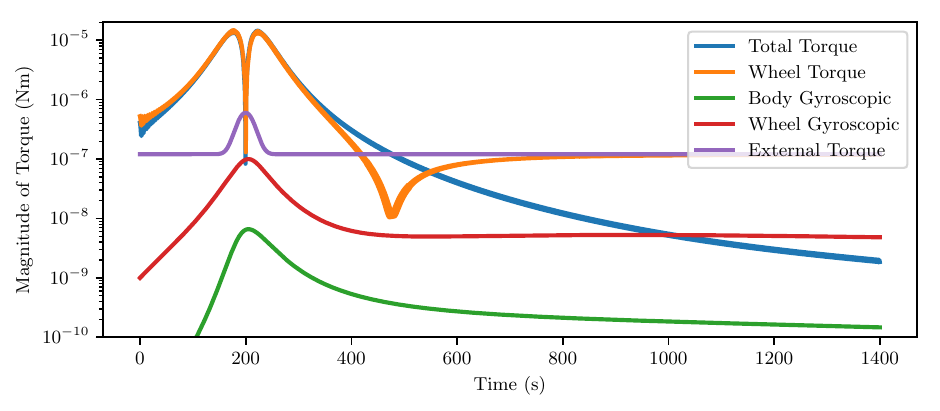}
  \caption{Longer time series plot of torque magnitudes during a flyover from
    directly above. A modelled external torque is added, directly opposing the
  direction of rotation, reaching a magnitude of 6$\times10^{-7}$\,Nm.}
  \label{fig:gyro_ext_torq}
\end{figure}

The gyroscopic terms depend on the angular velocity of the two components, whereas
the intentional torques come from their accelerations. This explains
why the plot for wheel torque drops close to zero when the
satellite is directly above the target. At this point the satellite's rotation
must begin slowing, resulting in a change of sign in its acceleration vector.
The torque never reaches zero since the target is also moving longitudinally,
requiring a small torque about a second axis.  
This change of sign
in the acceleration vectors correlates to the change in sign of the gradient of
the gyroscopic terms, since they depend on the integral of these acceleration
terms.
In the case that external torque is present, the magnitude of these gyroscopic terms can become
significant. Figure \ref{fig:gyro_ext_torq} shows the torque magnitudes from
(\ref{eq:labelled_torque}) with external torque present.
The external torque is applied in the
direction opposing the satellite's rotation as in Section
\ref{sec:external torq}. This torque is modelled to vary depending on the
orientation of the satellite as it interacts with the atmosphere. The
largest torque is felt when the satellite is directly above the target, with
its longest face presented to the oncoming drag force. At this point, the
external torque experienced by the satellite is 6$\times10^{-7}$\,Nm, which is the
maximum estimated value for this CubeSat design.
Although, at the peak of the flyover, the gyroscopic torque produced
by the flywheels is shown to become almost
as strong as their intentional torque, it is important to note that this is
being dealt with by the feedback loop in the controller. Due to gyroscopic
terms being a function of velocity, they will vary slowly, giving the
controller time to react and correct for the error they cause. This is why the
intentional wheel torque should always have a greater magnitude 
than the gyroscopic terms, as it is counteracting their torque in addition to
generating the desired change in angular velocity. This will only cease to be
the case when saturation occurs, preventing the necessary torques from being
produced. If desaturation could not occur at the start of the flyover, the
wheels would have some initial speed from counteracting external torques. This would 
result in larger gyroscopic terms and could cause the wheels to saturate during
a flyover. Hence the need for desaturation before each flyover starts. 

Throughout the orbit, the external torque both supports and opposes 
intentional wheel torque. On approach to the
target, the satellite must undergo angular acceleration, against the direction
of the external torque, thus requiring a larger magnitude of torque from the
flywheels. Once the satellite has passed directly above the
target, it must begin slowing its rotation speed, requiring an acceleration in
the opposite direction. During this period, the external torque is assisting
the torque produced by the flywheels, thus reducing their magnitude faster than
would occur in the case of no external torque.
After approximately 300 seconds of the two torques working cooperatively,
the total torque, which corresponds to the body's acceleration, dips below the magnitude of the
constant external torque. At this point, the constant external torque is
greater than that needed for the satellite's deceleration, prompting a change
of direction in flywheel torque.
Over a longer period, these values settle due to the
satellite's rotation slowing as its distance from the target increases. Towards
the end of the time series, satellite's acceleration is still decreasing. 
At this point, the satellite is now a quarter of the way
around the Earth from the target, requiring almost no change to its low
angular velocity. The flywheels equilibrate as they counteract the
constant external torque and gyroscopic terms.

Before this project, the gyroscopic effect caused by the flywheels within the
PROVE mission satellite was unknown. The magnitude of this torque is shown
to be less than the torque that can be created safely with the flywheels during
a flyover. This is helped by the slow rate at which they vary, since they
depend on the velocity of the components, instead of their acceleration.

\section{Control}
\label{sec:control}
With the equations of motion derived in Section \ref{sec:eqs_of_motion}, the
relationship between required rotation and necessary wheel torques can be
developed. In idealised conditions, this simply requires numerical integration
of the equations of motion, solving for the wheel speed acceleration at each
step. However, this rigid model would not work due to external factors, such as disturbance
torques and sensor noise. Therefore, a control system that
includes a feedback loop is necessary to produce a practical method of
calculating wheel torques.

The control law derived here builds off of previous work within the PROVE
mission, where a quaternion based, proportional-derivative controller was used.

For this project, an integration component was introduced. This comes from
adding the current attitude error, $\mathbf{q}_{\mathrm{err}}$, to the sum of
previous errors. Importantly, this sum is reduced each iteration by some
scaling factor, resulting in an exponential decay in the weighting of the errors as they
become older. Adding this term to the original control law forms a
propiortional-integral-derivative (PID) controller, taking account of the attitude
error, its integral over time, and the attitude's rate of change respectively. This gives the
full control law to be
\[
  \boldsymbol{\tau}_{\mathrm{W}} = -k_{\mathrm{p}} \mathbf{q}_{\mathrm{err}}
  - k_{\mathrm{i}} \mathbf{g}_{\mathrm{err}}
  - k_{\mathrm{d}} \boldsymbol{\omega}_{\mathrm{err}}, \numberthis
  \label{eq:control_law}
\]
where $\boldsymbol{\tau}_{\mathrm{W}}$ is the torque vector required from the
flywheels, equivalent to $\mathbf{E}_{\mathrm{W}}\dot{\mathbf{s}}$ from
(\ref{eq:final ode}), $\mathbf{g}_{\mathrm{err}}$ is the time-weighted sum of
$\mathbf{q}_{\mathrm{err}}$ terms, and the $k$ coefficients are the gains that
will be discussed in Section \ref{sec:linearisation}. Whilst solving for these
gains analytically, it became apparent that arranging the system's Jacobian
according to
axis would greatly reduce complexity when solving for its eigenvalues. For
this to be possible, the gains were split into three components, one for each
axis. This meant that the gains became diagonal matrices, giving the explicit control
law to be
\[
  \begin{bmatrix}\tau_{\mathrm{W}1}\\\tau_{\mathrm{W}2}\\\tau_{\mathrm{W}3}\end{bmatrix}
  =
  -\begin{bmatrix}k_{\mathrm{p}x}&0&0\\0&k_{\mathrm{p}y}&0\\0&0&k_{\mathrm{p}z}\end{bmatrix}\begin{bmatrix}q_{\mathrm{err}2}\\q_{\mathrm{err}3}\\q_{\mathrm{err}4}\end{bmatrix}
  - \begin{bmatrix}k_{\mathrm{i}x}&0&0\\0&k_{\mathrm{i}y}&0\\0&0&k_{\mathrm{i}z}\end{bmatrix}\begin{bmatrix}g_{\mathrm{err}1}\\g_{\mathrm{err}2}\\g_{\mathrm{err}3}\end{bmatrix}
  -
  \begin{bmatrix}k_{\mathrm{d}x}&0&0\\0&k_{\mathrm{d}y}&0\\0&0&k_{\mathrm{d}z}\end{bmatrix}\begin{bmatrix}\omega_{\mathrm{err}1}\\\omega_{\mathrm{err}2}\\\omega_{\mathrm{err}3}\end{bmatrix}.
\]
Note that the components of $\mathbf{q}_{\mathrm{err}}$ come directly from the
error quaternion describing the corrective rotation needed, the vector part of
this quaternion consists of $q_2$, $q_3$, and $q_4$.
\subsection{Terms}
\label{sec:control_terms}
The proportional term of the controller uses the attitude error vector,
$\mathbf{q}_{\mathrm{err}}$. This gives a correcting torque vector that is proportional and along
the same axis. This error vector comes from the quaternion that describes the rotation needed
from the current orientation to the target. This quaternion can be expressed
in the polar form
\[
  q_{\mathrm{err}} = \cos{\frac{\theta}{2}} +
  \hat{\mathbf{n}}\sin{\frac{\theta}{2}}. \numberthis \label{eq:polar_quat}
\]
From this quaternion, the unit vector, $\hat{\mathbf{n}}$, can be interpreted
as the axis about which to rotate in order to reach the target. The
$\sin{\frac{\theta}{2}}$ term scales this vector to increase with the angle of
error, $\theta$. This can be interpreted as the minimum required angle to reach
the target, where the maximum error of $\theta = 180^{\circ}$, results in
a scaling term of $1$. Hence, taking the last three terms of this error
quaternion gives the error vector used in the control law.

The integral term is used to correct for small constant errors that are too
small to be handled by the proportional controller. By summing the errors each
time the controller updates,
even small deviations can be registered and corrected for.
In order to avoid considering the entire error history equally, which could lead to
unpredictable behaviour, the history is continuously decayed as new terms are
added to the sum. This is done with scaling factor $t_0$, which decides the
duration over which significant consideration is given. The integral term is
therefore given to be
\[
  \mathbf{g}_{\mathrm{err}}(t) =
  \int_{0}^{t}\mathbf{q}_{\mathrm{err}}(t')\exp{\left(\frac{t'-t}{t_0}\right)}\mathrm{d}t',
\]
where $t$, represents the time up until which the integral term should
be calculated, and $t'$ is the integrated variable over this range. Note that the
convention here is that $t\geq0$, where $t=0$ refers to the same time at which
the inertial and rotating frames first align and the satellite's operation begins. 

The derivative term causes damping as the torque will be reduced if the body is
rotating towards the target too quickly. The magnitude of this of the error
vector comes from the difference
between current angular velocity, read from the satellite's gyro, and the
target angular velocity. The direction of the vector comes from $\mathbf{q}_{\mathrm{err}}$.
The magnitude of the target angular velocity depends on $\theta$ in
(\ref{eq:polar_quat}). In calculating
$\boldsymbol{\omega}_{\mathrm{tar}}$, an upper bound must be set on the speed
to avoid exceed the operational limits of the flywheels. 
An approximate value for this was calculated using M.
Tisaev's hardware performance values \cite{mansur_vals}.

The maximum angular speed of the flywheels is 10,000\,rpm, allowing
a safety margin, this will be treated as 8000\,rpm or 48000$^\circ$/s. The angular momentum of a
flywheel rotating with speed $s$ is given as
$m_{\mathrm{W}}R_{\mathrm{W}}^2\,s$.
With external torque omitted, the net angular momenta of the flywheels 
and the angular momentum of the body are equal.
Rotating about the body's $x$, or $y$, axis will be more difficult than
rotating about its $z$-axis since the CubeSat is longer about this dimension.
Using Tisaev's values for the hardware currently proposed, the satellite's moment of
inertia about its $x$-axis, will be 0.034\,kg\,m$^2$, whilst the moment of
inertia about the flywheel's axis of rotation will be $1.1\times
10^{-6}$\,kg\,m$^2$. The maximum angular speed
of the satellite can therefore be approximated as
\[
  \omega_{\mathrm{max}} =
  \frac{\frac{1}{2}m_{\mathrm{W}}R_{\mathrm{W}}^2}{I_x}= \frac{1.1\times
  10^{-6}}{3.4\times 10^{-2}}\, 48000^\circ/\mathrm{s}= 1.55^\circ/\mathrm{s}.
\]

With this limit on angular velocity, and a knowledge of the frequency of
compute cycles in the controller, a maximum angle per cycle can be calculated,
$\theta_{\mathrm{max}}$.
This scales the $\theta$ term from $\mathbf{q}_{\mathrm{err}}$ so that the
magnitude of the target angular velocity never exceeds $\omega_{\mathrm{max}}$. This
gives the target angular velocity to be
\[
  \boldsymbol{\omega}_{\mathrm{tar}} =
  \mathrm{min}\left\{\frac{\theta}{\theta_{\mathrm{max}}},
  1\right\}\,\omega_{\mathrm{max}}\,\mathbf{q}_{\mathrm{err}}.
\]

Revisiting how reaching the target was defined in Section
\ref{sec:definitions}, the body-fixed positive $z$-axis must rotate to align with the ground
target as this axis passes through the satellite's camera. 
The axis about which to
do this rotation most efficiently will always be perpendicular to both the
initial $z$-axis and the resultant $z$-axis. This indicates that the rotation
axis will always lie in the $x$-$y$ plane of satellite body frame. Hence the
$\mathbf{q}_{\mathrm{err}}$ vector will never have a $z$ component when
expressed in the body frame as shown in Figure \ref{fig:eg_error}. Because the direction of all three terms is
governed by $\mathbf{q}_{\mathrm{err}}$, the satellite will have 0
instantaneous angular velocity about the body frame's $z$-axis at all times.
The use of \textit{instantaneous} here is to account for the angular
displacement that will occur about the $z$-axis over time, but only due to
consecutive rotations about the other two axes.

\subsection{Gain Determination}
\label{sec:linearisation}
With the terms of the controller established, a method by which to acquire the
gains is required. Initially, this was done through manual tuning. 
However, an exploratory method of solving for the gains
analitically resulted in better performance and will be discussed here. This
method uses the set of differential equations that describe the system to
derive the optimal gains for controlling it. This is done by linearly
approximating the system about the equilibrium using a Jacobian matrix. 
By treating the target as stationary in the rotating body frame, the state
variables at this equilibrium can be easily determined. From here, the
characteristic polynomial of the Jacobian can be used to calculate the
relationship between the three gains in order to achieve repeated eigenvalues,
thus optimising for stability. Finally, a discretised version of the same Jacobian 
is used in order to take account of the undetermined update time of 
the controller. By setting the repeated eigenvalue of this matrix to zero,
the value of gains that produce critical damping can be attained.

Along with the two equations of motion derived in Section
\ref{sec:eqs_of_motion},
two additional differential equations
are required to represent all time dependent variables in the system. The
first equation describes the motion of the flywheels, which is governed by the
control law, and the second describes the
integrated element of the control law, $\mathbf{g}_{\mathrm{err}}$.

The integrated error is given as 
\[
    \mathbf{g}_{\mathrm{err}}(t) =
  \int_{0}^{t}\mathbf{q}_{\mathrm{err}}(t')\exp{\left(\frac{t'-t}{t_0}\right)}\mathrm{d}t',
\]
where the integration variable is $t'$. The derivative of
$\mathbf{g}_{\mathrm{err}}$, with respect to $t$, requires the Leibniz integral
rule. This allows for the derivative of an integral to be taken with respect
to the limit variable instead of the integration variable, giving
\begin{align*}
  \dot{\mathbf{g}}_{\mathrm{err}}(t) &= \dfrac{\mathrm{d}}{\mathrm{d}t}\left(\int_{0}^{t}\mathbf{q}_{\mathrm{err}}(t')
  \exp{\left(\frac{t'-t}{t_0}\right)}\mathrm{d}t'\right) \\[10pt]
  &= \mathbf{q}_{\mathrm{err}}(t) + \int_0^t \dfrac{\partial}{\partial
  t}\left(\mathbf{q}_{\mathrm{err}}(t')
  \exp{\left(\frac{t'-t}{t_0}\right)}\right)\mathrm{d}t'\\[10pt]
  &= \mathbf{q}_{\mathrm{err}}(t) - \frac{1}{t_0}\int_0^t 
  \mathbf{q}_{\mathrm{err}}(t')\exp{\left(\frac{t'-t}{t_0}\right)}\mathrm{d}t'\\[10pt]
  &= \mathbf{q}_{\mathrm{err}}(t) - \frac{1}{t_0}\mathbf{g}_{\mathrm{err}}(t).
\end{align*}

Some assumptions and simplifications have been made for this derivation in
order to reach an analytical solution that is feasible. 
The most significant of these is the omission of external torque form the
equations of motion. This is because it is an external parameter than can not be expressed
purely in terms of the system variables. This simplifies
(\ref{eq:final ode}) to become
\begin{align*}
  \dot{\boldsymbol{\omega}}_{\mathrm{B}} &=\mathbf{I}_{\mathrm{Tot}}^{-1}\left(
  \boldsymbol{\tau}_{\mathrm{Ext}} -
  \boldsymbol{\omega}_{\mathrm{B}}\times\mathbf{I}_{\mathrm{Tot}}
  \boldsymbol{\omega}_{\mathrm{B}} -  \mathbf{E}_{\mathrm{W}}\dot{\mathbf{s}} -
  \boldsymbol{\omega}_{\mathrm{B}}\times\mathbf{E}_{\mathrm{W}}\mathbf{s}\right)\\
  &= - \mathbf{I}_{\mathrm{Tot}}^{-1}\left(
  \boldsymbol{\omega}_{\mathrm{B}}\times\mathbf{I}_{\mathrm{Tot}}
  \boldsymbol{\omega}_{\mathrm{B}} +  \mathbf{E}_{\mathrm{W}}\dot{\mathbf{s}} +
\boldsymbol{\omega}_{\mathrm{B}}\times\mathbf{E}_{\mathrm{W}}\mathbf{s}\right).
\end{align*}
A further simplification is to treat the satellite
inertia matrix, $\mathbf{I}_{\mathrm{Tot}}$, as diagonal. This will prove to
simplify the characteristic equation significantly whilst maintaining a close
approximation of the real matrix. This is due to the symmetrical design of the CubeSat,
causing its principle axes of rotation almost align with the body-fixed axes.

The final simplification is to model the ground target as stationary, from the
satellite's perspective. This allows for the equilibrium values of the state
variables to be easily determined. This will prove to be an acceptable
simplification as the controller is able to maintain tracking the target,
treating its movement as another source of pointing error.

The full set of differential equations is given to be 
\begin{equation}
\begin{aligned}
  \dot{\boldsymbol{\omega}}_{\mathrm{B}} &= - \mathbf{I}_{\mathrm{Tot}}^{-1}\left( 
  \boldsymbol{\omega}_{\mathrm{B}}\times\mathbf{I}_{\mathrm{Tot}}
  \boldsymbol{\omega}_{\mathrm{B}} +  \mathbf{E}_{\mathrm{W}}\dot{\mathbf{s}} +
  \boldsymbol{\omega}_{\mathrm{B}}\times\mathbf{E}_{\mathrm{W}}\mathbf{s}\right)\\
  \dot{q} &= \frac{1}{2}q\odot\boldsymbol{\omega}_{\mathrm{B}}\\
  \mathbf{E}_{\mathrm{W}}\dot{\mathbf{s}} &= -\mathbf{K}_{\mathrm{p}} \mathbf{q}
  - \mathbf{K}_{\mathrm{i}} \mathbf{g}
  - \mathbf{K}_{\mathrm{d}} \boldsymbol{\omega}\\
  \dot{\mathbf{g}}_{\mathrm{err}} &= \mathbf{q}_{\mathrm{err}} -
    \frac{1}{t_0}\mathbf{g}_{\mathrm{err}}\label{eq:all_odes}.
\end{aligned}
\end{equation}

Note that the wheel speeds have remained multiplied by the inertia-like
tensor, $\mathbf{E}_{\mathrm{W}}$, which is done to simplify later
calculations. Since the two terms only occur as a product, they can be
expressed as a new vector, $\mathbf{s}_{\mathrm{E}}$, which is equivalent to
the net angular momentum generated by the flywheels, and its time derivative
equal to their net torque.

These multidimensional variables can now be represented by their individual
elements. Giving the scalar state variables to be
\[
  [\omega_{1} \quad \omega_{2} \quad \omega_{3} \quad q_{1} \quad q_{2} \quad
      q_{3} \quad q_{4} \quad s_{\mathrm{E}1} \quad s_{\mathrm{E}2} \quad s_{\mathrm{E}3} \quad g_{1} \quad g_{2} \quad
  g_{3}]^\top.
\]
The same decomposition is performed on the right side of
(\ref{eq:all_odes}) and is shown in Appendix
\ref{sec:decomposed_odes}. This set of equations can then be differentiated
with respect to each of the 13 state variables to form the Jacobian matrix of
the system. This Jacobian is shown in Appendix \ref{sec:jacobian}. 

With this Jacobian, the values for the state variables at equilibrium can be
substituted in. Since the target is modelled as stationary, this
system will reach equilibrium when the satellite is not rotating and there is
zero pointing error. This corresponds to each state variable becoming zero,
except for the pointing error quaternion, which becomes the identity
quaternion, $q=[1,0,0,0]^\top$, denoting no rotation.
Substituting in these terms yields a matrix in terms of just the principal 
moments of inertia and the controller gains.

Since the Jacobian is of shape $13\times13$, its characteristic polynomial will
be of degree 13.
This is not feasible for analysis as the analytical expression of the eigenvalues
would be too complex. However, rearranging the state variables by which axis
they correspond to yielded a block diagonal Jacobian, which is shown in 
Appendix \ref{sec:reordered_jacobian}.
A block diagonal matrix contains submatrices along its diagonal
and zeros everywhere else. The benefit of this is that the eigenvalues of the
whole matrix are the set of eigenvalues from each of the submatrices, resulting
in multiple characteristic polynomials with a lower degree. The Jacobian has
three $4\times4$ submatrices along its diagonal and a 0 for the final
element. The submatrices differ only in the axis that the gain and
inertia value correspond to. This allows for just one of the $4\times4$
submatrices to be analysed as the eigenvalues are the same for
the other two after changing the axis of the parameters. This is the
justification for distributing the gain values over the three dimensions,
otherwise this method would yield 3 equations for the same scalar gain
value. 

The first submatrix from the reordered Jacobian is
\[
  \begin{bmatrix}-\frac{k_{\mathrm{d}x}}{I_x}&-\frac{k_{\mathrm{p}x}}{I_x}&-\frac{k_{\mathrm{i}x}}{I_x}
  &0\\\frac{1}{2}&0&0&0\\0&1&-\frac{1}{t_0}&0\\-k_{\mathrm{p}x}&-k_{\mathrm{i}x}&-k_{\mathrm{d}x}&0\end{bmatrix}.
\]
The characteristic polynomial of this matrix is then calculated to be
\[
  \lambda\left(\lambda^{3} + \frac{I_x+k_{\mathrm{d}x}t_0}{I_xt_{0}}
  \lambda^{2} + \frac{2k_{\mathrm{d}x}+k_{\mathrm{p}x}t_0}{2I_{x} t_{0}}\lambda +
\frac{k_{\mathrm{i}x} t_0 + k_{\mathrm{p}x}}{2I_{x}t_0}\right).\numberthis
\label{eq:charpoly}
\]
Dividing by $\lambda$, whilst noting that one eigenvalue will be 0, gives a
cubic equation that can be solved for the three gains that act on the $x$-axis.
To achieve critical damping, the system must have 3 repeated eigenvalues,
meaning the characteristic polynomial will be of the form
\[
  (\lambda - a)^3 = \lambda^{3} - 3 a \lambda^{2}+ 3 a^{2}\lambda -
  a^{3}.\numberthis \label{eq:poly_pattern}
\]
Matching the coefficients from (\ref{eq:charpoly}) and
(\ref{eq:poly_pattern}) provides three equations that can be used to solve
for $a$, and two of the gains, in terms of the third. Choosing this gain to be
$k_{\mathrm{d}x}$ results in the simplest equations, giving
\begin{align*}
  a &= - \frac{I_{x} + k_{\mathrm{d}x} t_{0}}{3\, I_{x} \,t_{0}}\\[10pt]
  k_{\mathrm{p}x} &= \frac{2 \left(I_{x}^{2} - I_{x}
  k_{\mathrm{d}x} t_{0} + k_{\mathrm{d}x}^{2} t_{0}^{2}\right)}{3\, I_{x}\,
  t_{0}^{2}}\\[10pt]
  k_{\mathrm{i}x} &= \frac{2\left(k_{\mathrm{d}x} t_{0}- 2
  I_{x}\right)^{3}}{27\, I_{x}^{2} \,t_{0}^{3}}.
\end{align*}

These equations express the values that the proportional and integrated gains
must take on in terms of the derivative gain. They also show that the repeated
eigenvalue is equal to $a$, guaranteeing stability as 
$I_x$, $k_{\mathrm{d}x}$, and $t_0$ are all positive.

In order to find an analytical solution for $k_{\mathrm{d}x}$, the linearised
system must first be discretised. This is to represent the
discrete time step of the controller, which will play a significant role in the
performance of the controller. To achieve this, Euler's method
of numerical integration is used, representing the state variables in discrete
time steps that are adjusted by the jacobian matrix scaled by step size. This
is represented as
\begin{align*}
  \mathbf{x}_{n+1} &= \mathbf{x}_{n}+\Delta t \,\mathbf{J}\,\mathbf{x}_{n}\\
  &= (\mathbf{I}_{\mathbf{d}}+\Delta t\,\mathbf{J})\mathbf{x}_{n}, \numberthis
  \label{eq:discretised}
\end{align*}
where $\Delta t$ is the time step of the controller,
and $\mathbf{J}$ is the system's Jacobian matrix. 
Focussing again on just the first submatrix along the diagonal of the composite matrix shown in
(\ref{eq:discretised}) gives
\[
  \left[\begin{matrix}1 - \frac{\Delta t k_{\mathrm{d}x}}{I_{x}} & - \frac{2
      \Delta t\left(I_{x}^{2} - I_{x} k_{\mathrm{d}x} t_{0} + k_{\mathrm{d}x}^{2}
    t_{0}^{2}\right) }{3 I_{x}^{2} t_{0}^{2}} & - \frac{2 \Delta t\left(
     k_{\mathrm{d}x} t_{0}- 2 I_{x}\right)^{3}}{27 I_{x}^{3} t_{0}^{3}}  &
    0\\\frac{\Delta t}{2} & 1 & 0 & 0\\0 & \Delta t &  1- \frac{\Delta t}{t_{0}} &
    0\\- \Delta t k_{\mathrm{d}x} & - \frac{2 \Delta t\left(I_{x}^{2} - I_{x}
    k_{\mathrm{d}x} t_{0} + k_{\mathrm{d}x}^{2} t_{0}^{2}\right)}{3 I_{x}
    t_{0}^{2}} & - \frac{2 \Delta t \left(k_{\mathrm{d}x}
t_{0}- 2 I_{x}\right)^{3} }{27 I_{x}^{2} t_{0}^{3}}& 1\end{matrix}\right],
\]
where the gains, $k_{\mathrm{p}x}$ and $k_{\mathrm{i}x}$, have been expressed
in terms of $k_{\mathrm{d}x}$.

The eigenvalues of this matrix are
\begin{align*}
  \lambda _{1,2,3} &= 1 - \frac{\Delta t}{3t_0} - \frac{\Delta t k_{\mathrm{d}x}}{3
  I_x}\numberthis \label{eq:lamda_initial}\\
  \lambda _{4} &= 1,
\end{align*}
where there is a repeated eigenvalue as with the original system.

With discretised Jacobian matrices, stability occurs when the eigenvalues have
an absolute value less than 1, instead of the non positivity contraint present in
the original system. Critical damping occurs when $\lambda_{1,2,3}=0$.
Substituting this value into (\ref{eq:lamda_initial}) and rearranging gives the 
solution for $k_{\mathrm{d}x}$ to be
\[
  k_{\mathrm{d}x} = \frac{3I_{x}}{\Delta t}-
  \frac{I_{x}}{t_{0}}. \numberthis \label{eq:kdx}
\]
This result is reasonable
in that the strength of the gains should increase with the satellite's inertia, and
should decrease as the time step, $\Delta t$, increases. This avoids
overshooting before the orientation information has been updated. The value for
$t_0$ was found to have little effect on the magnitude of the gains when
compared to the other parameters. This is demonstrated by varying the value for
$t_0$ and plotting the resulting gains in one axis, shown in Appendix \ref{sec:t0}.

Although this value for $k_{\mathrm{d}x}$ would elicit critical damping
in a truly linear system, it will not be suitable for the nonlinear case. 
In practice, using the gains at this magnitude will cause the
controller to become unstable as it reacts to small nonlinear errors.
By reducing the magnitude of the gains, the controller will become less 
reactive and in turn become more stable. 
Hence, (\ref{eq:kdx}) acts as an upper limit on the gain magnitudes. A
scaling coefficient for $\mathbf{K}_{\mathrm{d}}$ is therefore introduced, allowing for all
three gain matrices to be scaled appropriately as
they are all expressed in terms the elements of $\mathbf{K}_{\mathrm{d}}$.

This gives the expressions for the automatic tuning of the control law gains to
be
\begin{align*}
  k_{\mathrm{d}i} &= \rho\left(\frac{3 I_{i}}{\Delta t}-
  \frac{I_{i}}{t_{0}}\right)\\[10pt]
  k_{\mathrm{p}i} &= \frac{2 \left(I_{i}^{2} - I_{i}
  k_{\mathrm{d}i} t_{0} + k_{\mathrm{d}i}^{2} t_{0}^{2}\right)}{3\, I_{i}\,
  t_{0}^{2}}\\[10pt]
  k_{\mathrm{i}i} &= \frac{2\left(k_{\mathrm{d}i} t_{0}- 2
  I_{i}\right)^{3}}{27\, I_{i}^{2} \,t_{0}^{3}},
\end{align*}
where $i$ represents the respective axis and $\rho$ is the scaling coefficient.

\begin{figure}
  \centering
\begin{subfigure}{0.5\textwidth}
  \includegraphics[width=\textwidth]{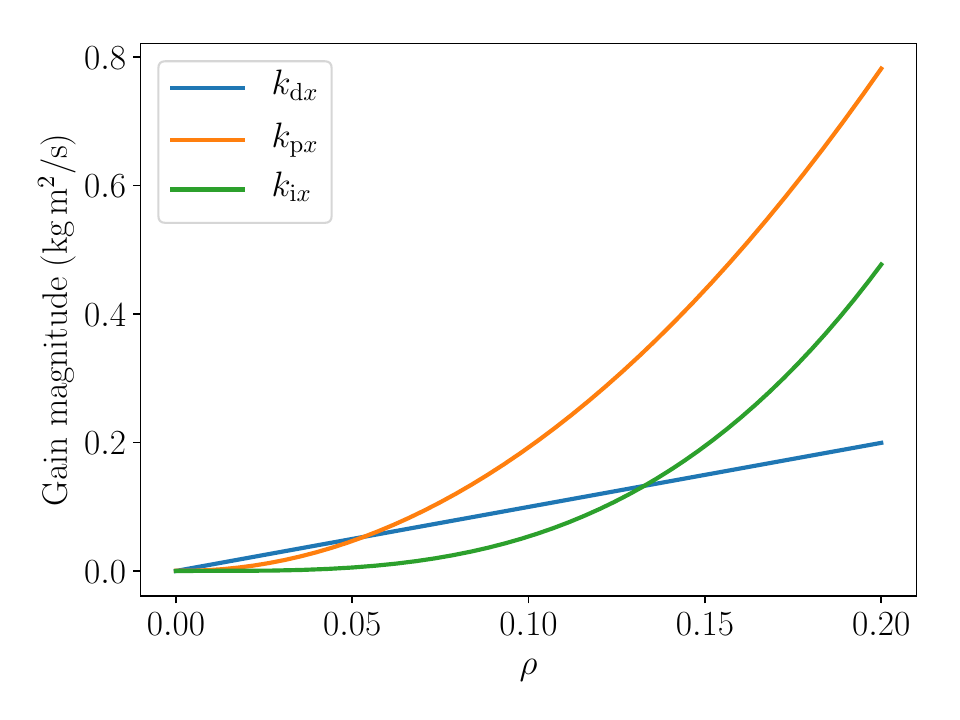}
  \caption{Magnitude of each gain's $x$ component as 
  $\rho$ is varied.}
  \label{fig:scaled_gains}
\end{subfigure}%
  \begin{subfigure}{0.5\textwidth}
  \includegraphics[width=\textwidth]{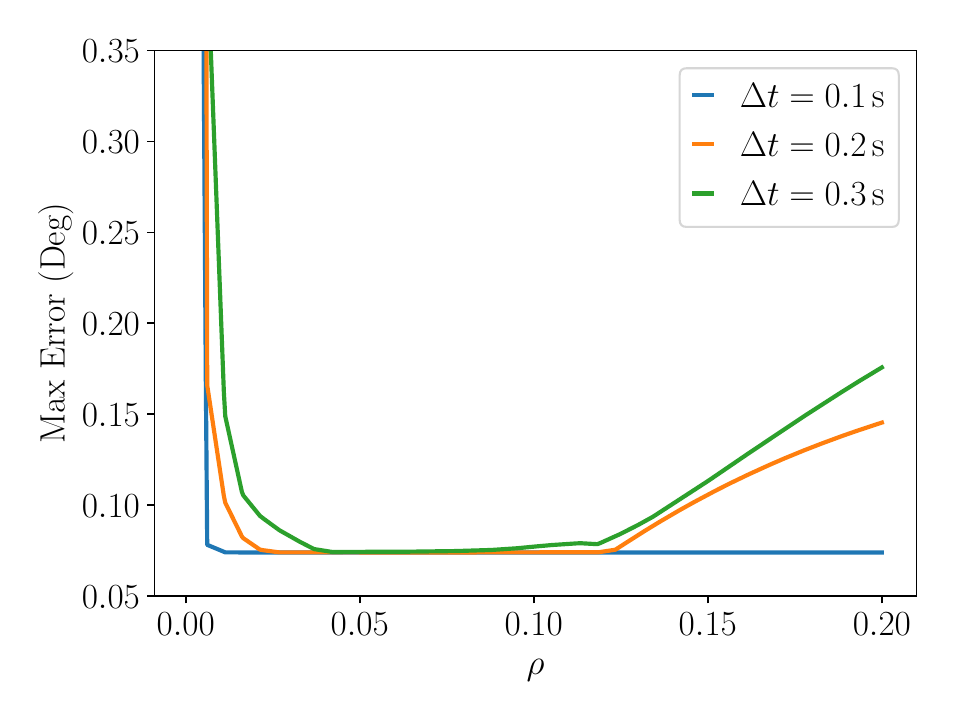}
  \caption{Maximum error as $\rho$ is varied.}
  \label{fig:gain_error}
\end{subfigure}%
\caption{The effect of varying the gain coefficient, $\rho$, on each of the gains and
  the resultant maximum pointing error to occur during a flyover from directly
above. Controllers with update times of 0.1, 0,2, and 0.3 seconds are compared.}
\label{fig:gain_coefficient}
\end{figure}

The scaling coefficient varies the elements of $\mathbf{K}_{\mathrm{d}}$
linearly whilst causing a nonlinear change in the $\mathbf{K}_{\mathrm{p}}$ and
$\mathbf{K}_{\mathrm{i}}$ elements in order to maintain optimal tuning. This
can be seen in Figure \ref{fig:scaled_gains}, showing the relationship between each
$x$-axis element of each gain matrix and the scaling coefficient. In order to select a value
for $\rho$, flyovers were simulated for a range of values whilst measuring the
maximum pointing error to occur. All other parameters were set to their default
values, shown in Appendix \ref{sec:params_table}. The results of this sweep are shown in Figure
\ref{fig:gain_error}. 

For values of $\rho$ close to zero, the gain magnitudes cannot generate a
torque large enough to rotate the satellite as necessary. This means that the satellite
cannot catch up with the target during the flyover, causing pointing error to
increase dramatically. For larger values of $\rho$, errors increase due to the
gains being too strong. Since these gains are based on a linear approximation
of the system, they only work for small perturbations from zero
pointing error. The stronger these gains are, the smaller this perturbation
must be, otherwise the satellite continuously overshoots the target, resulting
in sustained oscillations. This over shoot reaches 0.1$^\circ$ for 
the 0.3\,s controller when $\rho=0.2$, whilst the 0.1\,s
controller updates fast enough to avoid the effect. Since the update time of
the controller is not yet known, all three speeds will be considered when
analysing the controler. Therefore the gain coefficient of 0.05 was chosen, so as
to maximise the stability of all three controllers whilst maintaining the ability
to produce the necessary torque.

This coefficient safely minimises pointing error during a flyover
manoeuvre. This will also allow for attiude control during the
satellite's standby mode, when no ash cloud is present. However, 
when no ashcloud is present, the satellite will only need to maintain its orientation,
either relative to the sun or Earth. The attitude control system can therefore
prioritise power and use a controller with smaller gains, generating less
torque. 
This controller could be attained by lowering the gain coefficient to 0.01, at which point the integral
term becomes essentially zero, and the maximum torque from the controller will
be reduced. This would be adequate for the less stringent attitude requirements
during its standby mode.

\section{Analysis}
\label{sec:analysis}
In this section, the performance of the newly designed controler will be
assessed during flyovers of the ground based target.
The flyover manoeuvre will first be simulated under idealised conditions, for
which the sensor readings are precisely accurate, then with different
magnitudes of noise added to test the robustness of the controller.
From the PROVE mission specification, the control system should
maintain a pointing accuracy of around $0.1^\circ$ in perfect conditions, limited only
by operational constraints such as maximum wheel speed and controller update
frequency. This is with the intention of maintaining a
pointing accuracy of $1^\circ$ when sensor errors and external torques are
present.

The controller incorporates sensor readings into a feedback loop to correct for
errors in attitude. Calculating these errors and the resulting correction
relies on three inputs, which are defined in Table \ref{tab:cont_inputs}.
Although the satellite's position is only updated once every 12 hours, the
this is given in the form of a trajectory, meaining the position can be implicitly attained with
just a value of time. The other two inputs rely on explicitly calculated values
from on-board sensors. The frequency that these can be updated is not yet
known, as it depends on the capabilities of the chosen hardware.
Estimates for the minimum update times of each input are shown in Table
\ref{tab:cont_inputs}, however these will be varied in the simulations to
determine if longer update periods are feasible. This could allow for cheaper
components to be purchased, where possible, or reduce the power consumption
from the ADCS.

\begin{table}
  \centering
\begin{tabular}{l|l|l|l}
  \multicolumn{1}{c|}{Input Value} & \multicolumn{1}{c|}{Source}  &
  \multicolumn{1}{c|}{Standard Deviation}
  & \multicolumn{1}{c}{Minimum Update Time}    \\ \hline\hline
Position                        & Ground-based broadcast      &
$0$\,km\,-\,$2$\,km & 12\,h\\
Attitude                     & Sun sensor and magnetometer
&$0.1^\circ$\,-\,$0.25^\circ$                  & 1\,s      \\
Angular velocity                & Gyro sensor                 & $0.042^\circ /\mathrm{s}$                 & 0.1\,s
\end{tabular}
\caption{Showing the performance of the three input values for the attitude
  controller. The position values are estimed by K. Riesing \cite{tle}, whilst
  the orientation values, which use a Kalman filter for increased accuracy,
  come from R. Biggs' project \cite{kalman_star_tracker}.
  The gyro values are taken from the datasheet for the
  \textit{NanoMind A3200} \cite{gyro_data}, which is the computer that
contains the gyroscope being used for this satellite.}
\label{tab:cont_inputs}
\end{table}

\subsection{Idealised Conditions}
\begin{figure}
  \centering
  \vspace{-80pt}
  \includegraphics[width=0.9\textwidth]{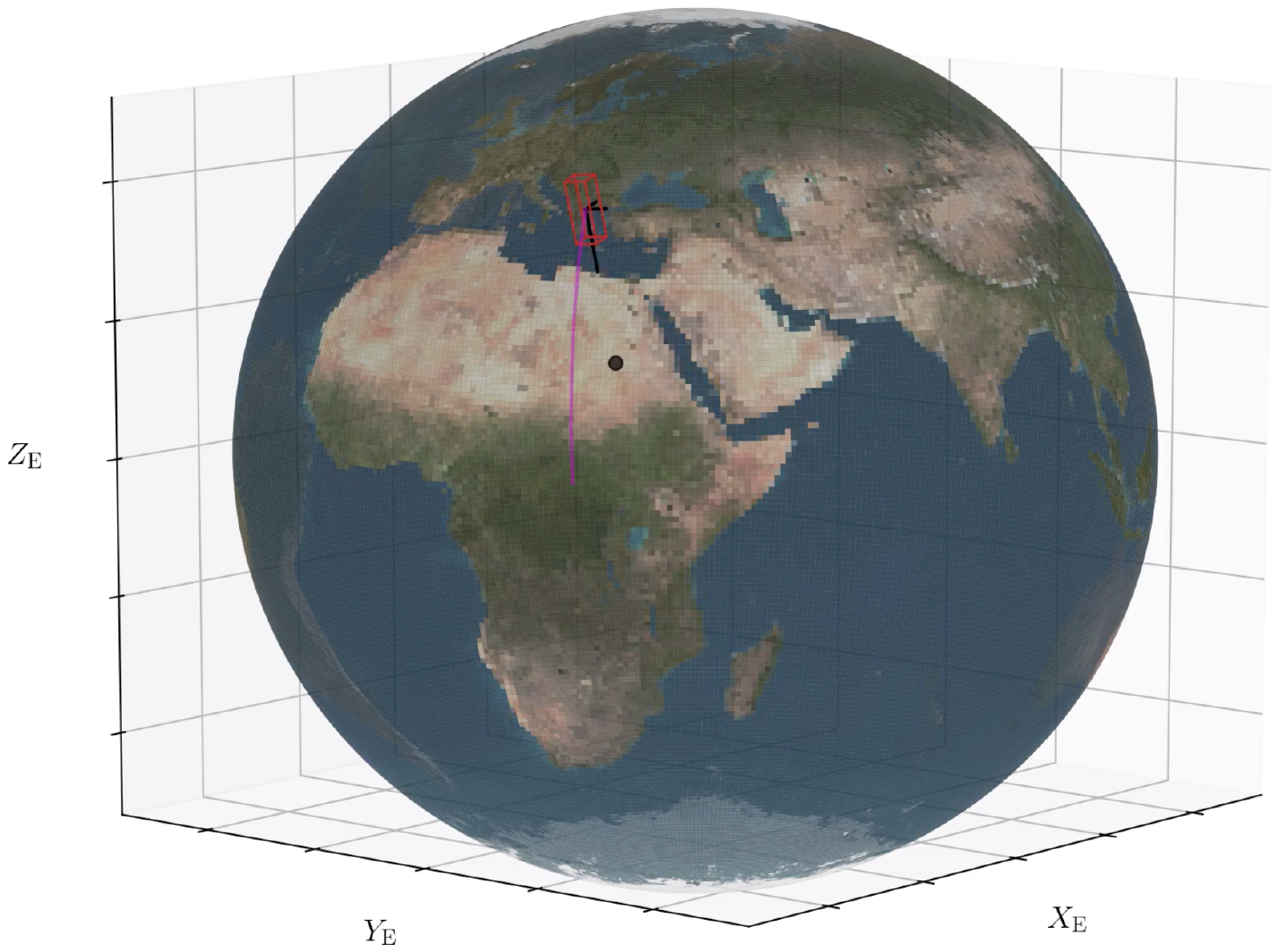}
  \caption{Example flyover manoevre, traced by the magenta line.}
    
  \label{fig:eg_flyover}
  
  \vspace{15pt}
\begin{subfigure}{.5\textwidth}
    \centering
    \includegraphics[width=\textwidth]{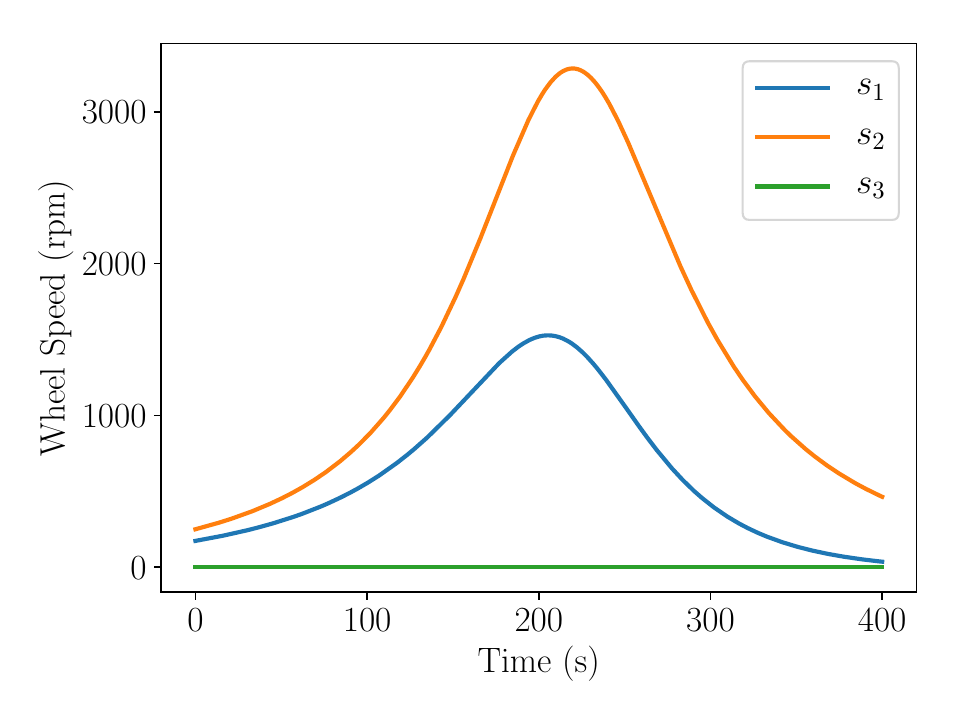}
    \caption{Speed profiles for the three wheels.}
    \label{fig:eg_speeds}
\end{subfigure}%
\begin{subfigure}{.5\textwidth}
    \centering
    \includegraphics[width=\textwidth]{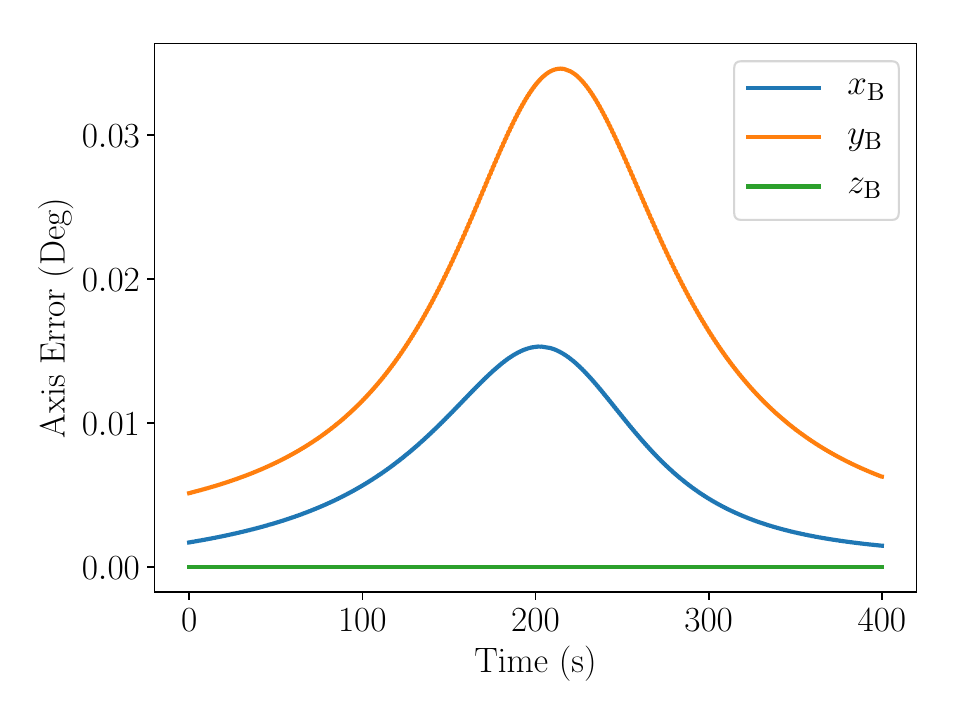}
    \caption{Pointing error in the body frame.}
    \label{fig:eg_error}
\end{subfigure}%
\caption{Time series plots of wheel speeds and body axis errors during the
flyover shown in Figure \ref{fig:eg_flyover}.}%
\label{fig:eg_time_series}
  
  \vspace{21pt}
\begin{subfigure}{.5\textwidth}
  \centering
  \includegraphics[width=\textwidth]{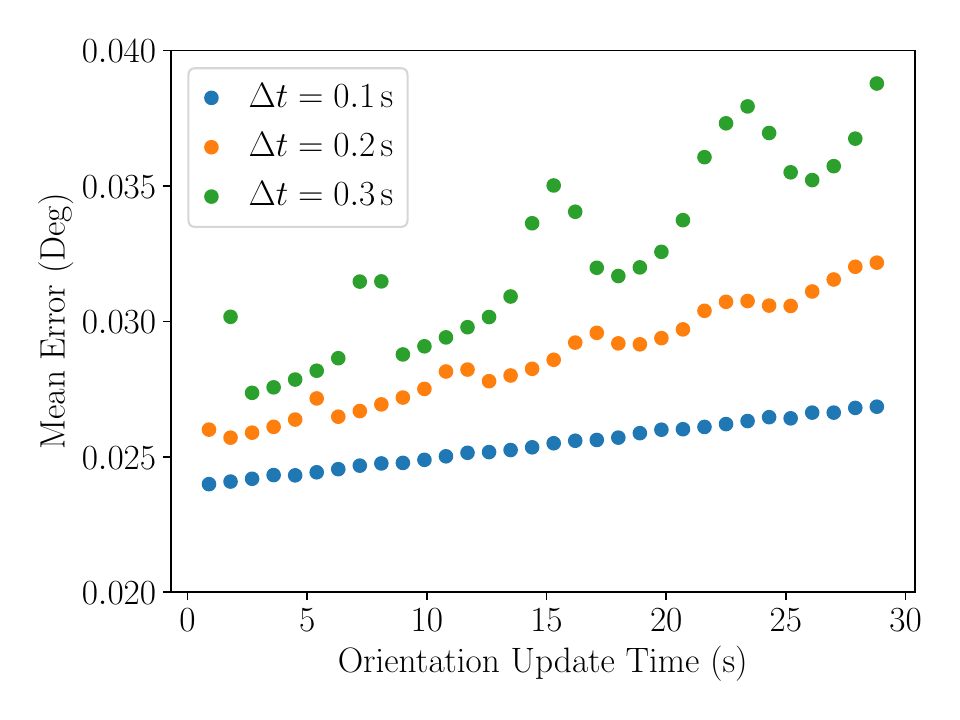}
  \caption{Average pointing error.}
\end{subfigure}%
\begin{subfigure}{.5\textwidth}
  \centering
  \includegraphics[width=\textwidth]{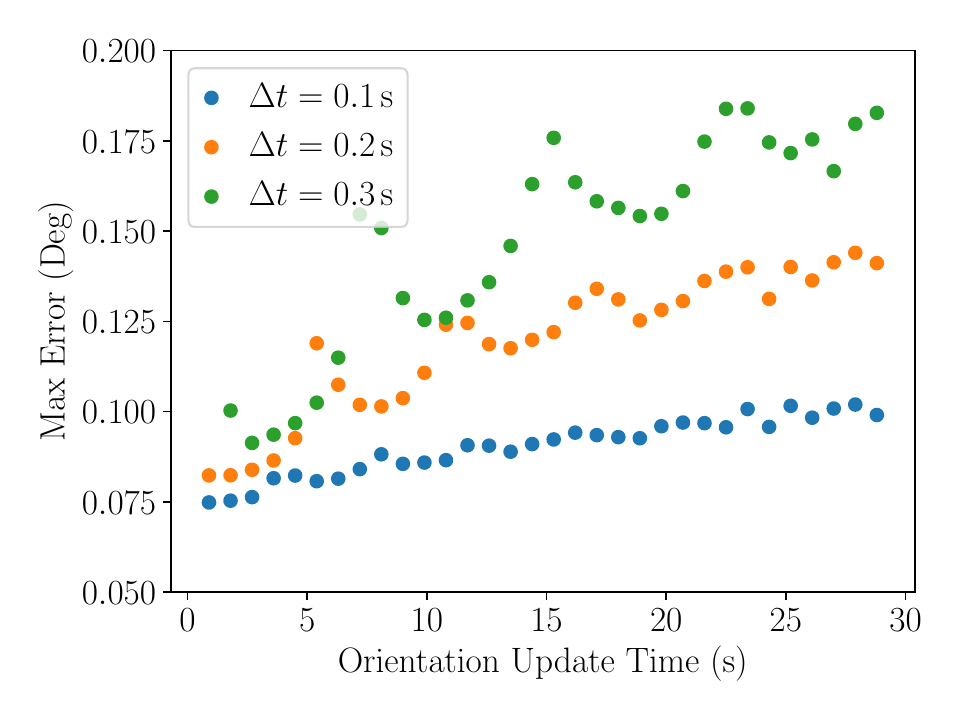}
  \caption{Maximum pointing error.}
  \label{fig:lag_err_sweep_max}
\end{subfigure}%
  \caption{Mean and maximum pointing errors during flyovers from
    directly above with varied orientation update rates. Controllers with
  compute cycles occurring every 0.1, 0.2, or 0.3 seconds are compared.}
  \label{fig:lag_err_sweep}
\end{figure}

For the following simulations, unless stated otherwise, the parameter values
that will be used are shown in Appendix \ref{sec:params_table}.
The satellite has been modelled to follow a polar orbit, with an altitude of
300\,km and a period of 1.5 hours. This means the satellite will coincide with
approximately the same location on Earth every 12 hours. Throughout the
entirity of its orbit, the satellite will remain pointing at the ash could,
requiring minimal torque when in the opposite hemisphere.
An example flyover manoeuvre is shown in Figure \ref{fig:eg_flyover}.
The error along each of the body axes and the corresponding wheel speeds from
the same flyover are shown in Figure
\ref{fig:eg_time_series}. No sensor noise has been introduced at this point.
This acts as a baseline from which to compare results with noise added in.
This example flyover spans 400 seconds, the approximate
period in which valid images of the ash cloud can be taken. During the
satellite's approach to the target, its orientation must change more rapidly the closer
it becomes, meaning it becomes increasingly hard to maintain pointing
accuracy. This is consistent with the pointing errors
shown in Figure \ref{fig:eg_error}, where the satellite is directly above the
target at $t=200$\,s. 

Although the axis error plot will strongly correlate to the measured pointing
error, they do not describe the same value. Axis error corresponds to the
components of the control law's $\mathbf{q}_{\mathrm{err}}$ term,
which is calculated at the start of each time step, $\Delta t$. The control law gives the
optimum torque to correct for this error in the time step. 
To gain an accurate measure of pointing error within this
discretised system, the recorded pointing error is the average of the
error at the start of the time step, when it is greatest, and the error after
the torque has been applied for $\Delta t$ seconds, when error is at its minimum. 
It is important to note that the $z$-axis
error will always be 0$^\circ$ because of how pointing error is defined,
as discussed at the end of Section \ref{sec:control_terms}.
The strong correlation between axis error and wheel speeds 
is due to the default wheel orientations being used whereby they rotate about
the body frame axes. Using a different set of rotation axes would result in
different speed profiles for the wheels. An example of this case can be seen in
Appendix \ref{sec:four_wheel_plot}.

Table \ref{tab:cont_inputs} shows that attitude is updated at most once per
second. This is because the orientation is
determined by collecting data from the light sensors on each surface of the
satellite, and the magnetometer, which is then passed through a Kalman filter. 
This process is computationally costly and should only be done as
frequently as necessary to save power. The gyro, however, is much more
efficient. It has a minimum update time of 0.1 seconds as it is only limited by
the update time of the controller that uses it. To account for this disparity
in update frequencies, the controller extrapolates the satellite's attitude
from its angular velocity. This provides more up to date values of attitude
than can be provided by the attitude sensors alone. This extrapolation is based
on the quaternion derivative shown in (\ref{eq:quat_deriv}) and uses
the Euler method such that
\[
  q_{n+1} = q_{n}+\frac{1}{2}\Delta t q_{n}\odot\boldsymbol{\omega}_{\mathrm{B}n}
\]
where $\Delta t$ is the update time of the controller. This was found to increase
pointing accuracy in all simulated cases. The impact of
attitude being calculated from two separate sources is explored in Section
\ref{sec:noise}. Even without noise, the performance of the
controller is affected by the size of $\Delta t$, as shown in Figure
\ref{fig:lag_err_sweep}. Here, the attitude update time is varied whilst
measuring both the mean and maximum pointing errors of three controllers.

\begin{figure}
  \begin{subfigure}{.5\textwidth}
  \centering
  \includegraphics[width=\textwidth]{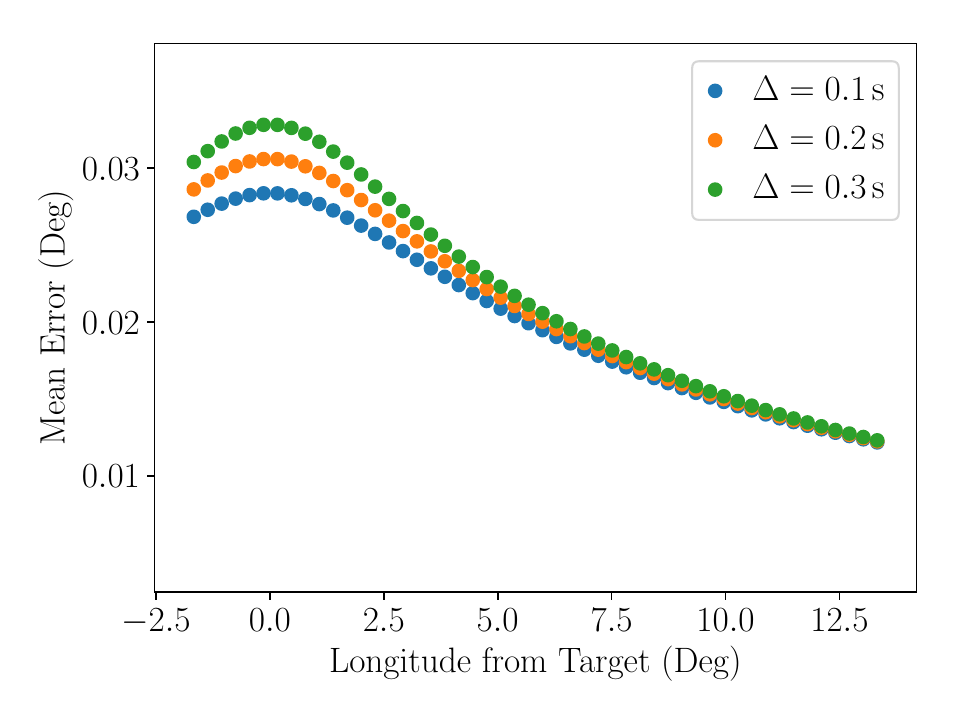}
  \caption{Average pointing error.}
  \end{subfigure}%
  \begin{subfigure}{.5\textwidth}
  \centering
  \includegraphics[width=\textwidth]{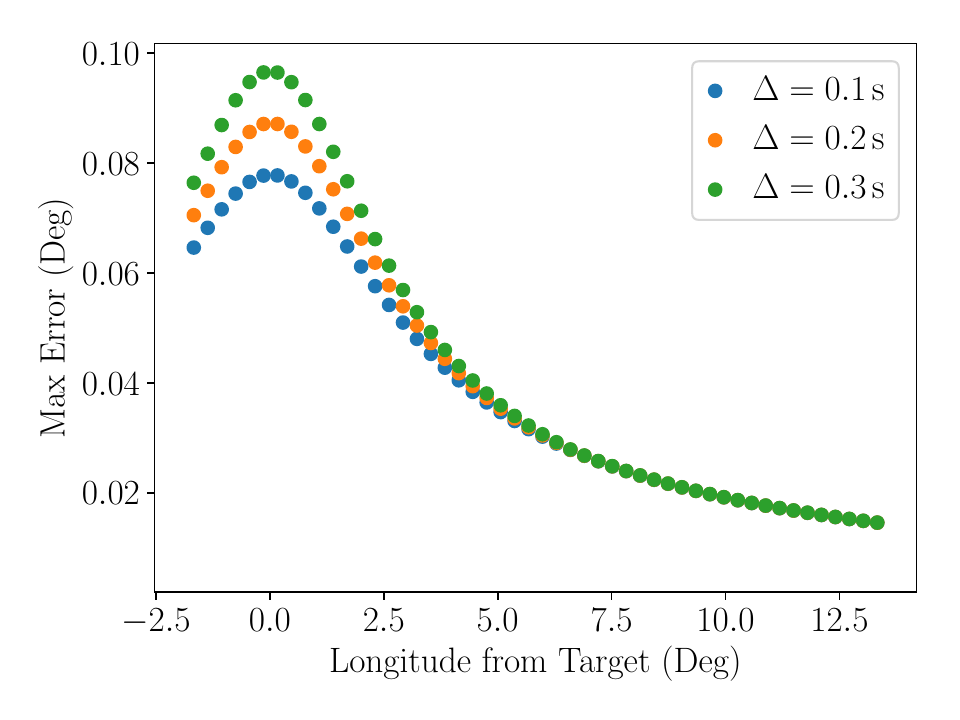}
  \caption{Maximum pointing error.}
  \end{subfigure}%
  \caption{The mean and maximum pointing error to occur during flyovers
    from varying angles. The maximum pointing error occurs when the satellite passes
    directly above the target, whilst the flyovers that require less sharp
  turning can maintain a better accuracy.}%
  \label{fig:flyover_angle}
\end{figure}

Each flyover in this plot lasts 400 seconds and had the satellite pass
directly above the ground target, which requires the sharpest turn and thus
causes the largest error. The best performance shown here achieves
0.075$^\circ$ maximum pointing error. Comparing this to the example flyover, where the
satellite passes 4$^\circ$ west of the target, the maximum error is just
0.041$^\circ$. The impact of flyover angle on pointing accuracy is shown in Figure
\ref{fig:flyover_angle}. With this in mind, the
worst-case scenario of flying directly overhead will be used for any
simulations used to test controller performance to confirm that the results are applicable to
any manoeuvre.

The max error values in Figure \ref{fig:lag_err_sweep} show that all
controllers can achieve approximately $0.1^\circ$ pointing accuracy as long as the
orientation update time is shorter than 5 seconds. This suggests that
the analytical approach used to tune the gains is valid. However, despite using
a low gain coefficient, the controller
can still respond chaotically to pointing errors that are in the
range of approximately 10$^\circ$ and above. This was solved
by gating the $\mathbf{q}_{\mathrm{err}}$ term to limit the magnitude of
pointing error that could be given to the controller. 
Therefore, considering the case that the satellite is pointing the
opposite direction to the target. Instead of accelerating to an unsafe
speed and overshooting, a constant angular velocity is used to slowly arrive
within safe proximity of the target. This works more as a precaution since the
controller should never reach a pointing error this large.

\begin{figure}
  \begin{subfigure}{.5\textwidth}
    \centering
    \includegraphics[width=\textwidth]{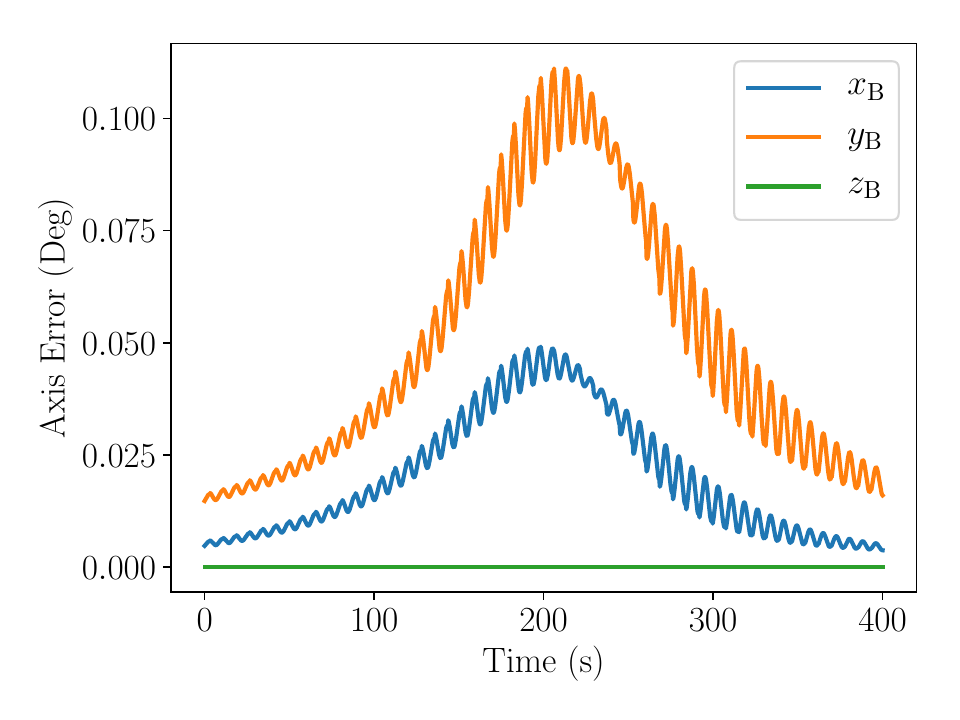}
    \caption{7.8 second orientation update time.}
  \end{subfigure}%
  \begin{subfigure}{.5\textwidth}
    \centering
    \includegraphics[width=\textwidth]{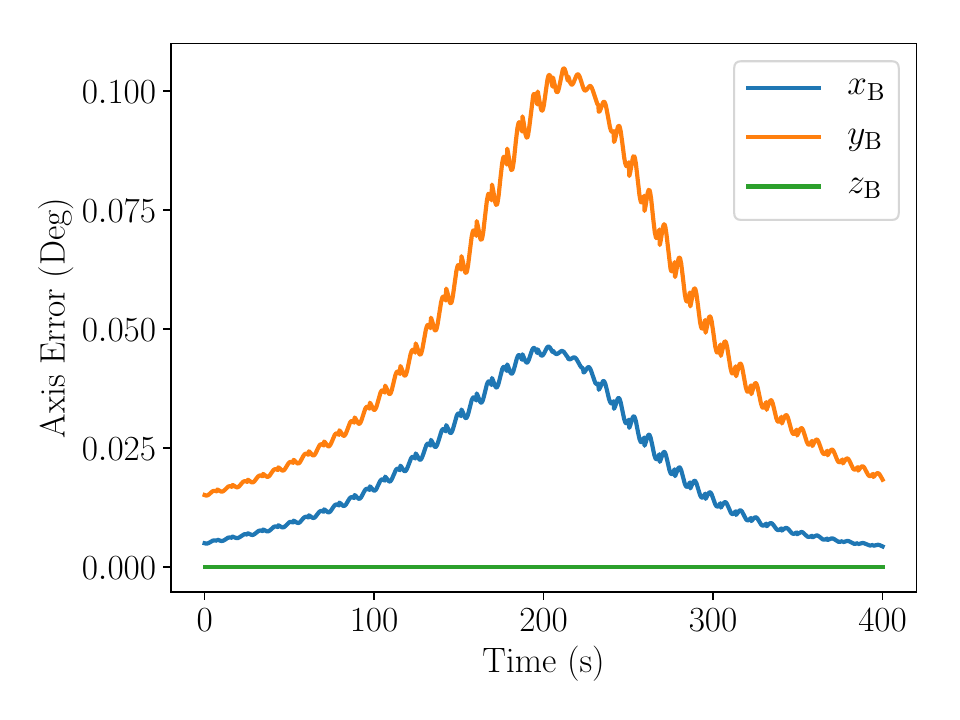}
    \caption{9 second orientation update time.}
  \end{subfigure}%
  \caption{Time series plots of axis errors during the same
flyover shown in Figure \ref{fig:eg_flyover}. Orientation update times are
compared to show the special case of when unstable oscillations are exacerbated 
by the timing of orientation updates.}%
\label{fig:err_slow_update}
\end{figure}

It is worth noting the undulations that occur in the error values of Figure
\ref{fig:lag_err_sweep}. These can be explained by analysing the small
oscillations exhibited by the satellite as
a result of the sparse orientation updates. In Figure
\ref{fig:lag_err_sweep}, the case of
$\Delta t=0.3$\,s, with orientation updated every 7.8 seconds,
shows a disproportionately large maximum pointing error when compared to the 9
second case plotted to its right.
Time series plots for these two cases are shown in
Figure \ref{fig:err_slow_update}. Looking first at the 7.8 second case, the
transient response to the orientation being updated shows the attitude
correcting and overshooting, then overshooting again in the other direction due
to the changing location of the target. The period of this transient
oscillation coincides with the 7.8 second update time of orientation, meaning
that the overshooting behaviour is encouraged by the poorly timed updates. Then
looking at the 9 second case, the period of the transient oscillation is the
same, however the longer update time avoids the oscillation being continuous,
leading to less significant overshooting.

It should be noted that this phenomenon only seems significant in idealised conditions. The largest
deviation from expected error as a result of this occurance is around
0.03$^\circ$, whereas variations in sensor noise will affect pointing error by
more than 10 times this amount.

Furthermore, simulations were run for orientation update times up to
30 seconds, in order to clearly show its effect on pointing
error. However, the value for this update time will realistically be in the
range of 1 to 5 seconds. Focussing on this range in the plots from Figure
\ref{fig:lag_err_sweep} shows that controller update times have a significantly
larger effect on the pointing error. This is to be expected since the gyro is
updated at this rate and can be used to approximate attitude. Consequently,
controller update time will be focussed on when assessing the controller's
performance in the presence of noise.

\subsection{Noise}
\label{sec:noise}
The error value of the gyro, in Table \ref{tab:cont_inputs}, comes from its
bias instability, which can be
interpreted as the standard deviation from the correct angular rate. This comes
from the datasheet for the gyro so is assumed to be accurate. Conversely,
the attitude sensor error is the least reliable of three. The values
were attained from R. Biggs' report which only used an update time of $\Delta t = 0.2$\,seconds. This means that
the noise added to the attitude sensors when $\Delta t\neq0.2$\,s may not
accurately represent the standard deviation they would experience. However,
controller performance will be assessed for a range of noise magnitudes,
meaning these standard deviations serve more as an approximate value from which
to vary noise.

Varying the noise magnitudes beyond their estimated value will
help develop safe margins of operation with respect to sensor error and provide
a measure of how sensitive the controller is to each input. It is important to
note that the errors in the attitude sensors are assumed to be independent of
each other, meaning they have zero mean. This is not the case for the position
and angular velocity readings. This is because the satellite's position will drift
over time from its estimated location due to small external forces. If the
position estimate is 1\,km north of the satellite's real position, then one
minute later, the position estimate will still be approximately 1\,km north and will
continue to diverge in the same approximate direction until a new TLE has
been attained. Given the long period over which this divergence happens, the position error
is modelled as constant over the 400 second flyover. Therefore, the
position of the satellite with simulated error is given as
\[
  \bar{\mathbf{r}}_{\mathrm{sat}} = \mathbf{r}_{\mathrm{sat}} +
  \mathbf{r}_{\mathrm{err}},
\]
where $\mathbf{r}_{\mathrm{err}}$ is constant for each flyover and points in a
random direction.

\begin{figure}
  \hspace{-62pt}
  \begin{subfigure}{.42\textwidth}
    \includegraphics[width=\textwidth]{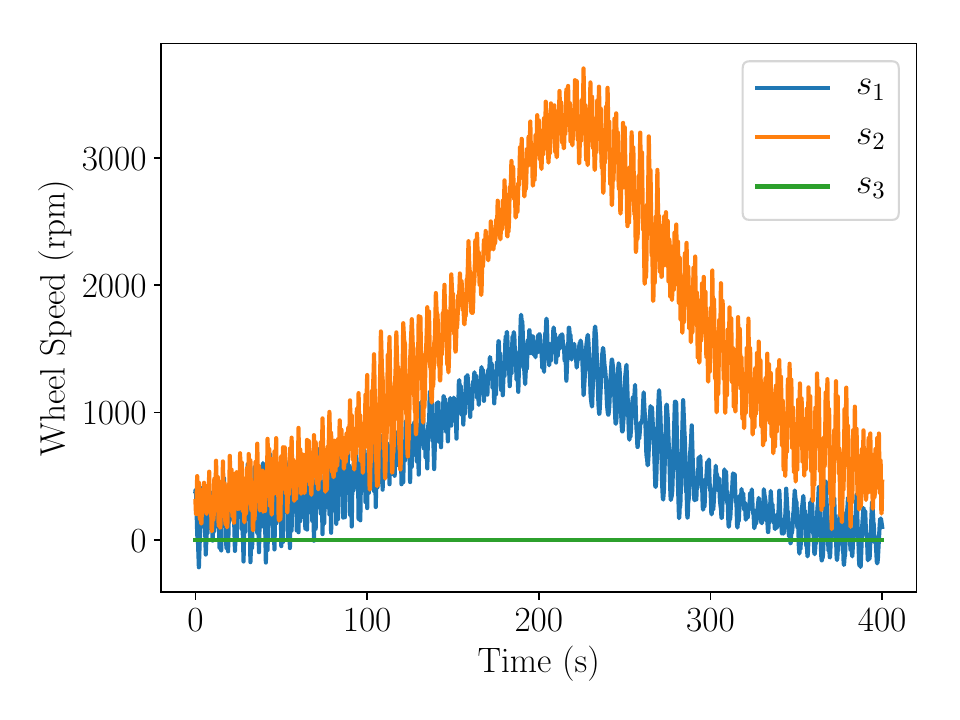}
    \caption{Wheel speed profiles.}
    \label{fig:eg_speeds_w_noise}
  \end{subfigure}%
  \begin{subfigure}{.42\textwidth}
    \includegraphics[width=\textwidth]{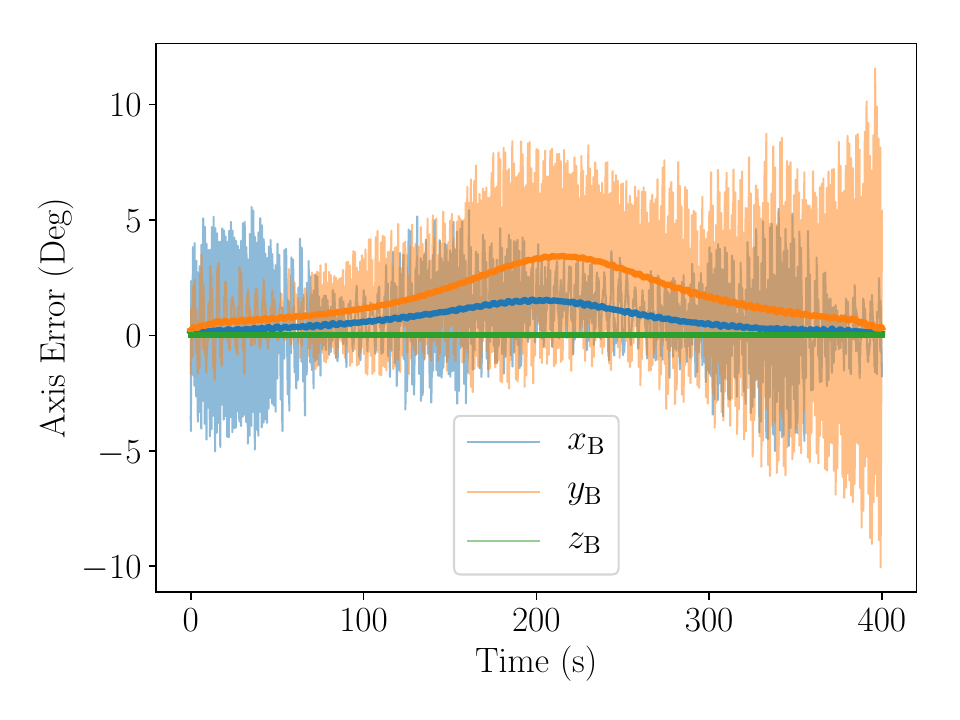}
    \caption{Pointing error in the body frame.}
    \label{fig:eg_error_w_noise}
\end{subfigure}%
\begin{subfigure}{.42\textwidth}
  \includegraphics[width=\textwidth]{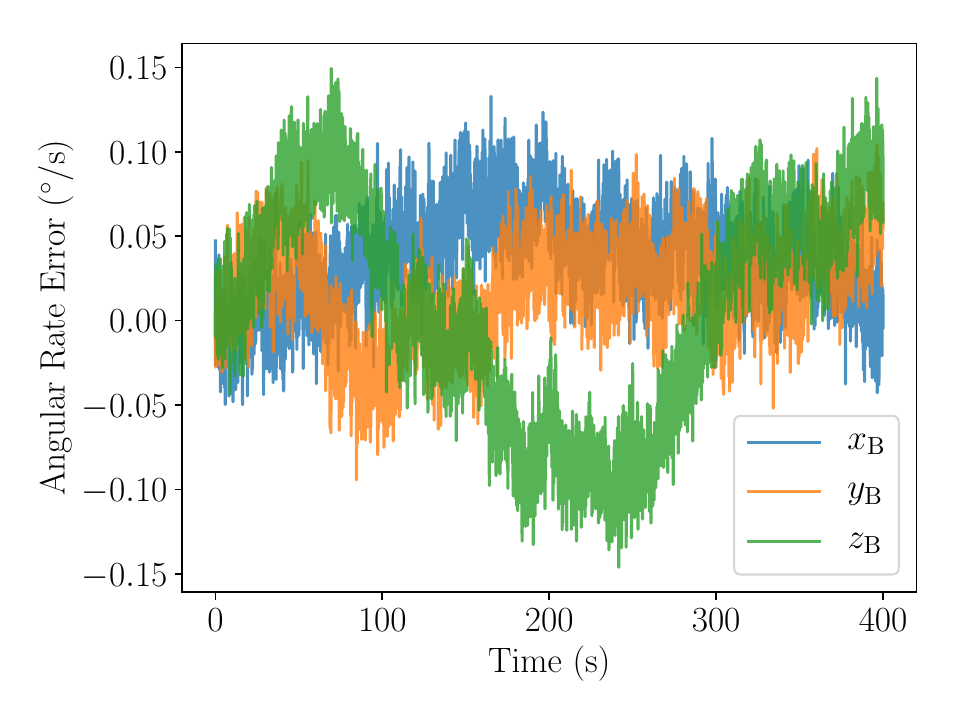}
  \caption{Isolated gyro noise.}
  \label{fig:gyro_noise}
  \end{subfigure}%
  \caption{Time series plots of the three wheel speeds and body axis errors with added
  gyro noise, shown isolated on the right. The standard deviation in this case
was 0.0437$^\circ$/s, similar to the expected value from Table
\ref{tab:cont_inputs}. This was achieved with
$\sigma_{\mathrm{g}}=0.0093^\circ$/s. The axis error plots have been
smoothed to show their averged effect.}
\label{fig:eg_time_series_w_noise}
\end{figure}

The gyro experiences two forms of error. The first is white noise, $\mathbf{w}$, which has a
high frequency and comes from thermo-mechanical events within the system.
However, this has a mean error of 0$^\circ$/s and so does not contribute to the
gyro's bias, which is the error over a long period. 
The second cause of error is bias instability, $\mathbf{b}$, which
comes from flicker noise in the electronics and is of a lower frequency. This
error depends on its previous values and so, like the position error, will not
give a mean error of 0$^\circ$/s. Using a common formulation for gyro noise \cite{w_err},
angular velocity with noise is given as
\[
  \bar{\boldsymbol{\omega}}_{n} = \boldsymbol{\omega}_{n}+
  \mathbf{w}_n+\mathbf{b}_n,
\]
where
\begin{align*}
  \mathbf{w}_n &= \sigma_{\mathrm{w}}\boldsymbol{\kappa}_n\\
  \mathbf{b}_n &= \mathbf{b}_{n-1}+\sigma_{\mathrm{b}}\boldsymbol{\kappa}_n\\
  \boldsymbol{\kappa}_n &\sim \mathcal{N}^3(0,1)\\
\sigma_{\mathrm{w}} &= \frac{\sigma_{\mathrm{g}}}{\sqrt{\Delta t}}\\
\sigma_{\mathrm{b}} &= \sigma_{\mathrm{g}}\sqrt{\Delta t},
\end{align*}
where noise coefficient, $\sigma_{\mathrm{g}}$, is varied to control the magnitude of gyro
noise, and $\Delta t$ is the time step used in the controller. This formulation
is beneficial as it allows for the use of just one parameter when controlling gyro
noise. The standard deviation of the gyro measurments from their correct value is given to be
$4.5\times\sigma_{\mathrm{g}}$. Hence, to model the expected standard deviation
in the gyro readings, $\sigma_{\mathrm{g}}=0.042/4.5=0.0093$. The example
flyover from Figure \ref{fig:eg_flyover} can now be repeated but
with gyro noise added in. Three time series plots from this flyover
are shown in Figure \ref{fig:eg_time_series_w_noise}. The
axes error values shown in Figure \ref{fig:eg_error_w_noise} are far more
chaotic than before. This is a result of the extrapolation of attitude using
the noisy gyro readings. The stronger lines plotted on top show the smoothed
curves of their respective axes. This demonstrates how, although the sensor may
provide very noisy data, most of this has zero mean, so will largely cancel out
over the orientation update period of 1 second, in this case.

To simulate noise in the attitude sensors, a rotation quaternion will be
constructed such that it rotates about a random axis, $\hat{\mathbf{n}}$, in the $x$-$y$ plane of
the body frame by an angle that follows a normal distribution with zero mean.
Recall that consecutive rotations are achieved by multiplying rotation quaternions
together, with the rightmost quaternion operating first.

This gives the orientation quaternion, with noise, to be
\begin{align*}
  \bar{q} &= q_{\mathrm{err}}\odot q\\
  q_{\mathrm{err}} &= \cos \frac{\theta_{\mathrm{err}}}{2} +
  \hat{\mathbf{n}}\sin \frac{\theta_{\mathrm{err}}}{2}\\
  \theta_{\mathrm{err}} &\sim \mathcal{N}(0,\sigma_{\mathrm{a}}),
\end{align*}
where $\sigma_{\mathrm{a}}$ is the standard deviation of the attitude value from its
correct value and the axis of rotation, $\hat{\mathbf{n}}$, is perpendicular to
$z_{\mathrm{B}}$.

\begin{figure}
  \vspace{-40pt}
  \centering
  \includegraphics[width=0.64\textwidth]{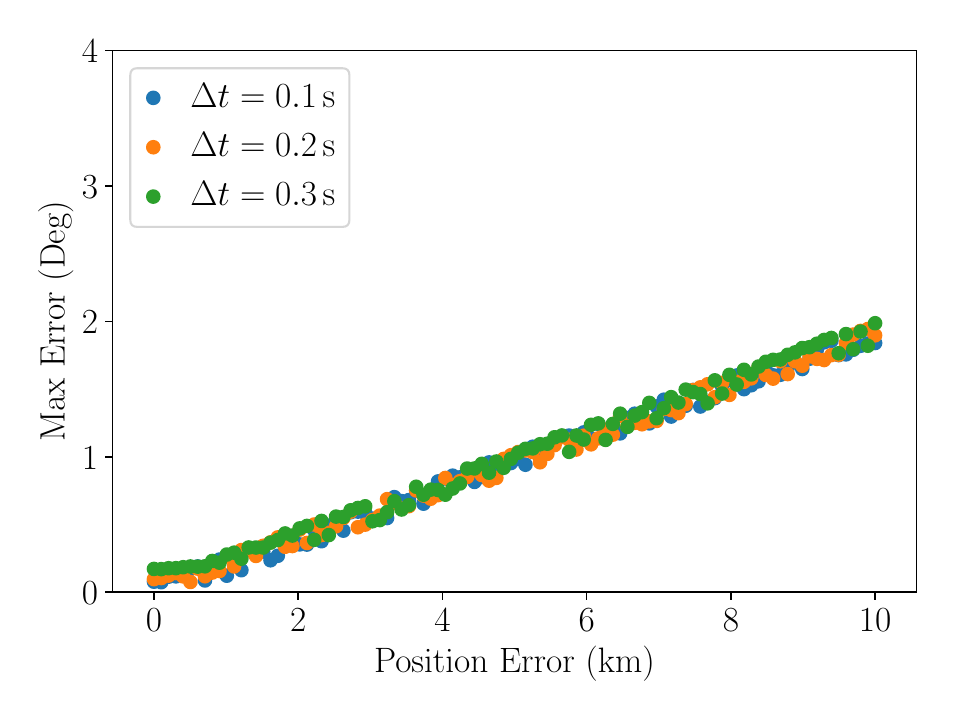}
  \includegraphics[width=0.64\textwidth]{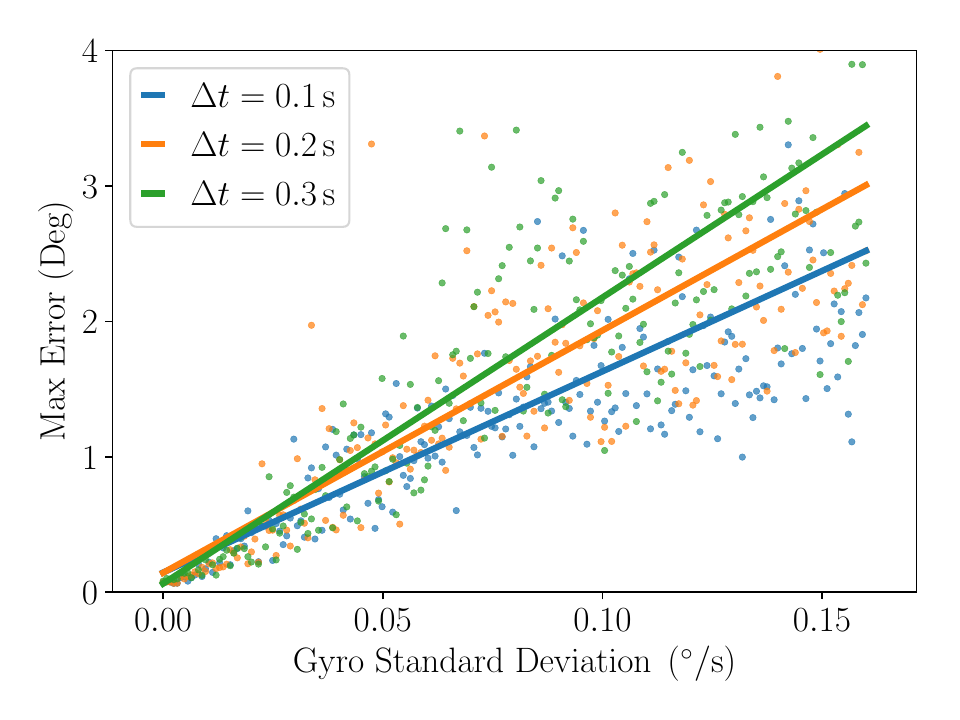}
  \includegraphics[width=0.64\textwidth]{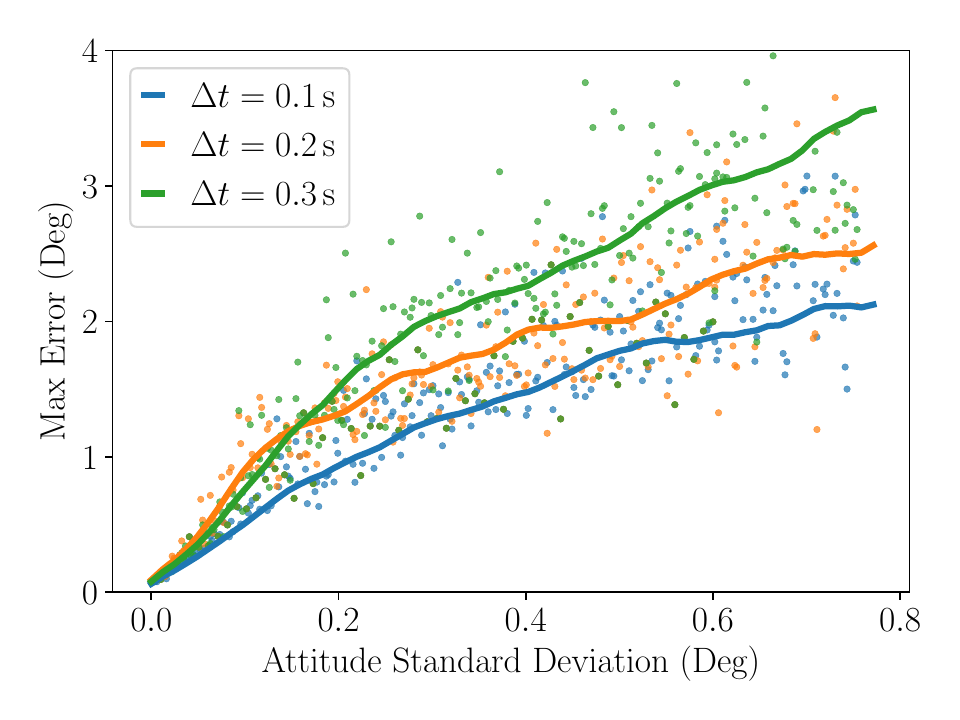}
  \caption{This shows the different effects that errors in one input have on
    the maximum pointing error during a flyover from
      directly above. Controllers with update times of 0.1, 0.2, or 0.3 
    seconds are compared. Due to the stochastic nature of gyro and attitude sensor noise, the
    data has been approximated to emphasise their effect on pointing error. In the case of gyro noise, a linear model could
  be fit, whereas for attitude noise, a nonlinear model was necessary.}
  \label{fig:1d_sweeps}
\end{figure}

With expressions for the input errors now established, their magnitudes can be
individually
varied to see how the controller performs. This will provide an understanding
of the safe margins in which the satellite can still accurately orientate itself
towards the ground target.
The relationship between error in each input and the maximum pointing error to
occur during a flyover is shown in Figure \ref{fig:1d_sweeps}

The top plot shows a strong linear relationship between position error and
maximum pointing error. This is due to the approximately constant value that
the error will have during the flyover duration, causing a constant offset to
the direction that the satellite points. 
Position error will have the greatest effect when it places the
satellite either directly ahead, or behind, its actual location in the orbit.
In this case, the angle discrepancy is largest when the estimated satellite position is
on the opposite side of the target to its actual position. This angle
discrepancy is then proportional to the magnitude of the
position error, using the small angle approximation.
The plot shows that position
error can become twice that of the estimated upper bound of 2\,km, whilst maintaining a
pointing error below 1$^\circ$. This plot also shows that the impact of
position error is independent of the update frequency of the controller.

Looking next at the impact of noise in the gyro, shown in the second plot of
Figure \ref{fig:1d_sweeps}. A line of best fit has been used to emphasise the
linear relationship between noise magnitude and
pointing error. The expected standard deviation in the gyro readings is
$0.042^\circ$/s. This value corresponds to a maximum pointing error of approximately
0.8$^\circ$ for all three controllers. With the target of achieving a pointing
error of 1$^\circ$, this leaves only a small margin of error for the other
two inputs. The significant effect that gyro noise has on the system's pointing
accuracy can be expected due to its use in extrapolating attitude between
sensor updates. This results in both the attitude, and the angular velocity, of
the satellite depending on the accuracy of the gyro. Despite this, it is
preferable to use extrapolation, since otherwise the controller 
is only using a time accurate value for attitude once per second at most.
This results in significant oscillations in
attitude throughout the flyover as the satellite periodically readjusts to the
sparsely updated attitude value.

Finally, looking at the impact of noise in the attitude sensors, shown in the
bottom plot of Figure \ref{fig:1d_sweeps}. In this plot, a
smoothing function was used to to emphasise the relationship between errors
whilst preserving the nonlinear shape of the plots. The
steeper gradient towards the start is due to the maximum angular rate
that can be achieved by the satellite. In Section \ref{sec:control_terms}, this
was estimated at
1.55$^\circ$/s, putting a limit on the angular displacement that can result
from a single incorrect attitude measurement. Therefore, in the case that
$\Delta t=0.2$\,s, the satellite could only rotate approximately 0.31$^\circ$
over one cycle, thus mitigating the effect of larger errors in particular. The
pointing error will still increases with noise, however, since the errors can
compound to produce a longer perturbation from the correct attitude.
The effect of attitude sensor noise is similar to that of gyro noise.
However, within the range of their expected values, 0.25$^\circ$ and 0.042$^\circ$/s
respectively, attitude noise is shown to have a stronger effect on pointing
error than gyro noise. 

Additionally, towards the upper limit of estimated
noise, the maximum pointing error is shown to exceed 1$^\circ$. In the
worst-case scenario, the controller with an update time of 0.3 seconds reaches
a maximum error of almost 2$^\circ$. Although the proposed pointing error has
been surpassed, it is important to note that these error terms are generated
using the direct standard deviations from the individual sensors. No
additional filtering has been performed using the combined inputs to reduce the
weight given to extreme values.
In practice, a Kalman filter equiped with a model of expected motion in
will be utilised. This allows for rapidly deviating readings to be smoothed about
an estimated average, causing less errors in the inputs given to the
controller. The use of unfiltered error values provides a worst case scenario
in which to understand their effect. By aiming to attain the target pointing accuracy
under these conditions provides a safe margin for unforseen error during actual
operation.

\begin{figure}
  \vspace{-30pt}
  \centering
  \includegraphics[width=0.64\textwidth]{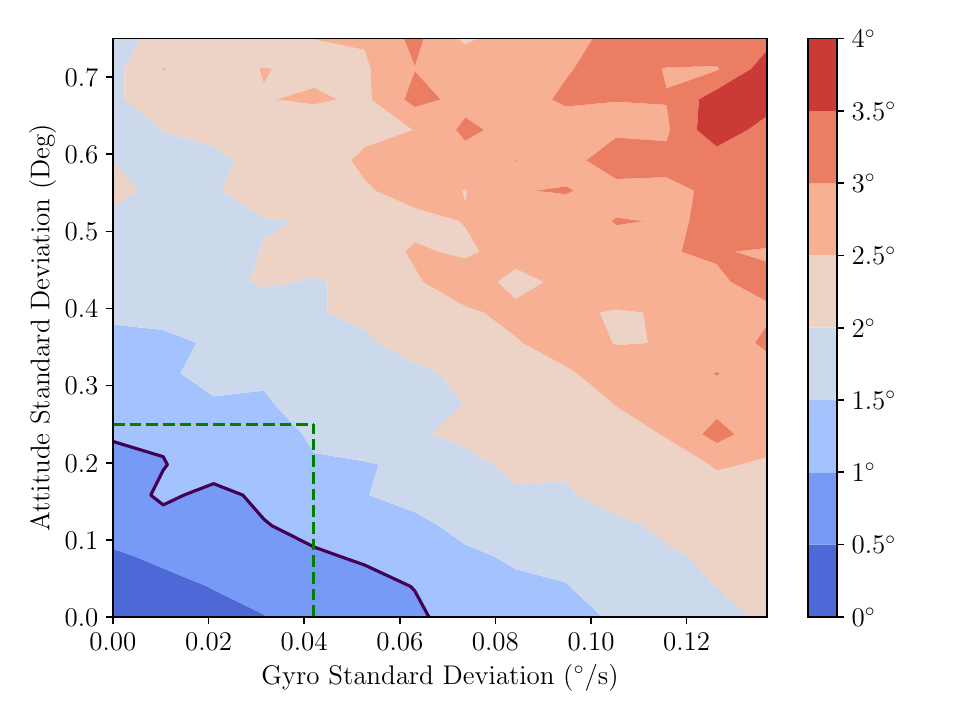}
  \includegraphics[width=0.64\textwidth]{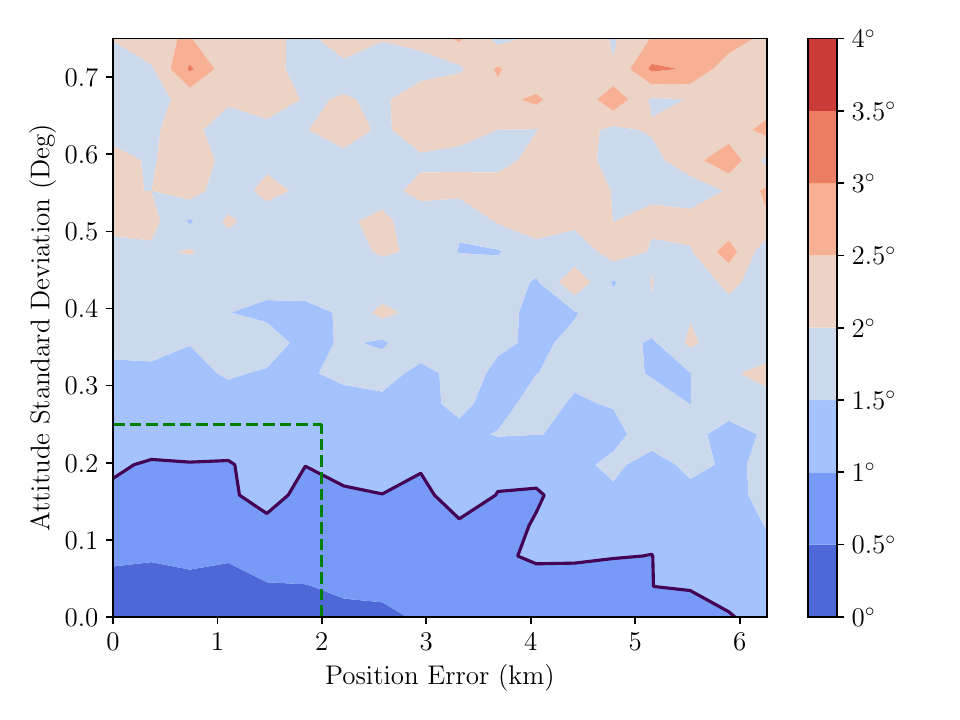}
  \includegraphics[width=0.644\textwidth]{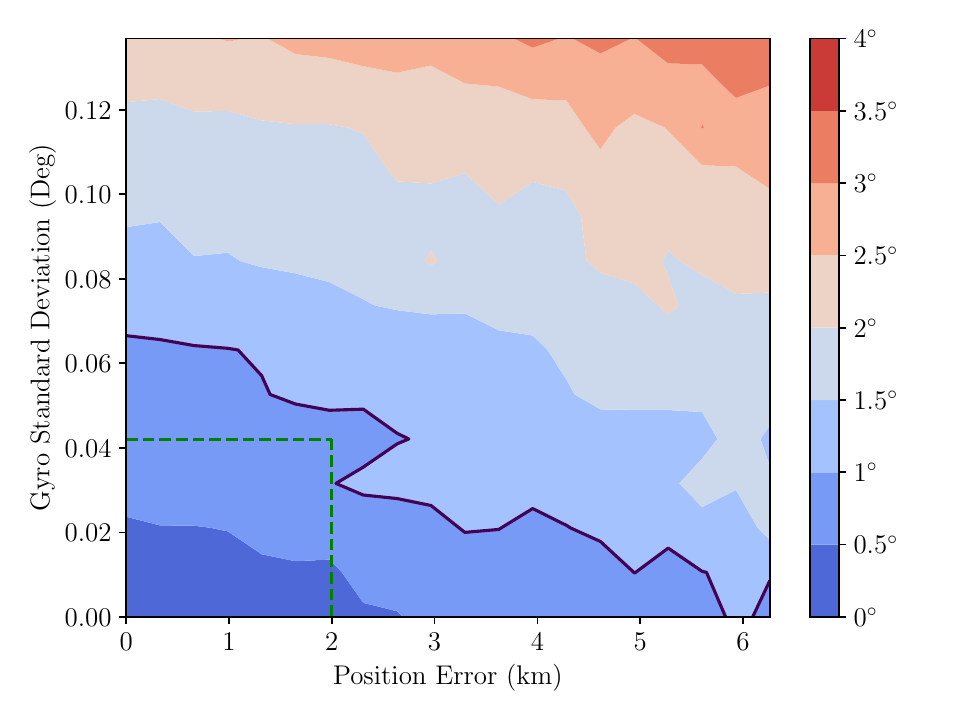}
  \caption{This shows the combined effects of errors present in two inputs at
    once. The maximum pointing error to occur during a flyover from directly
    above is represented by colour. The green box shows the expected range for
    the input errors, whilst the dark purple contour represents where the proposed
  maximum pointing error of 1$^\circ$ has been exceeded.}
\label{fig:2d_sweeps}
\end{figure}

With the effect of each individual sensor's noise quantified, their combined
effect wil now be assessed.
Figure \ref{fig:2d_sweeps} shows the maximum pointing error to occur during
flyovers from directly above as each pair of inputs is varied from zero up to
three times their estimated upper limit. This allows for the safe margins of sensor error
magnitudes to be determined, whilst taking account of the compounding
effect that may occur when multiple sources of error are present. 

The top plot of Figure \ref{fig:2d_sweeps} shows the maximum pointing error to occur
as gyro noise and attitude noise are varied. This
two dimensional sweep shows how a larger pointing error arises when noise is
present in both sensor inputs. As with the one dimensional sweep, attitude
sensor noise is shown to cause a steeper gradient in pointing error at lower 
magnitudes whilst pointing error increases linearly with gyro noise. The green
box shown in the bottom left corner indicates the range of noise that these
sensors are estimated to experience. Within this box, the maximum pointing errors
reach up to $1.54^\circ$, exceeding the proposed limit of $1^\circ$ shown by
the darkened contour. At the top
right corner of the plot, when both sensors experience three times that of their
expected noise, the
pointing error is shown to reach approximately $4^\circ$. This is similar to
the sum of pointing errors from each input's one dimensional sweep of noise
magnitude at these respective values. This suggests that there is no extra error caused
by the two inputs interacting with each other, giving a linear relationship
between net input error and resultant pointing error.

The combined effect of noise in the attitude sensors and
position error is shown in the middle plot of Figure \ref{fig:2d_sweeps}.
Firstly, the leftmost vertical slice of this plot shows
similar pointing errors to that of the top plot, which
corresponds to the single parameter sweep of attitude noise in Figure
\ref{fig:1d_sweeps}. However, moving horizontally, as position error is increased, 
the maximum pointing error increases only slightly. This relates to the
smaller effect that position error has on pointing accuracy,
when compared to the noise present in the other two inputs.

The third plot of Figure \ref{fig:2d_sweeps} shows the combined effect of 
gyro noise and position error.
A similar response is shown to that of the previous plot, with position error
causing only a small increase in pointing error when compared to the effect of
gyro noise. This
can again be attributed to the relatively small effect that position error has,
within its estimated range. Once again, the maximum value of pointing error here
is approximately equal to the sum of pointing errors 
from each inputs isolated noise sweep at their respective values.

\vspace{10pt}
These dual parameter sweeps show that pointing error is
proportional to the sum of input errors, as opposed to their product. A
positive relationship between input errors and pointing errors was always expected,
hence identifying this to be linear, within realistic ranges, is a positive
outcome. Although the controller update time used is these sweeps was set to
the default value of 0.1 seconds, the same linear relationship was shown for update
times $\Delta t = 0.2$\,s and $\Delta t = 0.3$\,s when using their respective
error values from the one dimensional sweeps.

The two plots that include attitude sensor noise show pointing error to exceed
the target of $1^\circ$ within the range of expected noise values. This was
expected, given the isolated plot of attitude error also exceeded the
target pointing accuracy. However, as mentioned before, these errors
represent the inputs before any filtering has been performed on the combined
values. The degree to which input error can be reduced has not yet been
quantified for the PROVE mission. Hence, in order to avoid over-reliance on
this filtering, it was omitted completely. The expected effect of this is to
attain larger pointing errors whilst still capturing the approximate behaviour
of the controller to the inputs, i.e., a linear relationship between the sum of
input errors and resultant pointing error.

\section{Conclusion}
The primary aim of this project was to develop the current system of equations
that were being used for attitude control within the PROVE mission. This
involved the incorporation of gyroscopic torques that arise from the
individually rotating flywheels. Additionally, a method by which to optimise
wheel accelerations in the case of a redundant fourth wheel was derived.

Improvements to the state-of-the-art quaternion based controller were then
explored. Firstly, an integral term was added to the control law. This allowed
for small residual errors to be recognised and reduced by integrating them over
time. The second improvement aimed to derive equations for optimal controller
gains such that tuning could be done automatically. By defining the gains in
terms of the satellite's parameters, this controller design could still be used
once the system architecture is finalised. 

To achieve this, the system was first linearly approximated about its
equilibrium. From there, the Jacobian was rearranged to attain a block
diagonal matrix, shown in Appendix \ref{sec:reordered_jacobian}. This meant
that eigenvalue analysis could be performed on
just one submatrix, since the result is transferable to the other two. Without
this, performing eigenvalue analysis on the entire 13$\times$13 Jacobian would not be
feasible.

By solving for the gains such that the characteristic polynomial of the system
yielded repeated eigenvalues, maximum stability could be achieved. This produced
equations for the necessary relationship between the gains in order to maintain
stability. The three gains could then be expressed in terms of just one.
Using this new representation, the system was then discretised, 
allowing for the controller's update time to be factored into the Jacobian. By
setting the repeated eigenvalue of this matrix to zero and solving for the
the gain value, a critically damped control law was attained. Treating the
system as linear, along with the assumption that the target is stationary, meant that the
controller was unstable in the presence of larger pointing errors. This was
fixed by reducing the gains proportionally using a gain coefficient.
With this adjustment, the controller performed better than could be
achieved with manual tuning methods.

After developing the controller, analysis could be performed on the
topics proposed for this project. Starting with external torque, each
potentially significant source
was assessed, arriving at an estimate for the maximum torque that will be
experienced during a flyover. Simulations were run to assess the effect this
has when at its estimated maximum value, working directly against the rotation of the satellite. The results
showed that, due to the relatively slow rate at which the external torque changes,
the controller was able to correct for it, maintaining the same pointing
accuracy as without external torque. The most notable effect that external torque had on
the system was the increase it caused in the maximum speed reached by the flywheels.
In the worst case, this was shown to result in wheel speeds reaching
8325\,rpm, increasing by 12\% from the case of no external torque. This
impacted the gyroscopic effect of the flywheels, as it depends on their
angular velocity.

The effect of gyroscopic torque within the system was previoulsy unexplored
within the PROVE mission. By isolating the torques during a flyover in which
maximum external torque was present, the gyroscopic effect was shown to reach a
magnitude of $10^{-7}$\,Nm at the peak of the flyover. This is approximately
two orders of magnitude smaller than the intentional torque from the
wheels during that period. Importantly, the controller maintained the
same pointing accuracy as in the idealised case when gyroscopic torque was 4
orders smaller than the intentional torque. This result showed that gyroscopic
reactions within the system are not significant enough to impact the pointing
accuracy of the satellite.

In order to validate that the automatically tuned controller could perform in
realistic conditions, a series of simulations were run to determine the
most significant factors impacting pointing error. Initial simulations, in
idealised conditions, showed that the update time of the controller had a greater effect on the pointing
error than the update time of the attitude sensors. This is because the
gyro readings can be used to extrapolate the attitude of the satellite in the
absence of an explicitly calculated value. The gyro is updated with
the controller, hence this update frequency becomes the limiting factor of performance
when sensor noise is omitted.

The controller depends on three variables to calculate flywheel torque. These
are the orientation, the angular velocity, and the position of the satellite.
These are calculated using light sensors on the exterior of the satellite, a gyro sensor,
and trajectory parameters from a TLE, respectively.
Each of the inputs will be subject to some
degree of error. A model with which to simulate the error for each input was
then developed, allowing for the magnitude of this error to be varied. From there, the effect
of each input error on pointing accuracy was compared. This showed attitude
sensor noise to be the most significant factor in causing pointing error.
Because this is updated sparsely, extrapolation from the gyro is required,
meaning any error in the attitude sensors will be propagated over the
extrapolation process until a new explicit value is attained from the attitude
sensors.

The effect of two input errors was then evaluated. The most important result
from these simulations was finding that pointing error was linearly related to
the sum of input errors, as opposed to a higher order relationship. This
allows for a linear model of total input error to be used in estimating the
maximum pointing error to occur. For example, if all inputs experience their
largest expected magnitude of error, the maximum pointing error is given to be
approximately 2.1$^\circ$. This uses the default simulation parameters shown in
Appendix \ref{sec:params_table}. 

Because no filtering was used on the inputs
before being fed into the controller's feedback loop, this acts as an upper limit
on expected pointing error. In reality, some degree of error reduction will be
possible by comparing the readings to a measure of expected variance for that
input, and attributing lower significance to values outside of its expectede
range.

This project has succesfully produced an improved model of the satellite's
dynamics. The model was then utilised to develop an improved control law along with a
method by which to automatically tune its gains. This controller was 
analysed through simulation, sourcing parameter values from other PROVE mission
projects and external research. A future improvement that could be made to this
controller would be to incorporate the known equations of motion into its
control law. This would consist of using the trajectory of the
satellite to calculate the expected rotation that will be needed during the
flyover. From here, sensor input could be less heavily weighted, acting more
for small corrections. This would reduce the effect of their errors
significantly. 

More research is also needed with respect to the external torques that 
will act on this CubeSat, since only approximate magnitudes have been
used so far. This is partly due to significant hardware choices in the satellite
that are unfinalised. For example, solar panels that extend from the
satellite are being considered, which  will significantly effect the
aerodynamic torque that acts on the satellite.

In addition to the physical design of the satellite, the performance of the
internal components will also need more research. By understanding the
limitations of the satellite's hardware, methods by which to reduce demand on
these components can be developed. For example, if the magnetorquer is found to be
more energy efficient than the flywheels, then it may be suitable for constant
torque generation during the satellite's standby mode in order to maintain fixed orientation.

\section{Appendices}
\appendix
\section{}
\subsection{Triple Vector Product to Matrix}
\label{app:A matrix}
Here is the full derivation through which the matrix vector product was
produced to represent the triple vector product from (\ref{eq:L_w with
A}). In the general case, when a vector triple product contains the same vector,
$\mathbf{a}$, sandwiching a different one, $\mathbf{b}$, the product can be
written as
\begin{align*}
  \mathbf{a}\times(\mathbf{b}\times\mathbf{a}) &=
  \begin{bmatrix}a_1\\a_2\\a_3\end{bmatrix}\times\left(
  \begin{bmatrix}b_1\\b_2\\b_3\end{bmatrix}\times\begin{bmatrix}a_1\\a_2\\a_3\end{bmatrix}\right)\\
  &=\begin{bmatrix}a_1\\a_2\\a_3\end{bmatrix}\times\begin{bmatrix}b_2 a_3 - b_3
  a_2\\b_3 a_1 - b_1 a_3\\b_1 a_2 - b_2 a_1\end{bmatrix}\\
  &=\begin{bmatrix}a_2(b_1 a_2 - b_2 a_1) - a_3(b_3 a_1 - b_1 a_3)\\a_3(
    b_3 a_1 - b_1 a_3) - a_1(b_1 a_2 - b_2 a_1)\\a_2(b_1 a_2 - b_2 a_1) -
  a_3(b_3 a_1 - b_1 a_3)\end{bmatrix}\\
  &=\begin{bmatrix}(a_2^2+a_3^2)b_1  -a_1 a_2 b_2  - a_1 a_3 b_3\\
                    -a_1 a_2 b_1 + (a_1^2+a_3^2)b_2  -a_2 a_3 b_3\\
                  -a_1 a_3 b_1 -a_2 a_3 b_2 + (a_1^2+a_2^2)b_3\end{bmatrix}\\
  &=\begin{bmatrix}(a_2^2+a_3^2) & -a_1 a_2  & - a_1 a_3 \\
                    -a_1 a_2  & (a_1^2+a_3^2) & -a_2 a_3 \\
                  -a_1 a_3 & -a_2 a_3  & (a_1^2+a_2^2)\end{bmatrix}
                  \begin{bmatrix}b_1\\b_2\\b_3\end{bmatrix}\\
  &=\mathbf{Ab}.
\end{align*}
\subsection{Quaternion Time Derivative}
\label{sec:quat_deriv}
This is a derivation of (\ref{eq:quat_deriv}), using the rules and conventions
discussed in Section \ref{sec:quaternions}. The orientation of the satellite at
time $t+\delta t$ is described by $q(t + \delta t)$. This is achieved by rotating from
orientation $q(t)$ by $\delta\theta=|\boldsymbol{\omega}|\delta t$
about the unit axis
$\hat{\boldsymbol{\omega}}=\frac{\boldsymbol{\omega}}{|\boldsymbol{\omega}|}$,
where $\boldsymbol{\omega}$ is the angular velocity of the satellite at time
$t$ in the inertial reference frame.
This rotation can be written as
\begin{align*}
  \delta q &= \cos\frac{\delta \theta}{2} + \hat{\boldsymbol{\omega}}
  \sin\frac{\delta \theta}{2}\\
  &= \cos\frac{|\boldsymbol{\omega}|\delta t}{2} + \hat{\boldsymbol{\omega}}
  \sin\frac{|\boldsymbol{\omega}|\delta t}{2}.
\end{align*}
Hence the final orientation of the satellite, at time $t+\delta t$, can be
written as the combined rotation $\delta q \odot q(t)$.
The derivative of a quaternion is defined, as with any other variable, as
\[
  \dot{q}(t) = \lim_{\delta t \to 0}\frac{q(t + \delta t) - q(t)}{\delta t}.
  \numberthis \label{eq:deriv_def}
\]
The numerator in (\ref{eq:deriv_def}) can be written as
\begin{align*}
  q(t + \delta t) - q(t) &= \delta q \odot q(t) - q(t)\\
  &= \left(\cos\frac{|\boldsymbol{\omega}|\delta t}{2} + \hat{\boldsymbol{\omega}}
  \sin\frac{|\boldsymbol{\omega}|\delta t}{2}\right)\odot q(t) - q(t)\\
  &= \left(\cos\frac{|\boldsymbol{\omega}|\delta t}{2}-1 + \hat{\boldsymbol{\omega}}
  \sin\frac{|\boldsymbol{\omega}|\delta t}{2} \right) \odot q(t)\\
  &= \left(-2\sin^2\frac{|\boldsymbol{\omega}|\delta t}{4} + \hat{\boldsymbol{\omega}}
  \sin\frac{|\boldsymbol{\omega}|\delta t}{2} \right) \odot q(t).
\end{align*}
The higher order term here can be neglected as $\delta t \to 0$, giving
\begin{align*}
\lim_{\delta t \to 0}\frac{q(t + \delta t) - q(t)}{\delta t} &= \lim_{\delta t
\to 0}\left(\frac{\hat{\boldsymbol{\omega}}
\sin\frac{|\boldsymbol{\omega}|\delta t}{2} \odot q(t)}{\delta t} \right)\\
&=\hat{\boldsymbol{\omega}}\lim_{\delta t
\to 0}\left(\frac{\sin\frac{|\boldsymbol{\omega}|\delta t}{2}}{\delta t}
\right) \odot q(t)\\
&=\hat{\boldsymbol{\omega}}\dfrac{\mathrm{d}}{\mathrm{d}t}
\left(\sin{\frac{|\boldsymbol{\omega}|t}{2}}\right)\bigg\rvert_{t=0} \odot
q(t)\\
&=\hat{\boldsymbol{\omega}}\left(\frac{|\boldsymbol{\omega}|}{2}\cos{\frac{|\boldsymbol{\omega}|t}{2}}\right)\bigg\rvert_{t=0}
\odot q(t)\\
&=\hat{\boldsymbol{\omega}}\frac{|\boldsymbol{\omega}|}{2}\odot q(t)\\
&=\frac{1}{2}\boldsymbol{\omega}\odot q(t)
\end{align*}
Finally, to attain the equation in the form that is used in Section
\ref{sec:eqs_of_motion}, the angular velocity vector must be transformed to the body
frame coordinate axes, giving $\boldsymbol{\omega}_{\mathrm{B}}$. The
relationship between $\boldsymbol{\omega}_{\mathrm{B}}$ and
$\boldsymbol{\omega}$ is described by the rotation quaternion $q(t)$ such that 
\[
  \boldsymbol{\omega} = q(t)\odot \boldsymbol{\omega}_{\mathrm{B}} \odot
  q^{-1}(t).
\]
Applying this to the quaternion time derivative gives
\begin{align*}
  \frac{1}{2}\boldsymbol{\omega}\odot q(t) &=
  \frac{1}{2}q(t)\odot\boldsymbol{\omega}_{\mathrm{B}}\odot q^{-1}(t)\odot q(t)\\
  &= \frac{1}{2}q(t)\odot\boldsymbol{\omega}_{\mathrm{B}},
\end{align*}
where the product, $q^{-1}(t)\odot q(t)$, gives the identity quaternion. The
final equation is therefore
\[
  \dot{q}(t) = \frac{1}{2}q(t)\odot\boldsymbol{\omega}_{\mathrm{B}}.
\]
\section{}
\subsection{Four Wheel Time Series}
\label{sec:four_wheel_plot}

\begin{figure}[h]
  \centering
  \includegraphics[width=\textwidth]{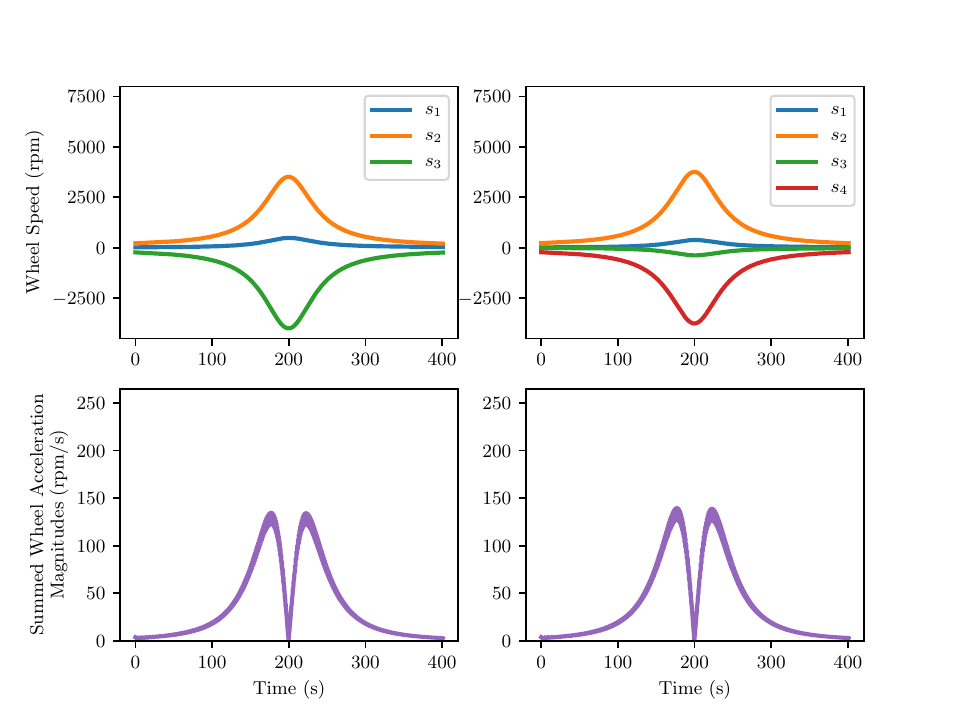}
  \caption{Time series plots from a flyover, showing wheel speeds and the
    sum of acceleration magnitudes. Three and four wheel configurations are
    compared, with the optimisation approach outlined in Section
  \ref{sec:four_wheels} applied to the four wheel case.}
\end{figure}

\section{}
\subsection{Integral Time Constant}
\label{sec:t0}
\begin{figure}[h]
  \hspace{-50pt}
  \begin{subfigure}{0.4\textwidth}
    \includegraphics[width=\textwidth]{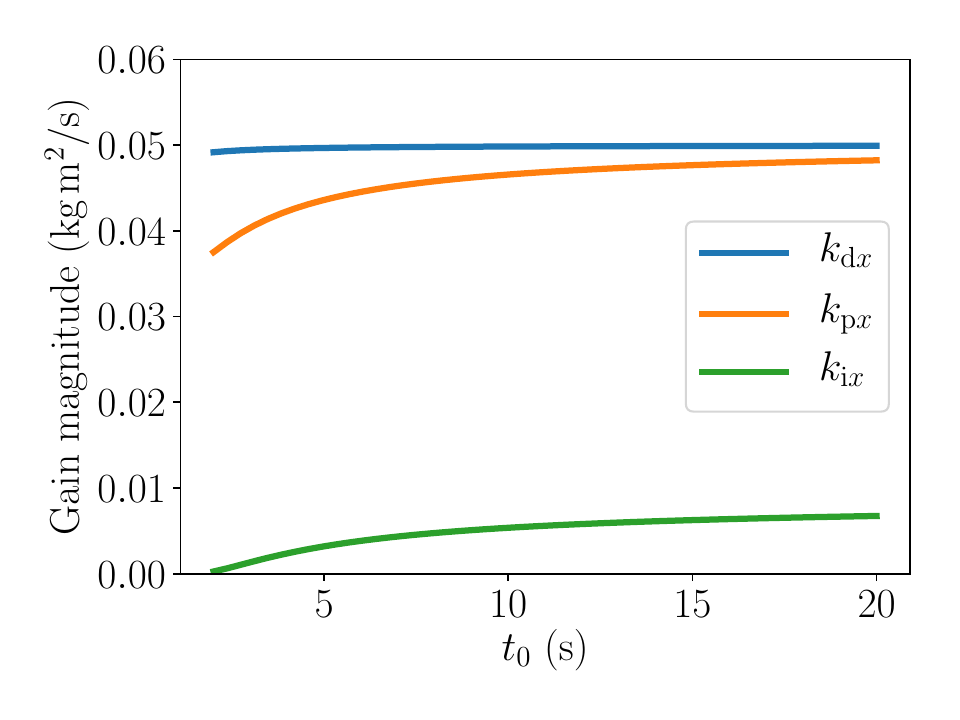}
    \caption{$\Delta t = 0.1$\,s}
  \end{subfigure}%
  \begin{subfigure}{0.4\textwidth}
    \includegraphics[width=\textwidth]{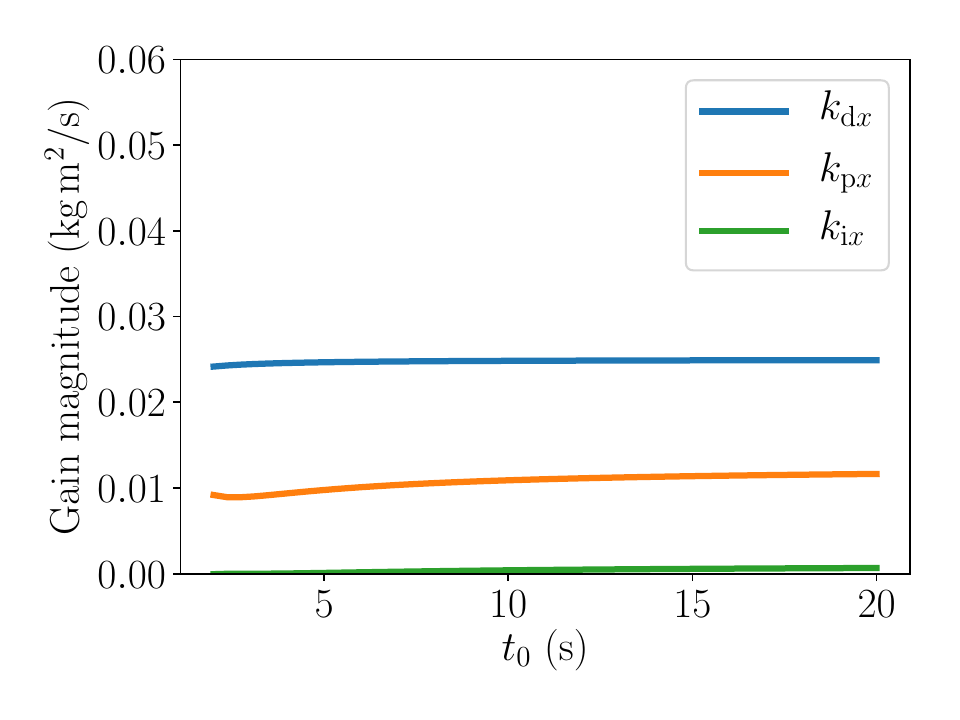}
    \caption{$\Delta t = 0.2$\,s}
  \end{subfigure}%
  \begin{subfigure}{0.4\textwidth}
    \includegraphics[width=\textwidth]{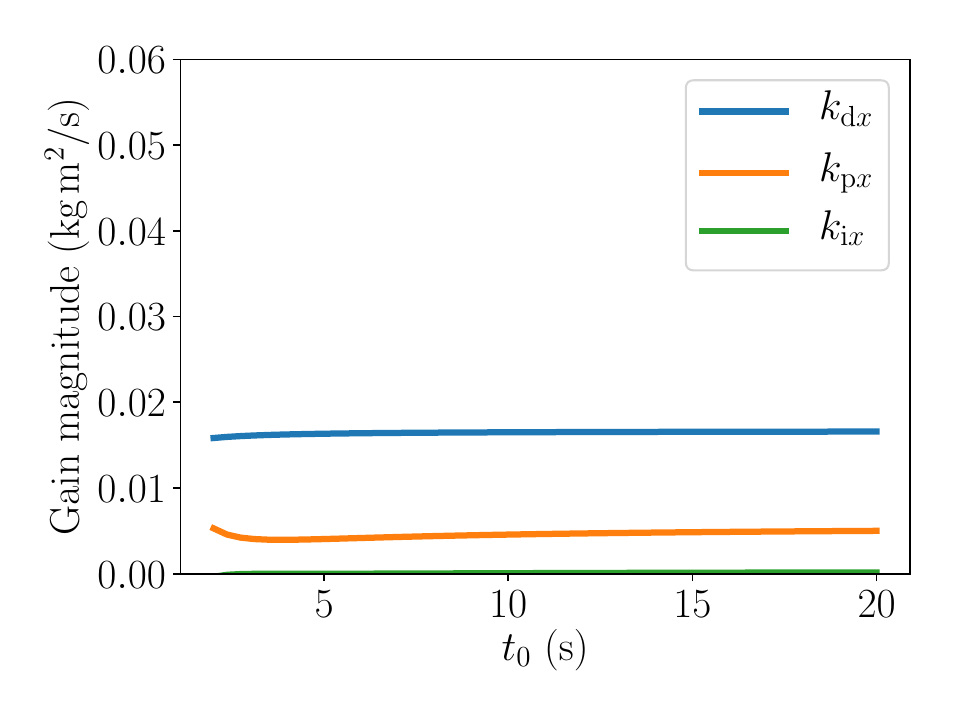}
    \caption{$\Delta t = 0.3$\,s}
  \end{subfigure}%
  \caption{Showing the minimal effect of the integral time constant, $t_0$, on
    the magnitude of the gains for the $x$-axis. Three controller update times are shown,  
  demonstrating the significant impact of $\Delta t$ in comparison.}
  \label{fig:int_time_const}
\end{figure}

Figure \ref{fig:int_time_const} shows the relatively small effect had by the integral time constant,
$t_0$, on the gains. Three controller update times are shown, demonstrating how
much larger the effect of $\Delta t$ is. The integral time constant was
chosen to be 10\,s, due to the time scale over which the satllite orientation
changes. However, similar results were found for different values.

\subsection{Decomposed System of ODEs}
\label{sec:decomposed_odes}
Below is the decomposed set of equations that are equivelent to
(\ref{eq:all_odes})

\[ 
  \left[\begin{matrix}\dot{w}_{1}\\\dot{w}_{2}\\\dot{w}_{3}\\\dot{q}_{1}\\\dot{q}_{2}\\\dot{q}_{3}\\\dot{q}_{4}\\\dot{s}_{\mathrm{E}1}\\\dot{s}_{\mathrm{E}2}\\\dot{s}_{\mathrm{E}3}\\\dot{g}_{1}\\\dot{g}_{2}\\\dot{g}_{3}\end{matrix}\right]
  = 
  \left[\begin{matrix}\frac{1}{I_x} \left(s_{\mathrm{E}2} \omega_{3} - s_{\mathrm{E}3} \omega_{2} + I_y
  \omega_{2} \omega_{3} - I_z \omega_{2} \omega_{3} - g_{1} k_{\mathrm{i}x} - k_{\mathrm{d}x} \omega_{1} - k_{\mathrm{p}x}
  q_{2}\right)\\\frac{1}{I_y} \left(- s_{\mathrm{E}1} \omega_{3} + s_{\mathrm{E}3} \omega_{1} - I_x \omega_{1}
  \omega_{3} + I_z \omega_{1} \omega_{3} - g_{2} k_{\mathrm{i}y} - k_{\mathrm{d}y} \omega_{2} - k_{\mathrm{p}y}
  q_{3}\right)\\\frac{1}{I_z} \left(s_{\mathrm{E}1} \omega_{2} - s_{\mathrm{E}2} \omega_{1} + I_x \omega_{1} \omega_{2}
  - I_y \omega_{1} \omega_{2} - g_{3} k_{\mathrm{i}z} - k_{\mathrm{d}z} \omega_{3} -
k_{\mathrm{p}z} q_{4}\right)\\-\frac{1}{2}\left(q_{2}
\omega_{1} + q_{3} \omega_{2} + q_{4} \omega_{3}\right)\\\frac{1}{2}\left(q_{1}
  \omega_{1} + q_{3} \omega_{3} - q_{4} \omega_{2}\right)\\\frac{1}{2}\left(q_{1}
  \omega_{2} - q_{2} \omega_{3} + q_{4} \omega_{1}\right)\\\frac{1}{2}\left(q_{1}
  \omega_{3} + q_{2} \omega_{2} - q_{3} \omega_{1}\right)\\ - k_{\mathrm{p}x} q_{2}- k_{\mathrm{i}x}g_{1}  - k_{\mathrm{d}x}
  \omega_{1}\\ - k_{\mathrm{p}y} q_{3}- k_{\mathrm{i}y}g_{2}  - k_{\mathrm{d}y} \omega_{2}\\ - k_{\mathrm{p}z} q_{4}- k_{\mathrm{i}z}g_{3}  - k_{\mathrm{d}z}
  \omega_{3}\\q_{2}- \frac{g_{1}}{t_{0}}\\
  q_{3}- \frac{g_{2}}{t_{0}} \\q_{4}- \frac{g_{3}}{t_{0}}\end{matrix}\right] 
\] 

\subsection{Jacobian}
\label{sec:jacobian}
Below is the Jacobian matrix that results from differentiating the above
equations by each of the state variables. 
\[
  \hspace{-50pt}
  \left[\begin{array}{ccccccccccccc}- \frac{k_{\mathrm{d}x}}{I_{x}} & \frac{- s_{\mathrm{E}3} + I_{y} \omega_{3} - I_{z} \omega_{3}}{I_{x}}
   & \frac{s_{\mathrm{E}2} + I_{y} \omega_{2} - I_{z} \omega_{2}}{I_{x}}
   & 0 & - \frac{k_{\mathrm{p}x}}{I_{x}}
  & 0 & 0 & 0 & \frac{\omega_{3}}{I_{x}} & - \frac{\omega_{2}}{I_{x}} & -
  \frac{k_{\mathrm{i}x}}{I_{x}} & 0 & 0\\\frac{s_{\mathrm{E}3} - I_{x} \omega_{3} +
  I_{z} \omega_{3}}{I_{y}}  & - \frac{k_{\mathrm{d}y}}{I_{y}} & \frac{- s_{\mathrm{E}1}
  - I_{x} \omega_{1} + I_{z} \omega_{1}}{I_{y}}  & 0 & 0 & - \frac{k_{\mathrm{p}y}}{I_{y}} & 0 & -
  \frac{\omega_{3}}{I_{y}} & 0 & \frac{\omega_{1}}{I_{y}} & 0 & - \frac{k_{\mathrm{i}y}}{I_{y}} &
  0\\\frac{- s_{\mathrm{E}2} + I_{x} \omega_{2} - I_{y} \omega_{2}}{I_{z}}  &
  \frac{s_{\mathrm{E}1} + I_{x} \omega_{1} - I_{y} \omega_{1}}{I_{z}}  & -
  \frac{k_{\mathrm{d}z}}{I_{z}} & 0 & 0 & 0 & - \frac{k_{\mathrm{p}z}}{I_{z}} &
  \frac{\omega_{2}}{I_{z}} & - \frac{\omega_{1}}{I_{z}} & 0 & 0 & 0 & -
  \frac{k_{\mathrm{i}z}}{I_{z}}\\- \frac{q_{2}}{2} & - \frac{q_{3}}{2} & -
  \frac{q_{4}}{2} & 0 & - \frac{\omega_{1}}{2} & - \frac{\omega_{2}}{2} & -
  \frac{\omega_{3}}{2} & 0 & 0 & 0 & 0 & 0 & 0\\\frac{q_{1}}{2} & - \frac{q_{4}}{2}
  & \frac{q_{3}}{2} & \frac{\omega_{1}}{2} & 0 & \frac{\omega_{3}}{2} & - \frac{\omega_{2}}{2}
  & 0 & 0 & 0 & 0 & 0 & 0\\\frac{q_{4}}{2} & \frac{q_{1}}{2} & -
  \frac{q_{2}}{2} & \frac{\omega_{2}}{2} & - \frac{\omega_{3}}{2} & 0 & \frac{\omega_{1}}{2} &
  0 & 0 & 0 & 0 & 0 & 0\\- \frac{q_{3}}{2} & \frac{q_{2}}{2} & \frac{q_{1}}{2}
  & \frac{\omega_{3}}{2} & \frac{\omega_{2}}{2} & - \frac{\omega_{1}}{2} & 0 & 0 & 0 & 0 & 0 &
  0 & 0\\- k_{\mathrm{d}x} & 0 & 0 & 0 & - k_{\mathrm{p}x} & 0 & 0 & 0 & 0 & 0 & - k_{\mathrm{i}x} & 0 &
  0\\0 & - k_{\mathrm{d}y} & 0 & 0 & 0 & - k_{\mathrm{p}y} & 0 & 0 & 0 & 0 & 0 & - k_{\mathrm{i}y} & 0\\0
  & 0 & - k_{\mathrm{d}z} & 0 & 0 & 0 & - k_{\mathrm{p}z} & 0 & 0 & 0 & 0 & 0 & - k_{\mathrm{i}z}\\0 & 0 &
  0 & 0 & 1 & 0 & 0 & 0 & 0 & 0 & - \frac{1}{t_{0}} & 0 & 0\\0 & 0 & 0 & 0 & 0
  & 1 & 0 & 0 & 0 & 0 & 0 & - \frac{1}{t_{0}} & 0\\0 & 0 & 0 & 0 & 0 & 0 & 1 &
  0 & 0 & 0 & 0 & 0 & - \frac{1}{t_{0}}\end{array}\right]
\]
\subsection{Jacobian at Equilibrium}
\label{sec:reordered_jacobian}
Below is the reordered Jacobian matrix from above, now with the equilibrium
values discussed in Section \ref{sec:linearisation} substituted in. The state
variables are now ordered by which axis they correspond to, given as

\[
  \left [ \omega_{1}, \quad q_{2}, \quad g_{1}, \quad s_{\mathrm{E}1}, \quad \omega_{2}, \quad
  q_{3}, \quad g_{2}, \quad s_{\mathrm{E}2}, \quad \omega_{3}, \quad q_{4}, \quad g_{3},
\quad s_{\mathrm{E}3}, \quad q_{1}\right ],
\]
which gives the Jacobian to be 
\[
  \left[\begin{array}{ccccccccccccc}- \frac{k_{\mathrm{d}x}}{I_{x}} & -
  \frac{k_{\mathrm{p}x}}{I_{x}} & - \frac{k_{\mathrm{i}x}}{I_{x}} & 0 & 0 & 0 & 0 & 0 & 0 & 0 & 0
  & 0 & 0\\\frac{1}{2} & 0 & 0 & 0 & 0 & 0 & 0 & 0 & 0 & 0 & 0 & 0 & 0\\0 & 1 &
  - \frac{1}{t_{0}} & 0 & 0 & 0 & 0 & 0 & 0 & 0 & 0 & 0 & 0\\- k_{\mathrm{d}x} & -
  k_{\mathrm{p}x} & - k_{\mathrm{i}x} & 0 & 0 & 0 & 0 & 0 & 0 & 0 & 0 & 0 & 0\\0 & 0 & 0 & 0 & -
  \frac{k_{\mathrm{d}y}}{I_{y}} & - \frac{k_{\mathrm{p}y}}{I_{y}} & - \frac{k_{\mathrm{i}y}}{I_{y}} & 0 &
  0 & 0 & 0 & 0 & 0\\0 & 0 & 0 & 0 & \frac{1}{2} & 0 & 0 & 0 & 0 & 0 & 0 & 0 &
  0\\0 & 0 & 0 & 0 & 0 & 1 & - \frac{1}{t_{0}} & 0 & 0 & 0 & 0 & 0 & 0\\0 & 0 &
  0 & 0 & - k_{\mathrm{d}y} & - k_{\mathrm{p}y} & - k_{\mathrm{i}y} & 0 & 0 & 0 & 0 & 0 & 0\\0 & 0 & 0 & 0
  & 0 & 0 & 0 & 0 & - \frac{k_{\mathrm{d}z}}{I_{z}} & - \frac{k_{\mathrm{p}z}}{I_{z}} & -
  \frac{k_{\mathrm{i}z}}{I_{z}} & 0 & 0\\0 & 0 & 0 & 0 & 0 & 0 & 0 & 0 & \frac{1}{2} & 0
  & 0 & 0 & 0\\0 & 0 & 0 & 0 & 0 & 0 & 0 & 0 & 0 & 1 & - \frac{1}{t_{0}} & 0 &
  0\\0 & 0 & 0 & 0 & 0 & 0 & 0 & 0 & - k_{\mathrm{d}z} & - k_{\mathrm{p}z} & - k_{\mathrm{i}z} & 0 & 0\\0
  & 0 & 0 & 0 & 0 & 0 & 0 & 0 & 0 & 0 & 0 & 0 & 0\end{array}\right].
\]

\section{}
\subsection{Simulation Parameters}
\label{sec:params_table}
This table describes the default values that are used for the parameters in
simulations of the controller. 
{\renewcommand{\arraystretch}{1.2}
\begin{table}[h]
  \hspace{-36pt}
\begin{tabular}{l|r|l|l}
Parameter                & Value                                          &
Unit    & Comment
\\ \hline \hline
Body mass                & 4                                              & kg      & Estimate taken from M. Tisaev's paper \cite{mansur_vals} \\
Body dimensions          & $0.1\times0.1\times0.3$                        & metres  & Exact dimensions of a 3U CubeSat                                           \\
Body centre of mass      & [0,0,0]$^\top$                                    & metres  & Expressed in body frame with origin at satellite centre of mass            \\
Flywheel mass            & 0.0166                                         & kg      & Taken from M. Tisaev's paper \cite{mansur_vals}          \\
Flywheel radius          & 0.0115                                         & metres  & Taken from M. Tisaev's paper \cite{mansur_vals}          \\
Flywheel height          & 0.02                                           & metres  & Taken from M. Tisaev's paper \cite{mansur_vals}          \\
Flywheel centres of mass & [0,0,0]$^\top$                                    & metres  & All three wheels are located at the centre of mass of the satellite        \\
Flywheel rotation axes   & $\boldsymbol{i},\boldsymbol{j},\boldsymbol{k}$ & none    & The 3 wheels rotate about the body frame axes                              \\
Orientation update time  & 1                                              & seconds & This is the proposed minimum update time                                   \\
Controller update time   & 0.1                                            & seconds & This is the proposed minimum update time                                   \\
Initial flywheel speeds  & 0,0,0                                          & rpm     & Before the flyover begins                                                  \\
Maximum flywheel speeds  & 8000                                           & rpm
& Estimated from M. Tisaev's paper \cite{mansur_vals}   \\
Satellite altitude       & 300                                            & km      & Average low Earth orbit altitude                                           \\
Orbit duration           & 1.5                                            & hours   & Corresponding orbit duration at 300\,km altitude           \\
Simulation update time   & 0.1                                            &
seconds & Must be at least that of the controller\\
Gain coefficient ($\rho$)  & 0.05& none & Decided in Section
\ref{sec:linearisation}\\
Integral time constant ($t_0$) & 10 & seconds & Duration over which
$\mathbf{q}_{\mathrm{err}}$ is considered significant\\
External torque & [0,0,0]$^\top$ & N\,m & For the duration of the flyover,
external torque is omitted. 
\end{tabular}
\end{table}
}
\end{document}